\documentclass[sigconf,authorversion,nonacm]{acmart}
\AtBeginDocument{%
  \providecommand\BibTeX{{%
    \normalfont B\kern-0.5em{\scshape i\kern-0.25em b}\kern-0.8em\TeX}}}

\usepackage{multirow}
\usepackage{makecell}
\usepackage{array}
\usepackage{natbib}
\usepackage{subcaption}
\usepackage{pdfpages}
\usepackage{enumitem}
\usepackage{soul}
\soulregister\cite7
\soulregister\ref7

\newcommand{\shortquote}[1]{“\textit{#1}”}
\usepackage{booktabs}

\begin{document}

\title[Benefits of Conversations in Neural Network Explainability]{May I Ask a Follow-up Question? Understanding the Benefits of Conversations in Neural Network Explainability}

\author{Tong Zhang}
\email{tong.zhang@ntu.edu.sg}
\affiliation{
  \institution{Nanyang Technological University}
  \country{Singapore}
}

\author{X. Jessie Yang}
\email{xijyang@umich.edu}
\affiliation{
  \institution{University of Michigan}
  \city{Ann Arbor}
  \state{Michigan}
  \country{USA}
}

\author{Boyang Li}
\email{boyang.li@ntu.edu.sg}
\affiliation{
  \institution{Nanyang Technological University}
  \country{Singapore}
}


\begin{abstract}
Research in explainable AI (XAI) aims to provide insights into the decision-making process of opaque AI models. To date, most XAI methods offer one-off and static explanations, which cannot cater to the diverse backgrounds and understanding levels of users. With this paper, we investigate if free-form conversations can enhance users' comprehension of static explanations in image classification, improve acceptance and trust in the explanation methods, and facilitate human-AI collaboration. We conduct a human-subject experiment with 120 participants. Half serve as the experimental group and engage in a conversation with a human expert regarding the static explanations, while the other half are in the control group and read the materials regarding static explanations independently. We measure the participants' objective and self-reported comprehension, acceptance, and trust of static explanations.
Results show that conversations significantly improve participants’ comprehension, acceptance \citep{davis1989perceived}, trust, and collaboration with static explanations, while reading the explanations independently does not have these effects and even decreases users’ acceptance of explanations. Our findings highlight the importance of customized model explanations in the format of free-form conversations and provide insights for the future design of conversational explanations.

\end{abstract}




\maketitle

\section{Introduction}
The rapid advancement of Artificial Intelligence (AI) is largely powered by opaque deep neural networks (DNNs), which are difficult to interpret by humans \citep{bodria2021benchmarking}. The lack of transparency prevents verification of AI decisions by human domain experts and is especially concerning in areas of high-stake decisions, such as healthcare and law enforcement, where erroneous algorithmic decisions could lead to severe consequences \citep{Caruana2015health, Cai2019health, zheng_designing_2023} and erosion of public trust \citep{quinn2021trust, Powles2017}.
To improve the explainability of AI models, numerous eXplainable Artificial Intelligence (XAI) methods have been proposed (for detailed reviews, we refer readers to \citet{yang2019evaluating, danilevsky-etal-2020-survey, bodria2021benchmarking}). 
It has been reported that explainability enhances user understanding \citep{bansal2021does} and trust \citep{gonzalez-etal-2021-explanations, luo_evaluating_2022} in AI models, improves human-AI collaboration in decision-making \citep{nguyen2022visual, lai2019on}, and helps AI developers identify and rectify model errors \citep{adebayo2020debugging, idahl2021towards}.
Despite these successes, a number of recent studies find that the explanations often do not resolve user confusion regarding the neural networks they are purported to explain \citep{forough2021manipulating, zhang2020effect, bansal2021does, wang2021are, lakkaraju2022rethinking, liao2020questioning, slack2023explaining, shen2023convxai}. These seemingly conflicting findings warrant further investigation. 

We postulate that two major factors contribute to the ineffectiveness of AI explanations. First, the explanations do not properly account for average users’ knowledge of machine learning, which may be insufficient to establish causal relations between the explanations and the model behaviors \citep{forough2021manipulating, ma2023who, springer2019progressive, xin2023what}. 
Communication theory posits that effective communication requires the senders and receivers to establish common ground~\citep{clark1981definite, clark1991grounding}. However, experts usually find it hard to accurately estimate what laypeople know \citep{wittwer2008underestimation, wilkesmann2011knowledge, miller2019explanation}. 
To make matters worse, underestimating and overestimating the receivers' knowledge level are equally detrimental to communication \citep{wittwer2008underestimation, lakkaraju2022rethinking}. 
As a result, the explanations designed by experts are almost always at a mismatch with the laypersons' actual knowledge level. 

Second, users of XAI have diverse intentions and information needs \citep{wang2021are, ehsan2021explainable, liao2020questioning, xin2023what}. For example, \citet{liao2021human} identifies five different objectives of users of explanations, including model debugging, assessing the capabilities of AI systems, making informed decisions, seeking recourse or contesting the AI, and auditing for legal or ethical compliance. One static explanation usually cannot satisfy all objectives and purposes. Therefore, researchers have suggested injecting interactivity to model explanations in order to establish common ground, address knowledge gaps, and create customized explanations that adapt to the users \citep{lakkaraju2022rethinking, Abdul2018trends, schmid2022missing, rohlfing2020explanation, cheng2019explaining, Mouadh2023interactive}. 

Existing work on interactive explanations can be broadly categorized into two types. The first type, interactive machine learning \citep{Amershi_2014, fails2003interactive}, allows users to provide feedback and suggestions to the machine learning model using model explanations. Their primary goal is to improve machine learning performances, rather than explaining model behaviors to layperson users. In this setting, explanations have been shown to improve user satisfaction \citep{smith2020no} and feedback quality \citep{liang-etal-2020-alice, kulesza2015principles}. The second type aims to elucidate model behaviors by allowing users to freely modify input features and observe how outputs change while showing feature attribution explanations \citep{cheng2019explaining, liu2021understanding, tenney-etal-2020-language, hohman2019gamut}. This type of interactivity has been shown to improve user understanding \citep{cheng2019explaining} and perceived usefulness \citep{liu2021understanding} of AI models. However, the effective use of these interactive approaches still requires a rudimentary understanding of machine learning, such as the generic relation between input and output, or what model properties the interpretations reveal. These interactive explanations cannot answer most types of follow-up questions laypeople may have.

Free-form conversations that accompany static explanations are arguably the most versatile mode of interaction as they allow users to ask arbitrary follow-up questions and receive explanations tailored to their backgrounds and needs \citep{liao2020questioning, feldhus2022mediators,lakkaraju2022rethinking}. 
Through interviews with decision-makers, \citet{lakkaraju2022rethinking} discover that they have a strong preference for explanations in natural language dialogue. They argue that conversational explanations satisfy five requirements of interactive explanations and are ideal for users with limited machine learning knowledge. 
With the progress in conversational characters \citep{zhang-etal-2022-history, ni2023recent, shuster2022blenderbot}, especially knowledge-based question answering \citep{ijcai2021p0611, luo-etal-2023-end, zhang-etal-2023-fc} powered by large language models \citep{touvron2023llama, ouyang2022training, LLMSurvey}, AI systems that can answer questions about their own decisions appear to be within our reach in the near future. However, before investing effort to develop such a chatbot, it would be beneficial to empirically quantify the effects of conversational explanations.

In the current study, we conduct Wizard-of-Oz experiments to investigate how conversations assist users in understanding static explanations of image classification models, improving acceptance and trust in XAI methods, and selecting the best AI models based on explanations.
Specifically, a total of 120 participants join our experiments. We first present them with static explanations for an image classification task and measure their objective understanding and subjective perceptions of static explanations. After that, half of the participants, who are assigned to the experimental group, seek to clarify any doubts with an online textual conversation with an AI system, played by human XAI experts. The other half of the participants, assigned to the control group, read materials about the static explanations independently. After the conversation or reading session, participants complete the same pre-session measurements. From the results, we estimate the effects of conversational explanations.

The experimental measurements include both an objective component and a subject component of the users' understanding and perception. In the objective evaluation,  from three candidate neural networks, the users need to choose one network that would be the most accurate on test data so far unobserved, using information from the static explanations. This task, known as model selection, is one of the most fundamental tasks for machine learning practitioners \citep{anderson2004model}. By design, the three candidate networks make exactly the same predictions on the same inputs but have different rationales for the predictions, as revealed by the static explanations. Hence, the only way for the users to make the right choice is to correctly understand the explanations.   
The subjective evaluation contains 13 questions requiring users to self-report three aspects of their perceptions of the static explanations: comprehension, acceptance, and trust. 

Results show that free-form conversations with XAI experts in the Wizard-of-Oz setting significantly improve comprehension, acceptance, trust, and collaboration with static explanations. Our study underscores the effects of free-form conversations on neural network explainability in practice and provides insights into the future development of conversational explanations. To the best of our knowledge, this is the first study of how free-form conversations may facilitate neural network explainability in practice.

\section{Related Work}
In this section, we review three bodies of research that motivate our study. First, we explore the existing work of static Explainable Artificial Intelligence (XAI). Second, we discuss interactive explanations, especially the limitations of existing methods and the need for conversations to enhance explainability. Lastly, we examine different types of human-AI collaboration and the design of the subjective evaluation during collaboration.

\subsection{Static Explanation}
Explainable Artificial Intelligence (XAI) refers to those models that can explain either the learning process or the outcome of AI predictions to human users~\citep{yang2019evaluating}.  Static XAI involves models that provide a fixed, one-time explanation, without the capability for further user interaction or exploration. They are usually categorized into two groups: self-explanatory models and post-hoc methods. Post-hoc methods can be categorized into feature attribution methods and example-based methods.
Self-explanatory models are inherently transparent, offering clarity in their decision-making processes and facilitating explainability \citep{danilevsky-etal-2020-survey, bodria2021benchmarking}. 
Examples of such models include linear regression, logistic regression, decision trees, Naive Bayes, attention mechanism \citep{bahdanau2014neural}, decision sets \citep{lakkaraju2016interpretable}, rule-based models \citep{rudzinski2016multi,yang2017scalable}, among others.
However, the requirements of self-explanatory models place constraints on model design, which may cause them to underperform in complex tasks. Conversely, the majority of recent XAI methods are post-hoc XAI methods, which can be used for an already developed model that is usually not inherently transparent \citep{selvaraju2017grad, ribeiro2016should, chen2021hydra,verma2020counterfactual, adadi2018peeking, bodria2021benchmarking}. These methods often do not attempt to explain how the model works internally, but instead, employ separate techniques to extract explanatory information. Post-hoc XAI methods can be viewed as reverse engineering processes that employ other independent explanatory models or techniques to extract explanatory information without altering, elucidating, or even understanding the inner workings of the original black-box model. There are two main groups of methods to generate post-hoc XAI explanations, i.e., feature attribution methods and example-based methods.

\subsubsection{Feature Attribution Methods} Feature attribution methods \citep{sundararajan2017axiomatic, selvaraju2017grad, cortez2013using,simonyan2013deep, lundberg2017unified, ribeiro2016should,hu2018locally,alvarez2017causal,liu2018interpretation,shih2018symbolic,ignatiev2019abduction} explain model predictions by investigating the importance of different input features to final predictions. There are two main types of feature attribution methods, gradient-based methods \citep{cortez2013using, sundararajan2017axiomatic, selvaraju2017grad, simonyan2013deep, lundberg2017unified} and surrogate methods \citep{ribeiro2016should,hu2018locally, alvarez2017causal, liu2018interpretation, shih2018symbolic, ignatiev2019abduction}. Gradient-based methods use gradients/derivatives to evaluate the contribution of a model input on the model output. An example method is Grad-CAM \citep{selvaraju2017grad}. It superimposes a heatmap on the regions of important input features by weighting the activations of the final convolutional layer by their corresponding gradients and averaging the resulting weights spatially. Besides directly calculating the importance score of input features, several methods propose to use a simple and understandable surrogate model, e.g., a linear model, to locally approximate the complex deep neural model. Surrogate models can explain the predictions from the complex deep neural model due to their inherent interpretable nature. LIME and its variants are typical methods for generating local surrogate models. LIME \citep{ribeiro2016should} builds a linear model locally around the data point to be interpreted and generates an instance-level explanation for the output.

\subsubsection{Example-based Methods.} Example-based methods \citep{chen2021hydra,verma2020counterfactual, tran2021counterfactual,mothilal2020explaining,poyiadzi2020face,jeyakumar2020can} refer to those that explain predictions of black-box models by identifying and presenting a selection of similar or representative instances. Those examples can be selected or generated from different perspectives, such as training data points that are the most influential to the parameters of a prediction model or the predictions themselves \citep{chen2019looks, chen2021hydra, yoon2020data}, counterfactual examples that are similar to the input query but with different predictions \citep{verma2020counterfactual, wachter2017counterfactual,sharma2019certifai,karimi2020model,tran2021counterfactual,mothilal2020explaining,poyiadzi2020face}, or prototypes that contain semantically similar parts to input instances \citep{croce2019auditing,bien2011prototype,jeyakumar2020can,mikolov2013distributed,doshi2015graph,kim2016examples}.

In this work, we mainly focus on feature attribution methods as they directly highlight the importance of input features, making the decision-making process of models more intuitive \citep{kim2023help} than example-based methods for laypeople. Specifically, we select Grad-CAM from gradient-based methods and LIME from surrogate methods to conduct conversational explanations with participants.

\subsection{Interactive Explanation}
Several studies emphasize the need for interactivity in XAI methods \citep{lakkaraju2022rethinking, Abdul2018trends, schmid2022missing, rohlfing2020explanation}. For instance, \citet{lakkaraju2022rethinking} find that decision-makers strongly prefer interactive explanations. Similarly, a literature analysis by \citet{Abdul2018trends} suggests that interactions can help users progressively explore and gather insights from static explanations. \citet{rohlfing2020explanation} reason that explanations should be co-constructed in an interaction between the explainer and the explainee, adapting to individual differences since the human understanding process is dynamic. From an interdisciplinary perspective, \citet{schmid2022missing} underscore the necessity of user-XAI interactions to adapt to diverse information requirements.

To integrate interactivity and explainability, two primary methodologies emerge. One group of methods focuses on using explanations to help users provide feedback about improving machine learning models. In these methods, the interactivity lies in the cycle of model explanation, user feedback, and model improvement. Explanations aim to help users better understand model decisions and provide valuable feedback. As a result, machine learning models can be incrementally trained with additional loss terms from explanatory feedback \citep{kulesza2015principles, lertvittayakumjorn-etal-2020-find, liang-etal-2020-alice, ijcai2017p371, schramowski2020making, smith2020no} or with added data points \citep{alkan2022frote, biswas2013simultaneous, teso2019explantory, teso2021interactive}. However, these methods are aimed at machine learning practitioners who can well understand and utilize explanations.
Another group focuses on enhancing user understanding of explanations by allowing them to modify the model input and observe changes in the corresponding output. Such interactivity has been shown to improve user comprehension and the perceived utility of AI models \citep{cheng2019explaining, liu2021understanding}. For instance, \citet{tenney-etal-2020-language} and \citet{hohman2019gamut} propose different user interfaces that allow for debugging and understanding machine learning models by examining input-output relationships. However, a rudimentary understanding of machine learning is still required for effective utilization of these interfaces, such as the generic relation between input and output, or what model properties the interpretations reveal.

HCI researchers have recently proposed that XAI methods should align with the ways humans naturally explain mechanisms. Specifically, \citet{lombrozo2006structure} argues that an explanation is a byproduct of a conversational interaction process between an explainer and an explainee. \citet{miller2019explanation} argues that explanations should contain a communication process, where the explainer interactively provides the information required for the explainee to understand the causes of the event through conversations. 
Building on this perspective of human explanations, recent works envision "explainability as dialogue" to provide explanations suitable for a wide range of layperson users \citep{liao2020questioning, feldhus2022mediators, lakkaraju2022rethinking}. While there is much theoretical analysis about the significance of conversations in explainability, practical investigations into their impact on users remain limited. In this context, two previous works have investigated the practical effect of conversations for explainability \citep{slack2023explaining, shen2023convxai}. \citet{shen2023convxai} apply conversational explanations to scientific writing tasks, observing improvements in productivity and sentence quality. \citet{slack2023explaining} design dialogue systems to help users better understand machine learning models on diabetes prediction, rearrest prediction, and loan default prediction tasks. Despite these advances, the conversations in these studies are generated based on templates and cope with limited predefined user intentions. In this study, we explore the role of free-form conversations in enhancing users' comprehension of static explanations, and how they affect users' acceptance, trust, and collaboration with these explanations.

\subsection{Human-AI Collaboration}
Human-AI collaboration is an emerging research area, which explores how humans and AI systems can work together to achieve shared goals \citep{Sarita2023using, wei2023Transitioning, kim2023help}. Prior studies within this domain have investigated collaborations between humans and various AI systems, from robots \citep{gero2020mental, supportingTrust2017, Claudia2023, li2022Application, bhat_evaluating_2024} to virtual agents \citep{ashktorab2020humanai, Cai2019health, Avella2022, numata2020achieving}. The tasks involved span a broad scope, including text \citep{bansal2021does} and image \citep{kim2023help} classifications, medical diagnosis \citep{Cai2019health}, deception detection \citep{lai2019on} and cooperative games \citep{gero2020mental, ashktorab2020humanai, feng2019what}. An area of particular interest within these collaborations is the role of explanations in influencing human-AI decision-making \citep{nguyen2021effectiveness, nguyen2022visual, lai2019on, bansal2021does}.

Our study aligns with existing work on human-AI collaboration \citep{nguyen2021effectiveness, nguyen2022visual, lai2019on, bansal2021does, feng2019what}. In our work, participants need to collaborate with explanations to choose the most accurate neural networks among others. Instead of exploring the role of explanations in collaboration, we mainly examine the potential of conversations in aiding users to effectively use explainability techniques and understand their outputs.

\section{Method}
\label{sec:method}
Our study aims to investigate the impact of conversations on the explainability of AI models by observing participants' comprehension, acceptance, trust of the static explanations, and ability to use the explanations to select the most accurate neural networks before and after the conversation. Our study has received approval from the Institutional Review Board at Nanyang Technological University (\#IRB-2023-254).

\begin{table}[H]
    \caption{Academic disciplines of our participants and the number of participants in each group. There are 120 participants from 4 different discipline groups.}
    \centering
    \begin{tabular}{l c}
        \toprule[1pt]
        \textbf{Academic Discipline} & \textbf{Number of Participants} \\
        \hline
        Business & 23\\
        Engineering & 16\\
        Humanities & 55\\
        Science & 26\\
        \bottomrule[1pt]
    \end{tabular}
    \label{tab:user_domain}
\end{table}
\subsection{Participants}

A total of 120 participants joined our study.  All were 21 years old or older, fluent in English, and had not been involved in research about XAI previously. 
We recruited our participants in two ways: by posting advertisements on an online forum and by emailing students and staff across various departments and schools.
They are from a wide range of disciplines to promote diversity. For ease of reporting, we categorize their disciplines into four groups:
\begin{itemize}
    \item Business, including Business and Accountancy.
    \item Engineering, including Civil and Environmental Engineering, Computer Science, Electrical and Electronics Engineering, Maritime Studies, and Food Science.
    \item Humanities, including Psychology, Economics, Communication Studies, Linguistics and Multilingual Studies, and Sociology.
    \item Science, including Biology, Chemistry, Chemical Engineering and Biotechnology, Sport Science \& Management, Mathematics, Medicine, and Physics.
\end{itemize}
Table \ref{tab:user_domain} shows statistics of the academic disciplines that the participants enrolled in.

\subsection{Experimental Task} 

\begin{figure*}[t]
\centering
\resizebox{0.9\linewidth}{!}{
\begin{subfigure}[t]{.3\textwidth}
  \centering
  \includegraphics[width=1\linewidth]{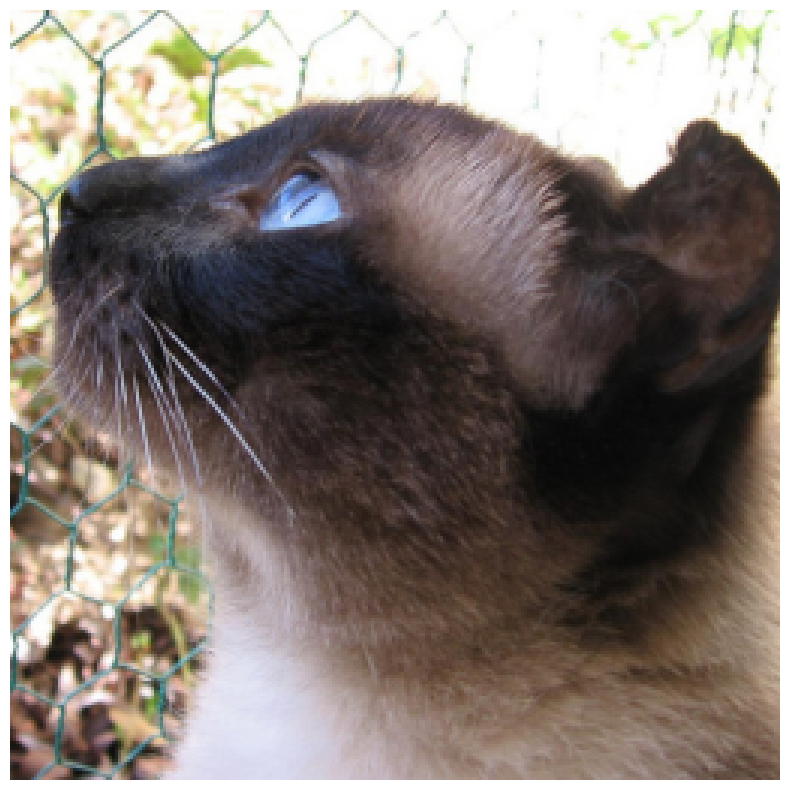}
  \caption{Input to the classification model (Swin Transformer)}
  \label{fig:example_input}
\end{subfigure}\hspace{0.04\textwidth}
\begin{subfigure}[t]{.3\textwidth}
  \centering
  \includegraphics[width=1\linewidth]{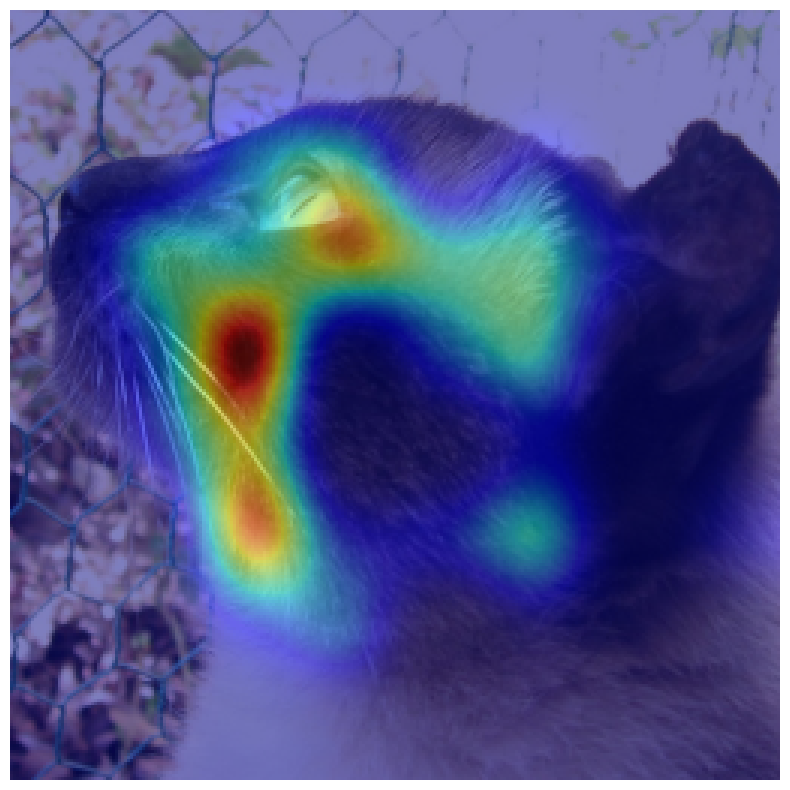}
  \caption{Explanation generated by Grad-CAM}
  \label{fig:grad_cam_example2}
\end{subfigure}\hspace{0.04\textwidth}
\begin{subfigure}[t]{.3\textwidth}
  \centering
  \includegraphics[width=1\linewidth]{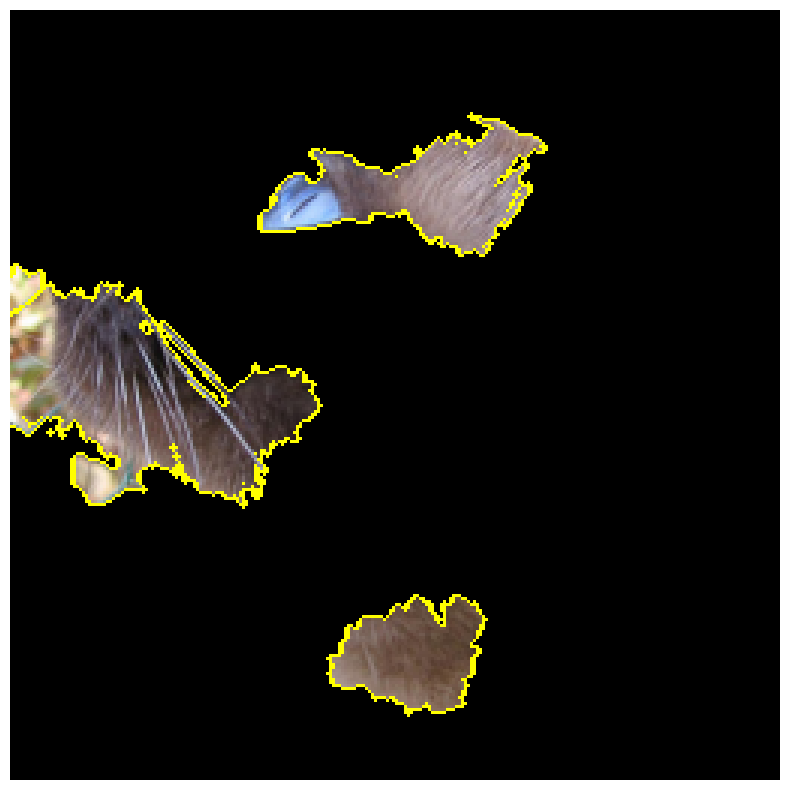}
  \caption{Explanation generated by LIME}
  \label{fig:lime_example2}
\end{subfigure}
}
\caption{Example explanations generated by Grad-CAM and LIME. (a) is the input to the classification model (Swin Transformer), (b) is the explanation generated by Grad-CAM, and (c) is the explanation generated by LIME. The predicted class of the model is "Siamese cat".}
\label{fig:explanation_example2}
\end{figure*}

In our study, we focus on the image classification task on the ImageNet dataset \citep{deng2009imagenet}. Image classification task is a cornerstone in the field of computer vision (CV) that has been the subject of various human-AI collaborative studies \citep{taesiri2022visual, jeyakumar2020can}. We train three classification models with different top-1 classification accuracies: Swin Transformer \citep{liu2021swin} (84.1\%), VGG-16 \citep{vgg16Z14a} (71.6\%), and AlexNet \citep{alex2012imagenet} (56.5\%).  
To generate explanations for model predictions, we select two explanation techniques from two main categories of feature attribution explanation methods: LIME \citep{ribeiro2016should} (a surrogate method) and Grad-CAM \citep{selvaraju2017grad} (a gradient-based method). We focus on feature attribution explanations as we believe the relationship between input features and model predictions is more intuitive to understand than example-based methods for laypeople \citep{kim2023help}. Figure \ref{fig:explanation_example2} displays example explanations generated by these two explanation methods.

\begin{figure*}[t]
    \centering
    \includegraphics[width=0.9\textwidth]{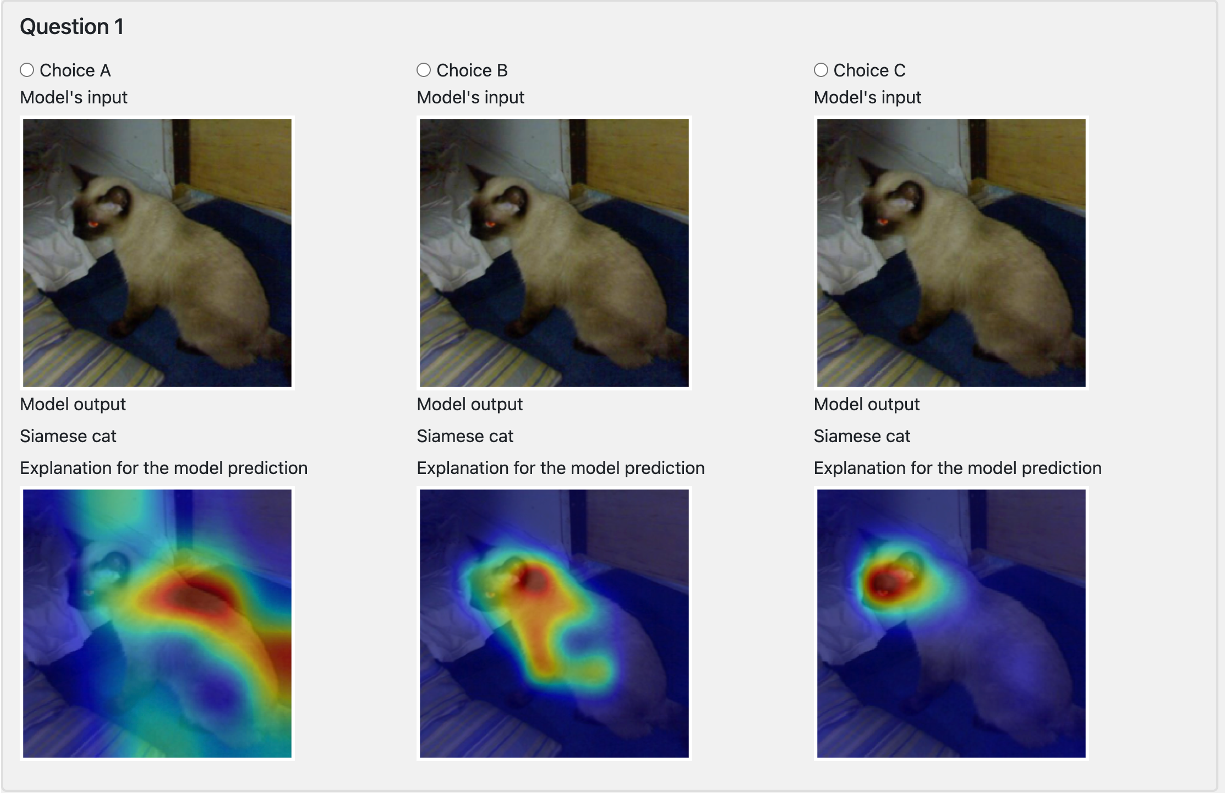}
    \caption{An example of the objective evaluation. The objective evaluation aims to objectively measure participants' comprehension of static explanations. Each choice contains a prediction from a different classification model, paired with its respective static explanation. Participants need to choose the best model based on the explanations.}
    \label{fig:q1_example}
\end{figure*}

To conduct the study, we design and build a web-based platform where participants can remotely finish the whole procedure of the experiment. After users log into the platform, we first evaluate their objective and subjective understanding of static explanations. The objective explanations require participants to choose, from three classification models, the most accurate on unobserved test data. 
The three classification models yield identical decisions on 5 images. The only differences between the three networks lie in their explanations. Hence, to select the best model, the participants must rely on the explanations. Figure \ref{fig:q1_example} presents an example question, including the original image, the model outputs, and the explanations. The full set of questions used in the study can be found in Appendix~\ref{appendix_section}.

\begin{table*}[t]
    \caption{Detailed questions in the subjective evaluation. The user will respond to each question using a 7-point Likert scale.}
    \centering
    \resizebox{0.9\linewidth}{!}{
    \begin{tabular}{c|p{13cm}}
        \toprule[1pt]
        \textbf{Aspect} & \textbf{Question} \\
        \cmidrule(lr){1-2}
        \multirow{1}*[-0.5em]{Comprehension} & How much do you think you understand the explanations provided for predictions of deep learning models? \\
        \cmidrule(lr){1-2}
        \multirow{3}*[-2em]{\makecell{Perceived\\Usefulness}} & Using explanations would improve my understanding of deep learning models’ predictions. \\
        \cmidrule(lr){2-2}
        & Using explanations would enhance my effectiveness in understanding predictions of deep learning models.\\
        \cmidrule(lr){2-2}
        &  I would find explanations useful in understanding predictions of deep learning models.\\
        \cmidrule(lr){1-2}
        \multirow{3}*[-1.2em]{\makecell{Perceived\\Ease-of-Use}} & I become confused when I use the explanation information. \\
        \cmidrule(lr){2-2}
        & It is easy to use explanation information to understand predictions of deep learning models.\\
        \cmidrule(lr){2-2}
        & Overall, I would find explanation information easy to use.\\
        \cmidrule(lr){1-2}
        \multirow{2}*[-1.5em]{\makecell{Behavioral\\Intention}} & I would prefer getting explanation information as long as it is available when getting predictions from deep learning models. \\
        \cmidrule(lr){2-2}
        & I would recommend others use explanation information to understand predictions of deep learning models.\\
        \cmidrule(lr){1-2}
        \multirow{4}*[-3.5em]{Trust} & How would you rate the competence of the explanation method? - i.e. to what extent does the explanation method perform its function properly?\\
        \cmidrule(lr){2-2}
        & How would you rate the dependability of the explanation method? - i.e. to what extent can you count on the explanation method to explain predictions of deep learning models? \\
        \cmidrule(lr){2-2}
        & How would you rate your degree of faith that the explanation method will be able to explain predictions of deep learning models in the future? \\
        \cmidrule(lr){2-2}
        & How would you rate your overall trust in the explanation method?\\
        \bottomrule[1pt]
    \end{tabular}
}
    \label{tab:q2_questions}
\end{table*}

The subjective evaluation measures participants' self-reported perception of the static explanations, including their comprehension \citep{hoffman2018metrics, cheng2019explaining}, acceptance \citep{davis1989perceived, davis1989user, diop2019extension, Christopher2023Acceptance}, and trust \citep{davis1989perceived, davis1989user, diop2019extension, Yang:2017:EEU:2909824.3020230, guo_enabling_2023}. Based on an in-depth review of existing literature, we chose the questions from those that have been validated in prior research. The subjective evaluation contains a total of 13 questions, each utilizing a 7-point Likert scale for responses.
Table~\ref{tab:q2_questions} lists all the questions we used. Labels of the 7-point Likert scale are listed in Appendix~\ref{appendix_section}.

\begin{figure*}[t]
    \centering
    \includegraphics[width=\textwidth]{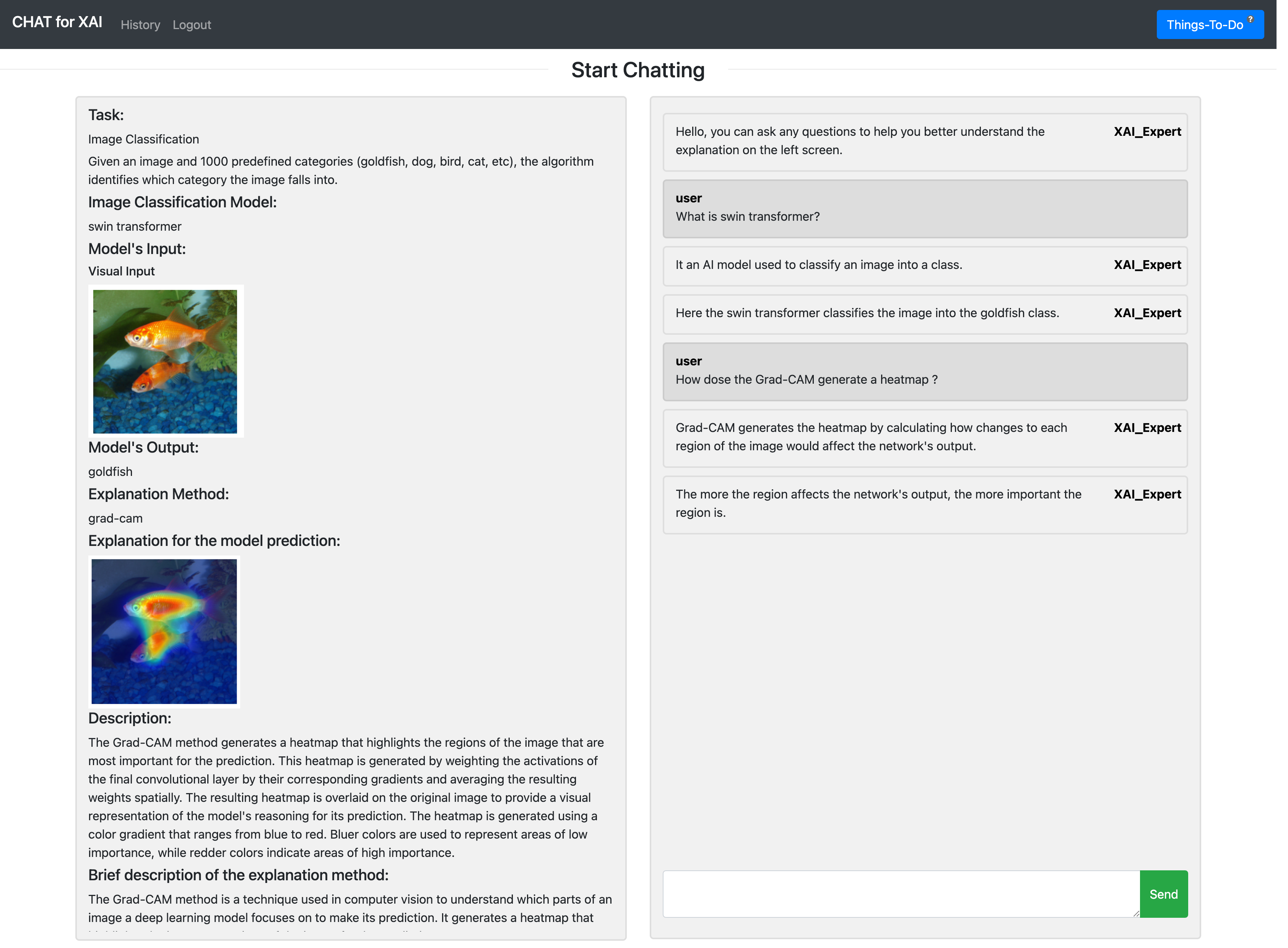}
    \caption{The web page where users can discuss static explanations with an expert.}
    \label{fig:chat_page}
\end{figure*}

After these two evaluations, participants are divided into two groups, i.e., the control group and the experimental group. Participants in the control group read static explanations for 15 minutes. Participants in the experimental group conduct conversational explanations with participants in the Wizard-of-Oz (WoZ) setting \citep{kelly1984wizard-of-oz}. They interact with a dialogue system that they believe to be autonomous but is actually operated by a human expert on machine learning. 

To support the WoZ experiment, we built a conversation page with a two-section structure, as depicted in Figure \ref{fig:chat_page}. On the left, the page shows a task description, a textual description of the prediction model, a textual description of the explanation technique, an example input image, the model prediction on the input image, a static explanation for the prediction, and a textual description of the explanation. On the right, the interface enables users to converse with XAI experts, seeking clarifications and posing questions about the explanation. For the users in the control group, we replace the textual chat user interface with a 15-minute timer. Once the timer reaches zero, users are allowed to proceed to the post-evaluations.
Users from both groups receive the same post-evaluations, which are identical to the pre-evaluations. We discuss the evaluations below.

\subsection{Experimental Design}

There are two independent variables and two categories of dependent variables. The independent variable in the experiments is the explanation method: LIME or Grad-CAM and the method of understanding static explanations: conversation with human experts or reading static explanations. As we devise both subjective and objective evaluations before and after conversations or readings, two categories of dependent variables were collected in the experiment: the model selection accuracy and the self-reported perception scores.

\subsubsection{Objective Evaluation -- Selection of Classification Models}

The evaluation aims to objectively evaluate participants' understanding of the static explanations.
Participants are presented with 5 input images, on which the three neural networks make the same decisions. The only differences between the three networks lie in their explanations.
Participants need to choose the one that would be the most accurate on unobserved test data. Hence, to make the correct selection, the participants must understand the explanations. We use the accuracy of selecting the correct model to measure participants' objective understanding of static explanations.

We recognize that existing explanation techniques are not always faithful to the underlying model \citep{Adebayo2018SanityChecks, Kindermans2019, jacovi2020towards} and do not always provide actionable information for model selection. As our goal is to test if the users can understand the static explanations \emph{when} they do provide actionable information, rather than evaluating the static explanations themselves, we selected input images where better classification models indeed have more reasonable and intuitive explanations. This approach allows users to easily pick the best classification models if they understand the static explanations well. 
We deem an explanation more reasonable when it focuses more on discriminative features that are unique to the predicted class and less on spurious features that are irrelevant to the class. In addition, good models should have explanations that rely on multiple types of discriminative features. 
This is because a model relying on multiple features is robust and makes the correct decision even if some discriminative features are missing or occluded. In the example in Figure \ref{fig:q1_example}, Model B is better than Model A or Model C as Model B utilizes both the head and the body of the cat for classification. In addition, unlike Model A, Model B does not focus on the background, which is irrelevant to the predicted class, Siamese Cat.

\subsubsection{Subjective Evaluation}
\label{q2_section}
We also measure participants' subjective perception of static explanations, including their comprehension, acceptance, and trust. The subjective evaluation contains a total of 13 questions listed in table \ref{tab:q2_questions}. All questions utilize a 7-point Likert scale for responses.
\begin{itemize}[leftmargin=3.5ex, topsep=4pt]
    \item Comprehension \citep{hoffman2018metrics, cheng2019explaining}: Participants' subjective perceptions of their understanding of explanations. It complements the objective evaluation, providing a holistic perspective on participants' understanding of static explanations.
    \item Perceived Usefulness \citep{davis1989perceived, davis1989user, diop2019extension}: The degree to which participants feel that the explanations enhance their experience with deep learning models. Along with \textit{perceived ease of use} and \textit{behavioral intention}, these three aspects measure participants' acceptance of static explanations. They are derived from the Technology Acceptance Model (TAM) \citep{davis1989perceived, davis1989user, diop2019extension},  a widely applied theory for understanding individual acceptance and usage of information systems.  As the explanations are used by end-users, investigating their acceptance of the explanations is very important. 
    \item Perceived Ease of Use \citep{davis1989perceived, davis1989user, diop2019extension}: Participants' assessment of the simplicity and clarity of the explanations.
    \item Behavioral Intention \citep{davis1989perceived, davis1989user, diop2019extension}: The tendency of participants to utilize the explanation information in the future.
    \item Trust \citep{muir1996trust, Tita2022Systematic}: Participants' confidence in the explanation methods keeping functioning as intended. Trust has been recognized as an important factor in human-AI collaboration as it mediates the human’s reliance on AI models, thus directly affecting the effectiveness of the human-AI team \citep{sebo2020influence, seaborn2021voice, doshi2017towards, vorm2022Integrating, Andrew2023Explainable}.

    The literature demonstrated that static explanations have inconsistent effects on users’ trust in AI systems. On one hand, several studies have demonstrated that detailed explanations \citep{Glass2009towards, taehyun2023improving, Andrew2023Explainable}, contrastive explanations \citep{oro70421}, and example-based explanations \citep{yang2020how} can enhance user trust in systems. On the other hand, studies showed that static explanations do not have strong effects on user trust in AI systems \citep{cheng2019explaining, kunkel2019let, wang2021are, zhang2020effect}. 

    One main reason for these inconsistent reports is that trust is mediated by the users’ understanding of the static explanations \citep{kunkel2019let, wang2021are, zhang2020effect}, and such understanding is often absent. According to theories of trust \citep{mcknight1998initial, lim2009why, hoffman2018metrics}, the ability to build a mental model of AI systems is the key for user trust in AI. Unsurprisingly, studies on the effects of static explanations for laypersons show that users with limited knowledge of machine learning struggle to understand static explanations and the decision-making processes they are supposed to explain. Consequently, these users do not exhibit increased trust in AI systems after receiving static explanations \citep{zhang2020effect, wang2021are}.

    With this paper, we quantitatively investigate whether customized conversations about static model explanations can enhance user understanding and improve trust. The conversational approach toward explanations has been advocated by previous studies \citep{Glass2009towards, pieters2011explanation, james2019ican, feldhus2022mediators, lakkaraju2022rethinking} but never experimentally verified. For example, through interviews with decision-makers, \citet{lakkaraju2022rethinking} found that decision-makers strongly prefer conversational explanations that allow them to ask follow-up questions.

\end{itemize}

\subsection{Detailed Study Procedure}
Before participation, individuals are required to sign an informed consent form that outlines the objectives and procedures of the study. The form also clarifies compensation details and assures both the anonymity and confidentiality of data collected during the study. Upon signing the consent, participants receive an email that guides them to access the study platform. 

After logging in, a pop-up prompt provides an overview of the tasks ahead. Participants then complete pre-experiment objective and subjective evaluations of the static explanations. The objective evaluation measures participants' understanding of static explanations by letting them choose, from three classification models, the most accurate on unobserved test data. There are 5 explanation examples in the objective evaluation. The subjective evaluation, with 13 self-reporting questions, probes the perceived comprehension, acceptance, and trust towards the static explanations. Following these evaluations, participants in the experimental group engage in a WoZ discussion about static explanations. During the conversation, one example image is displayed on the screen. The example image is different from those used in the evaluations; however, the explanation methods remain the same. Participants are motivated to understand the explanations as they need to select the best-performing classification model using explanations only when doing objective evaluation. Our XAI experts faithfully answer the user's questions based on their knowledge, trying to help the user gradually understand the explanation. For participants in the control group, they read the static explanation for 15 minutes which is the average conversation time of the experimental group. After the conversation or 15-minute reading, participants complete the same set of evaluations as before. All evaluation outcomes and conversation records are documented. Upon study completion, each participant receives a \$10 reward.

\begin{table*}[t]
    \caption{Results of the experimental group before and after conversations, and the control group before and after 15-minute reading. Each score is presented as mean $\pm$ standard deviation and the change $\delta$ before and after. $^{*}$ $p < 0.001$
    }
    \centering
    \resizebox{\linewidth}{!}{
    \begin{tabular}{c|c|c|c|c|c|c|c|c}
        \toprule[1pt]
        \makecell{Explanation\\Methods}& Group & \makecell{Evaluation\\Timing} & \makecell{ Objective\\Understanding\\(Decision-Making\\Accuracy)} & \makecell{Subjective\\Understanding} & \makecell{Perceived\\Usefulness} & \makecell{Perceived\\Ease of Use} & \makecell{Behavioral\\Intention} & \makecell{Trust}\\
        \midrule
        \multirow{4}*{LIME} & \multirow{2}*{experimental} 
                            & before & 0.38 $\pm$ 0.20 & 4.03 $\pm$ 1.35 & 5.09 $\pm$ 1.07 & 4.48 $\pm$ 0.94 & 5.25 $\pm$ 0.95 & 4.15 $\pm$ 0.88\\
                            & & after  & 0.53$^{*}$ $\pm$ 0.16 & 5.30$^{*}$ $\pm$ 0.88 & 5.92$^{*}$ $\pm$ 0.66 & 5.28$^{*}$ $\pm$ 0.84 & 5.83$^{*}$ $\pm$ 0.81 & 4.92$^{*}$ $\pm$ 0.73 \\ 
                            \cmidrule(l){2-9}
                            & \multirow{2}*{control}
                            & before & 0.37 $\pm$ 0.17 & 4.57 $\pm$ 1.43 & 5.67 $\pm$ 0.95 & 4.87 $\pm$ 1.26 & 5.73 $\pm$ 0.69 & 4.37 $\pm$ 0.90\\
                            & & after& 0.40 $\pm$ 0.20 & 4.60 $\pm$ 1.16 & 5.33 $\pm$ 0.96 & 4.48 $\pm$ 1.26 & 5.27 $\pm$ 1.08 & 4.36 $\pm$ 1.05\\ \midrule
        \multirow{4}*{Grad-CAM} & \multirow{2}*{experimental} 
                            & before & 0.82 $\pm$ 0.21 & 4.17 $\pm$ 0.91& 5.49 $\pm$ 0.97 & 4.71 $\pm$ 0.95 & 5.52 $\pm$ 0.65 & 4.40 $\pm$ 1.00\\
                            & & after  & 0.92$^{*}$ $\pm$ 0.11 & 5.43$^{*}$ $\pm$ 0.97 & 6.12$^{*}$ $\pm$ 0.60 & 5.58$^{*}$ $\pm$ 0.82 & 6.08$^{*}$ $\pm$ 0.79 & 5.19$^{*}$ $\pm$ 0.80\\ 
                            \cmidrule(l){2-9}
                            & \multirow{2}*{control}
                            & before & 0.81 $\pm$ 0.20 & 4.07 $\pm$ 1.34 & 5.58 $\pm$ 0.59 & 4.36 $\pm$ 1.15 & 5.45 $\pm$ 0.71 & 4.22 $\pm$ 0.96 \\
                            & & after& 0.79 $\pm$ 0.19 & 4.40 $\pm$ 1.28 & 5.46 $\pm$ 0.69 & 4.70 $\pm$ 1.21 & 5.33 $\pm$ 0.83 & 4.42 $\pm$ 0.87\\ 
        \bottomrule[1pt]
    \end{tabular}
}
    \label{tab:result}
\end{table*}

\section{Results \& Discussion}
Table \ref{tab:result} tabulates the mean and standard deviation (SD) for all the measures. 
As explanation methods (LIME vs. Grad-CAM) and group (experimental vs. control) are between-subjects variables and time (before vs. after) is a within-subject variable, we conduct a three-way Analysis of Variance (ANOVAs).

\subsection{Effects of explanations on objective decision accuracy and subjective measures}

\begin{figure}[!ht]
\centering \includegraphics[width=0.9\linewidth]{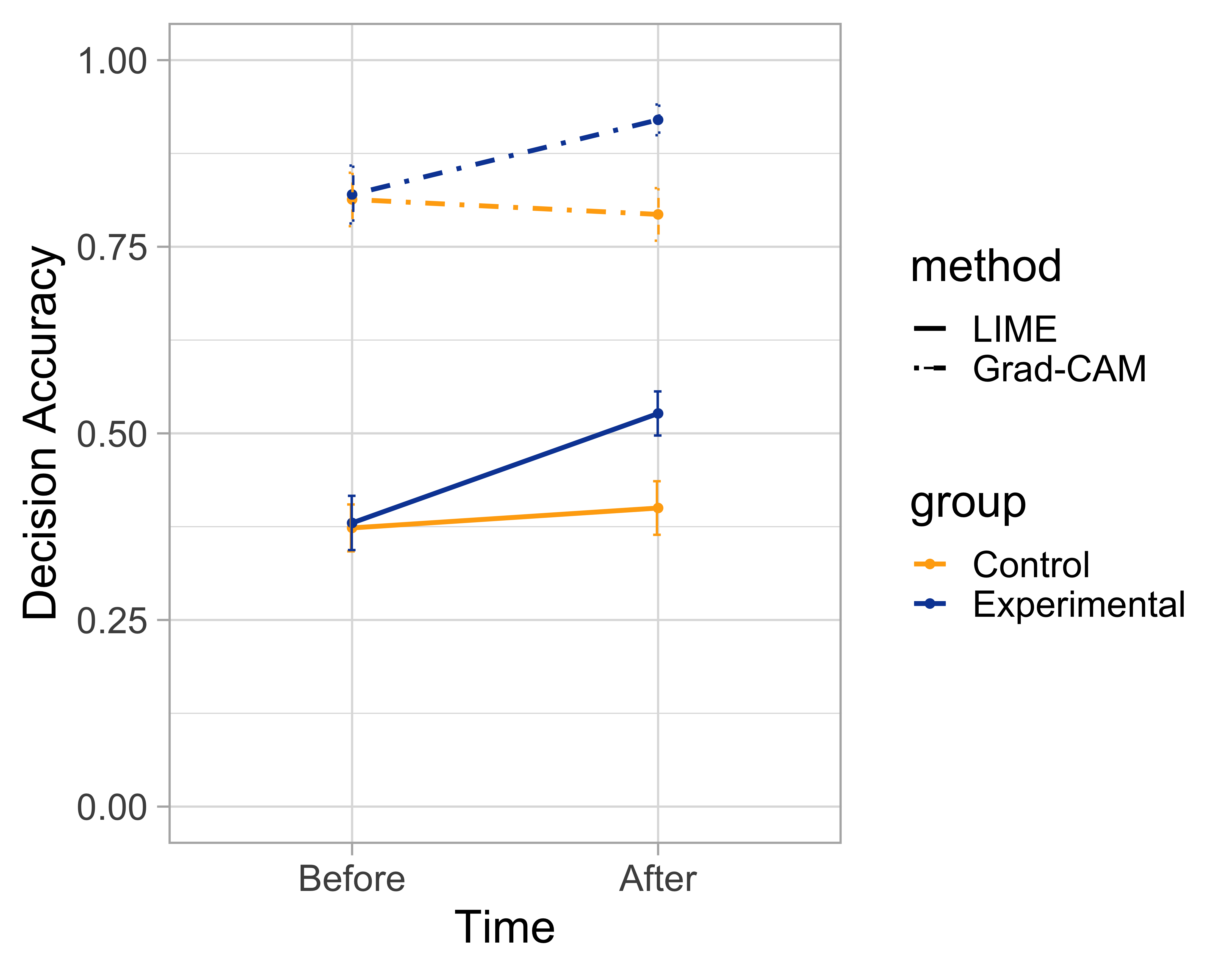}
\caption{Objective decision accuracy for different groups before and after conditions.}
\label{fig:interaction_da}
\end{figure}

Results show significant main effects of group ($F(1,116) = 5.60, p=.02$), method ($F(1,116) = 218, p<.001$) and time ($F(1,116) = 12.51, p<.001$). The experimental group, the Grad-CAM method, and the after-conversation condition display a higher objective decision accuracy. We also find a significant interaction effect between group and time ($F(1,116) = 11.3, p=.01$), as displayed in the figure \ref{fig:interaction_da}. In the participant's initial decision, there were no significant differences between the experimental and control conditions. During participants' final decision, those who interact with the XAI expert~(i.e., experimental condition) have better decision accuracy. This phenomenon highlights the effectiveness of conversational explanations in enhancing the objective understanding of static explanations of users.

We observe varied objective performance between LIME and Grad-CAM ($F(1,116) = 218, p<.001$). Grad-CAM has a higher accuracy of objective decision accuracy compared to LIME. A potential reason might be the inherently intuitive nature of the explanations produced by Grad-CAM compared to LIME.

\begin{figure}[!ht]
\centering\subfloat[LIME]{\includegraphics[width=0.9\linewidth]{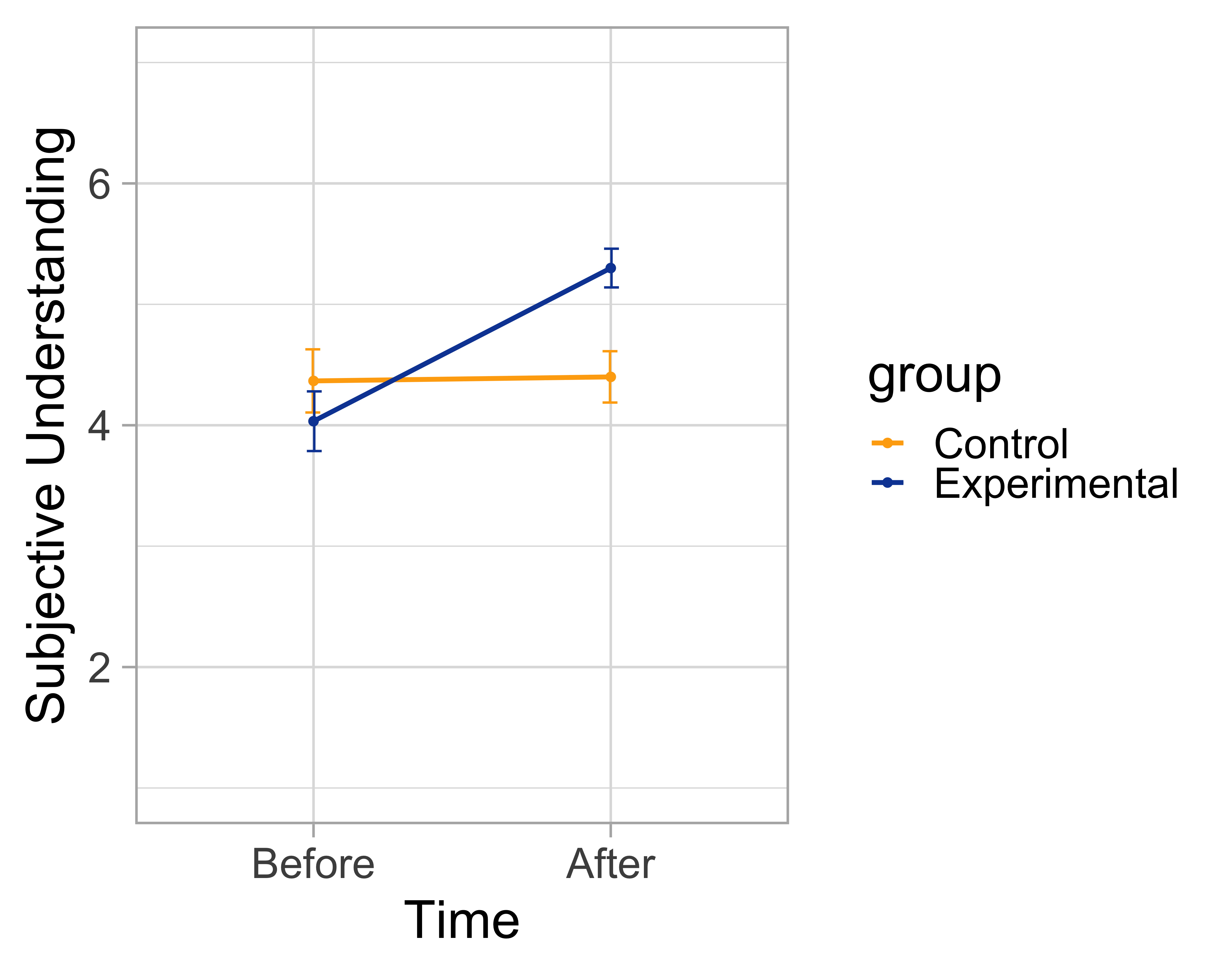}}
\hspace{0\textwidth}
\subfloat[Grad-CAM]{\includegraphics[width=0.9\linewidth]{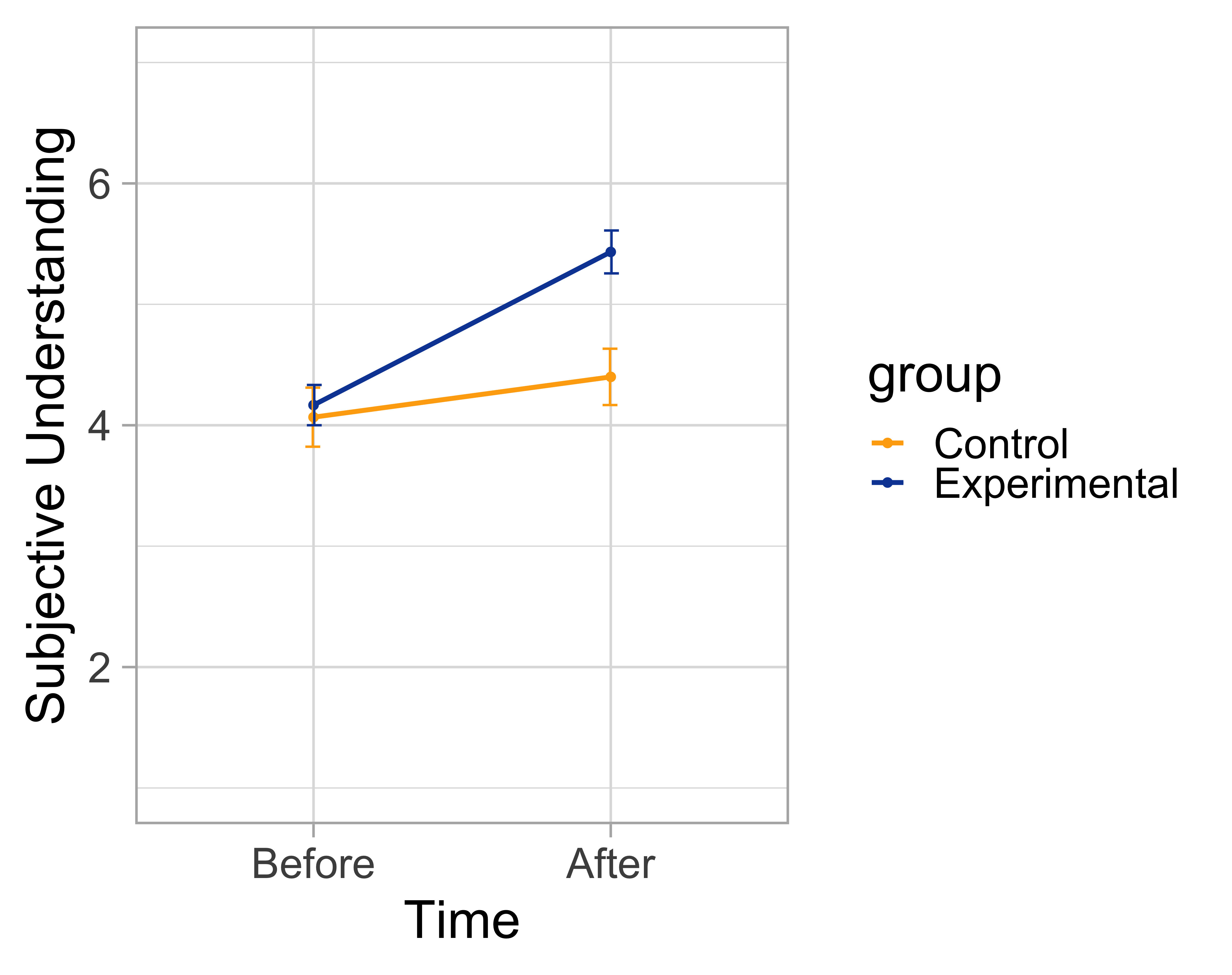}}
\caption{Subjective understanding score for (a) LIME and (b) Grad-CAM before and after conditions. }
\label{fig:su_interaction}
\end{figure}

In terms of participants' subjective understanding, we find a significant main effect of the evaluation timing ($F(1,116) = 4.08, p<.001$). Participants receiving conversational explanations have a significantly larger improvement in subjective understanding. We also observe a significant interaction effect between group and time ($F(1,116) = 37.3, p<.001$), shown in figure \ref{fig:su_interaction}. Initially, there is no significant difference in the participants' self-reported understanding of static explanations between the experimental and control groups. After the conditions, participants in the experimental group demonstrate a higher self-report understanding compared to those in the control group.

The main effect of the explanation method ($F(1,116) = .72, p=.40$) is not significant for participants' subjective understanding, contrasting with its effect on objective understanding. Even though participants can intuitively choose the best classification model based on the heatmap in the objective evaluation, participants' initial self-reporting understanding score of Grad-CAM is just slightly larger than 4 (average understanding). This shows that participants still feel confused about how Grad-CAM works and how it explains models' predictions, even though they can perform well in the objective evaluation. 
This also demonstrates that subjective and objective evaluations measure participants' understanding of static explanations from complementary aspects. Self-reporting scores can be influenced by personal biases, while the objective evaluation might not capture users' feelings about understanding. Combining both methods can provide a comprehensive view of participants' understanding of static explanations.

\begin{figure}[!ht]
\centering\subfloat[LIME]{\includegraphics[width=0.9\linewidth]{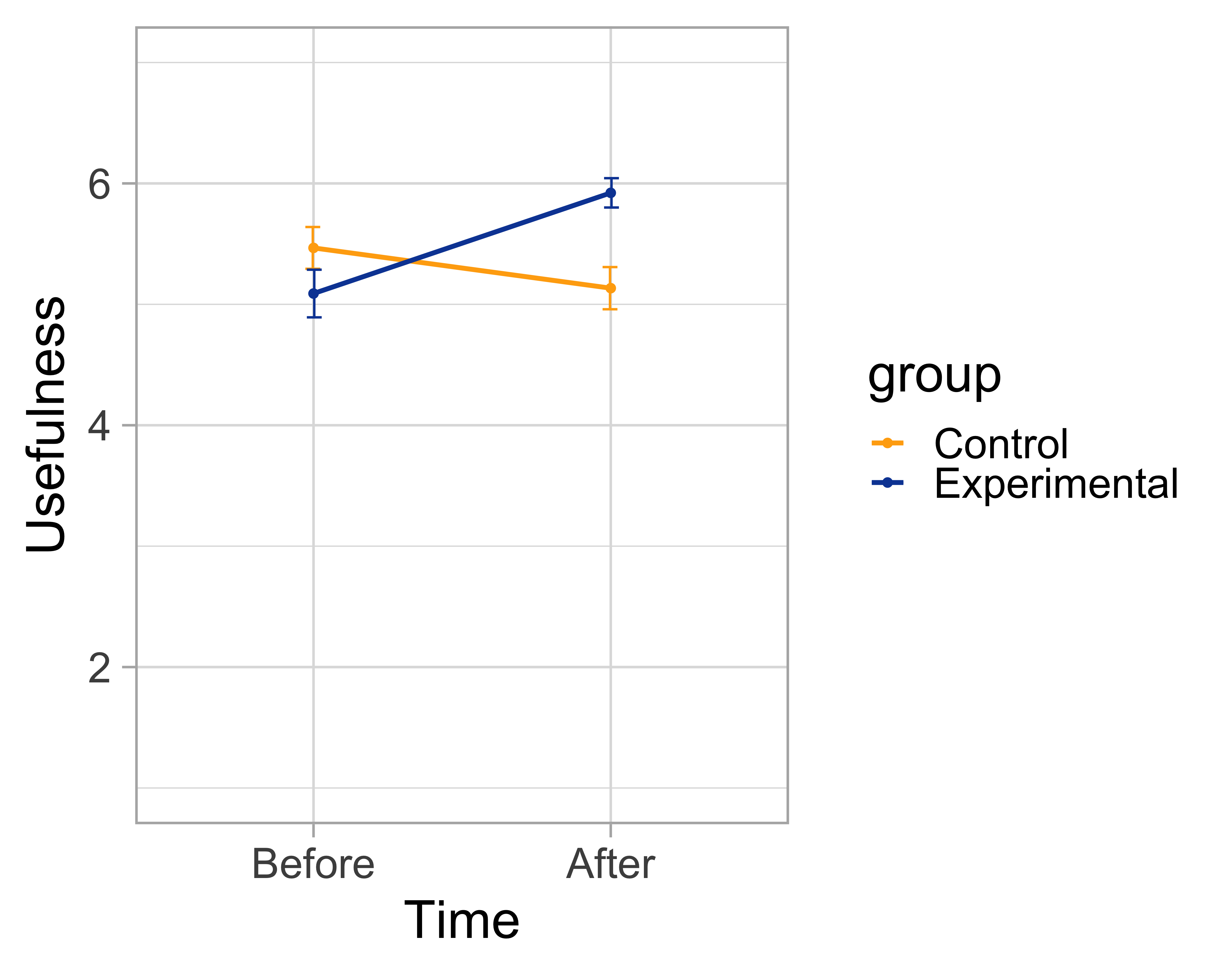}}\hspace{0.0\textwidth}
\subfloat[Grad-CAM]{\includegraphics[width=0.9\linewidth]{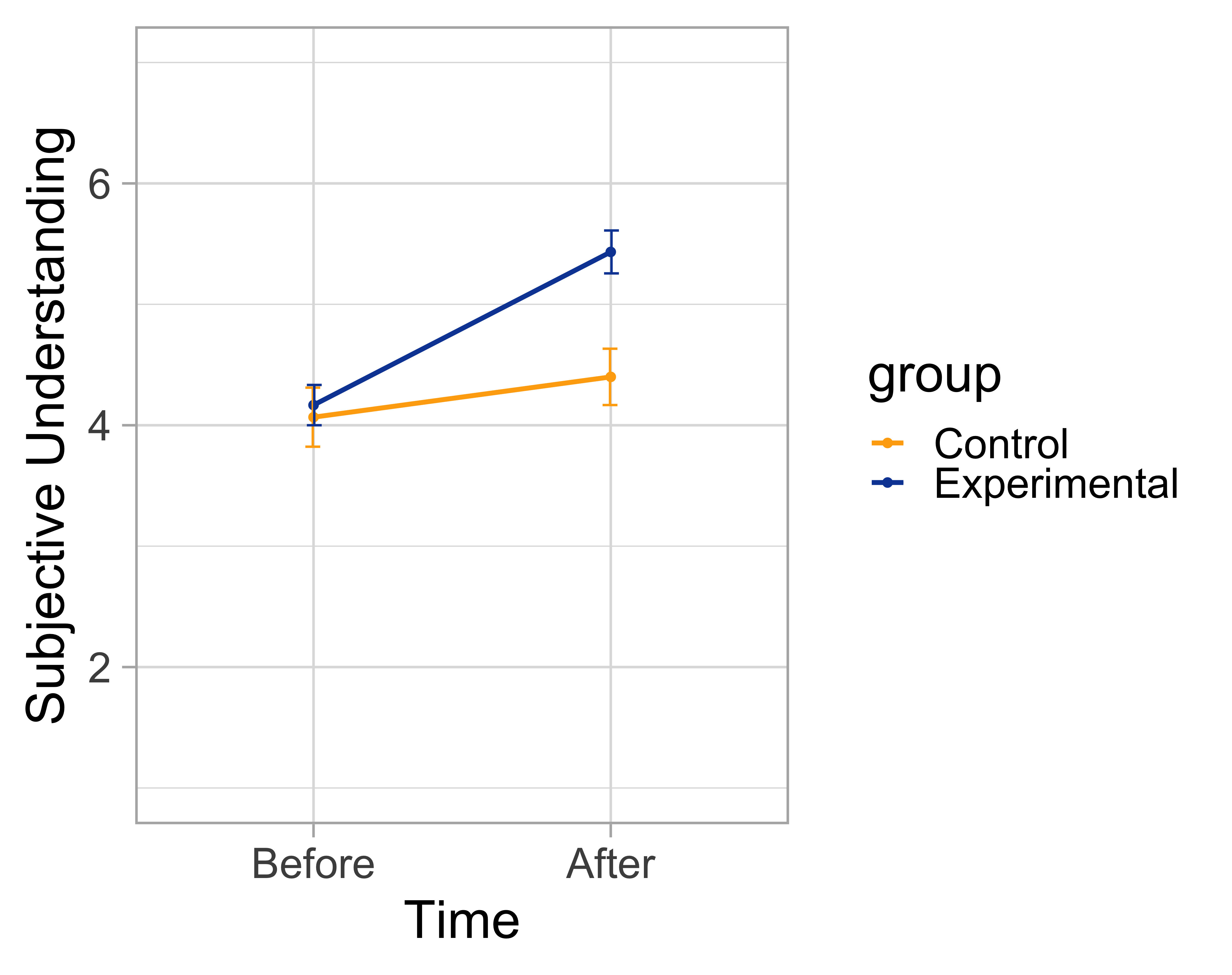}}
\caption{Participants' self-report usefulness score for (a) LIME and (b) Grad-CAM before and after conditions. }
\label{fig:u_interaction}
\end{figure}

For the perceived usefulness, results show a significant main effect of time ($F(1,116) = 14.6, p<.001$), as well as a significant interaction effect between group and time ($F(1,116) = 52.9, p<.001$), as depicted in figure \ref{fig:u_interaction}. The experiment group (i.e., receiving conversational explanation) results in a larger increment of perceived usefulness. For the control group, the Grad-CAM method increases perceived ease of use when participants are given more time to view the static explanation. However, a reversed trend is observed for the LIME method in the control group -- the perceived ease of use drops after additional time is provided. 

\begin{figure}[!ht]
\centering\subfloat[LIME]{\includegraphics[width=0.9\linewidth]{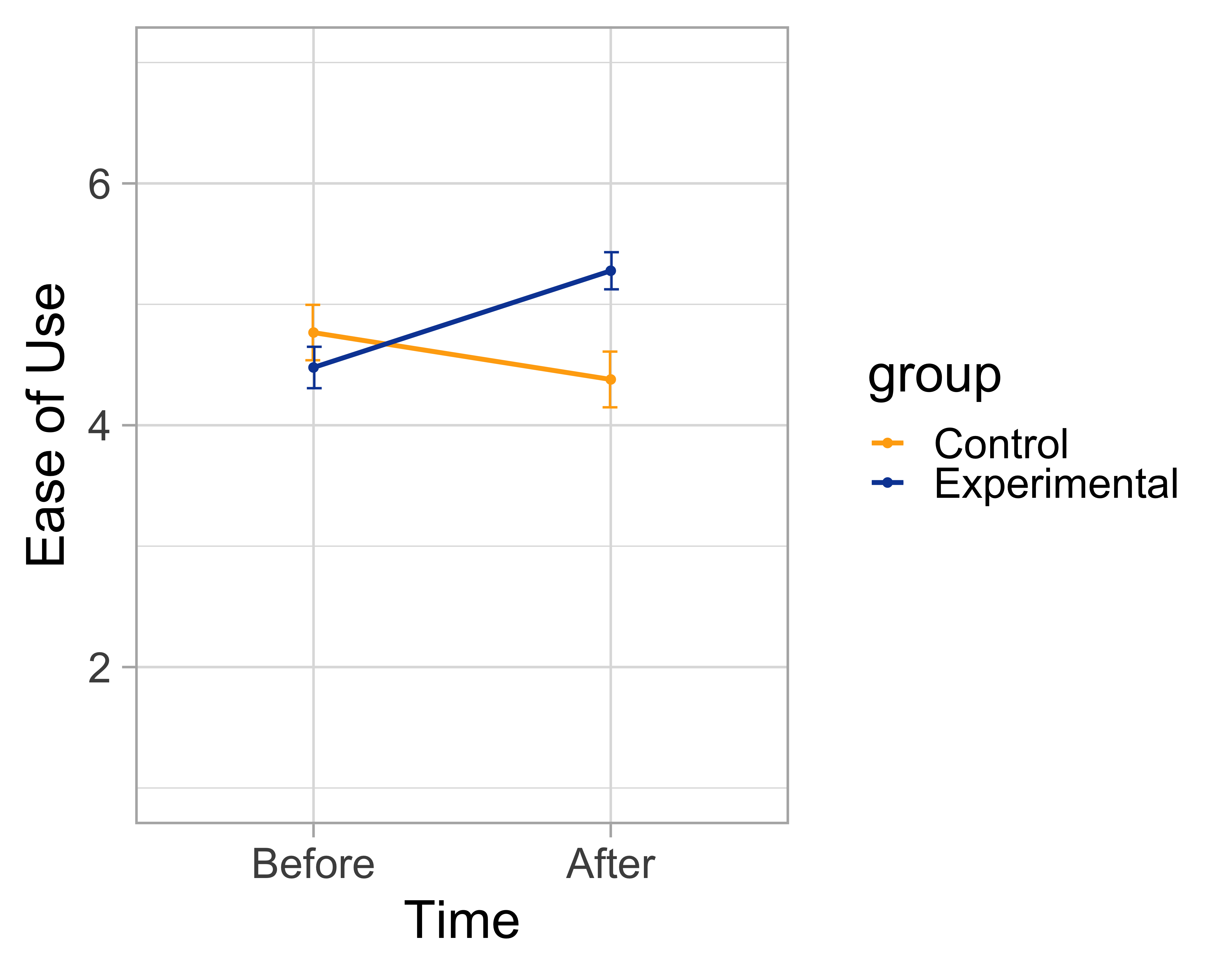}}\hspace{0.0\textwidth}
\subfloat[Grad-CAM]{\includegraphics[width=0.9\linewidth]{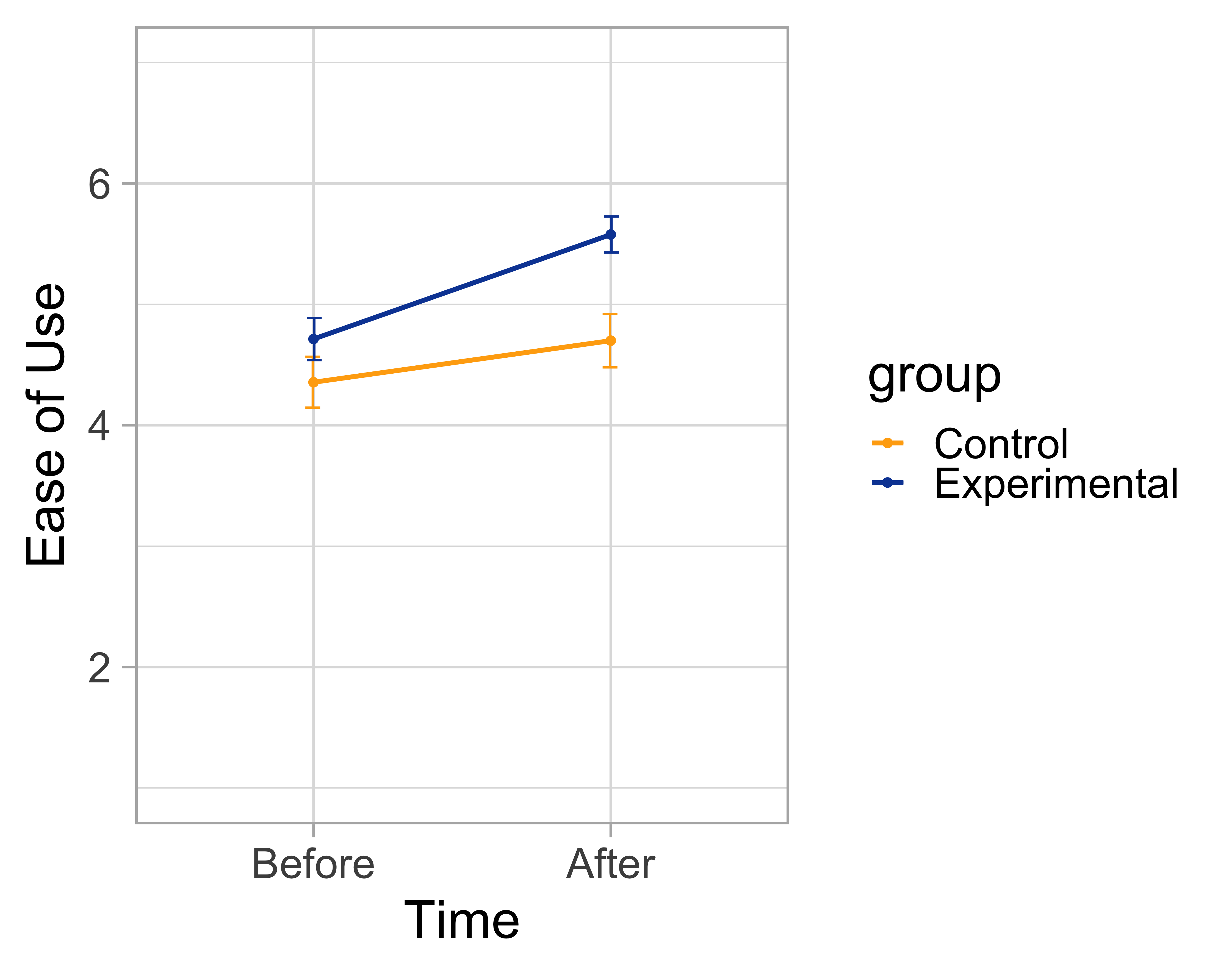}}
\caption{Participants' self-report ease of use score for (a) LIME and (b) Grad-CAM before and after conditions. }
\label{fig:eoe_interaction}
\end{figure}

Similar results are observed for participants' perceived ease of use. There are significant main effects of group ($F(1,116) = 5.19, p=.002$) and of time ($F(1,116) = 30.3, \: p<.001$), as well as a significant interaction effect between group and time ($F(1,116) = 33.7, p<.001$). The perceived ease of use increases largely for the experiment group after interacting with XAI experts. For the control group, the Grad-CAM method increases perceived ease of use while LIME methods decrease it when giving participants more time to view the static explanation.

\begin{figure}[!ht]
\centering\subfloat[LIME]{\includegraphics[width=0.9\linewidth]{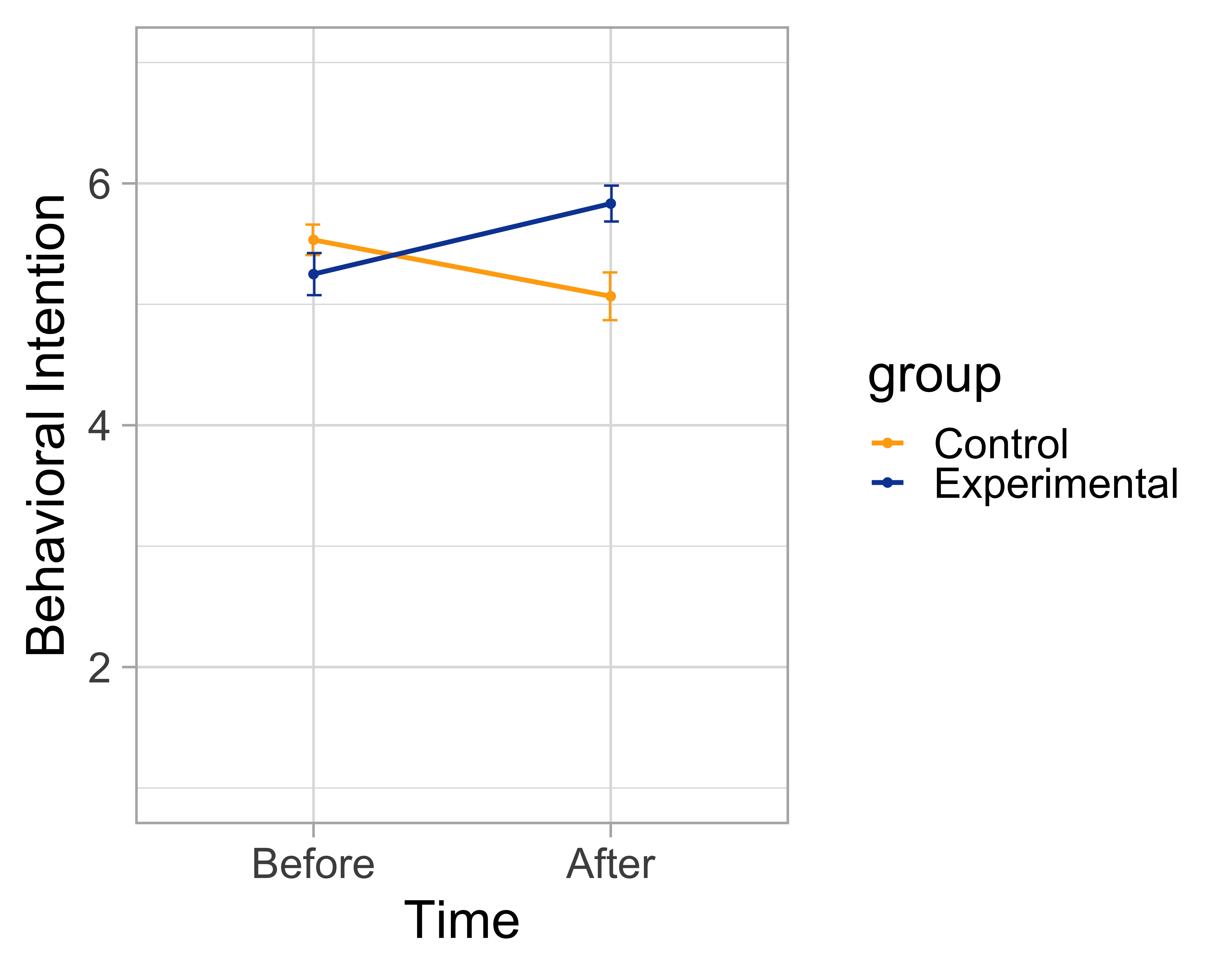}}\hspace{0.0\textwidth}
\subfloat[Grad-CAM]{\includegraphics[width=0.9\linewidth]{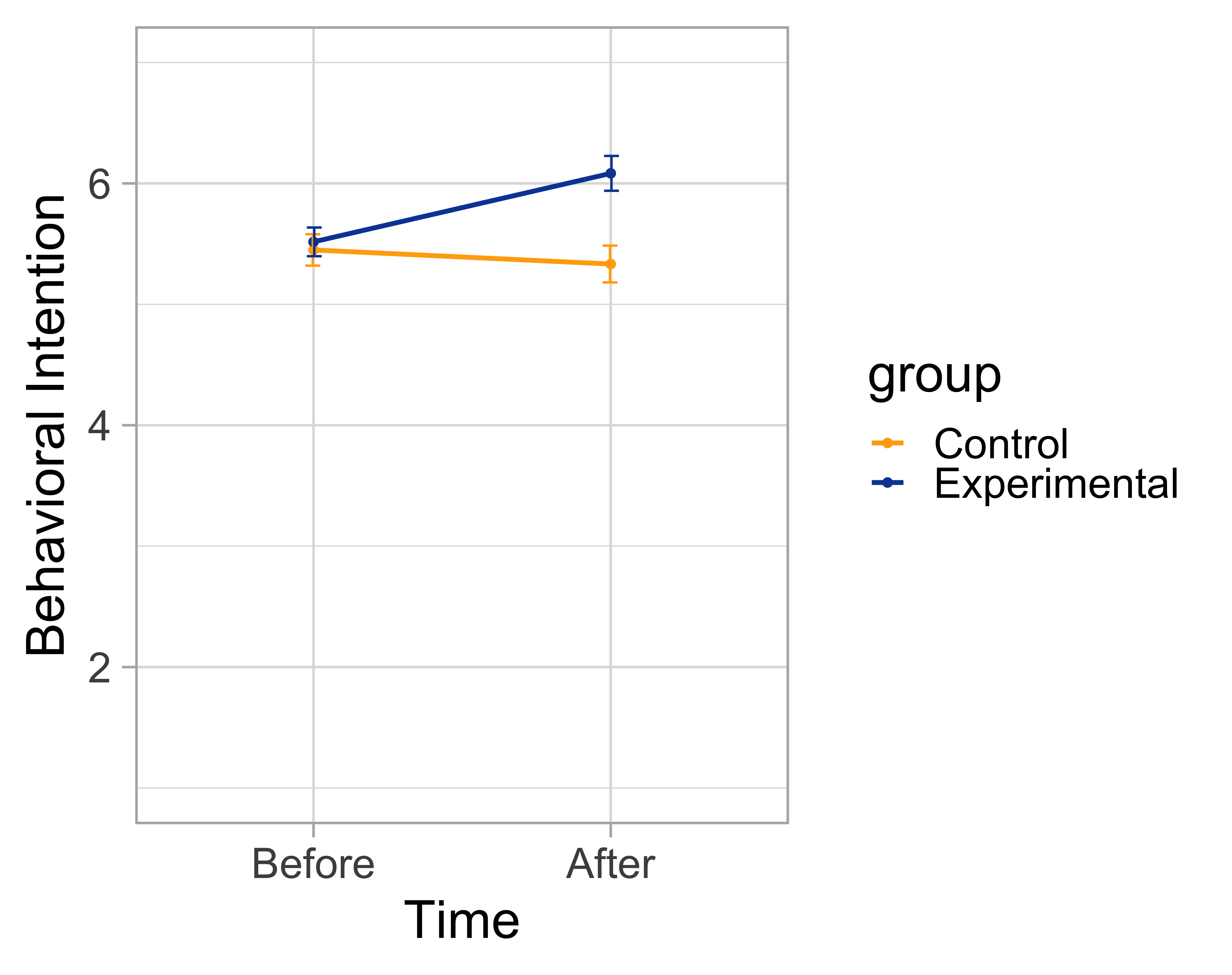}}
\caption{Participants' self-report behavioral intention score for (a) LIME and (b) Grad-CAM before and after conditions. }
\label{fig:bi_interaction}
\end{figure}

For the behavioral intention, results show a significant main effect of the time ($F(1,116) = 3.92, p=.005$) and a significant interaction effect between group and time ($F(1,116) = 3.92, p<.001$) as shown in figure \ref{fig:bi_interaction}. Participants increase their behavioral intention and are more inclined to use explanations in future scenarios after receiving conversational explanations.
On the contrary, the behavioral intention of the control group decrease for both Grad-CAM and LIME. 

The boost in usefulness, ease of use, and behavioral intention for the experimental group can be attributed to the increased understanding of static explanations. Prior to the expert interactions, participants might have had limited knowledge or even misconceptions about the explanation methods. Experiment results show that participants gain a clearer understanding of how the XAI methods function, after the participants’ questions are addressed in the conversations. Consequently, they report perceiving the static explanations as more useful and easier to use, and report higher inclination to use the static explanations in future tasks.

The perceived usefulness, ease of use, and behavioral intention of the control group all decrease after reading static explanations for a longer time. This trend suggests a decreased willingness to utilize explanations in future scenarios. The reluctance may be attributed to the frustration the control group faced in attempting to comprehend the static explanations on their own. Research by \citet{Carolin2023ExplainableAI} on the impact of cognitive fit and misfit in the acceptance of AI system usage highlights this phenomenon. They found that users experiencing a cognitive misfit with the AI system often report negative moods, which in turn, reduce their perceived usefulness, ease of use, and behavioral intention of the AI systems. The contrary results of the control group and the experimental group also underscore the importance and effectiveness of conversations in enhancing user behavioral intentions of static explanations.

\begin{figure}[!ht]
\centering\subfloat[LIME]{\includegraphics[width=0.9\linewidth]{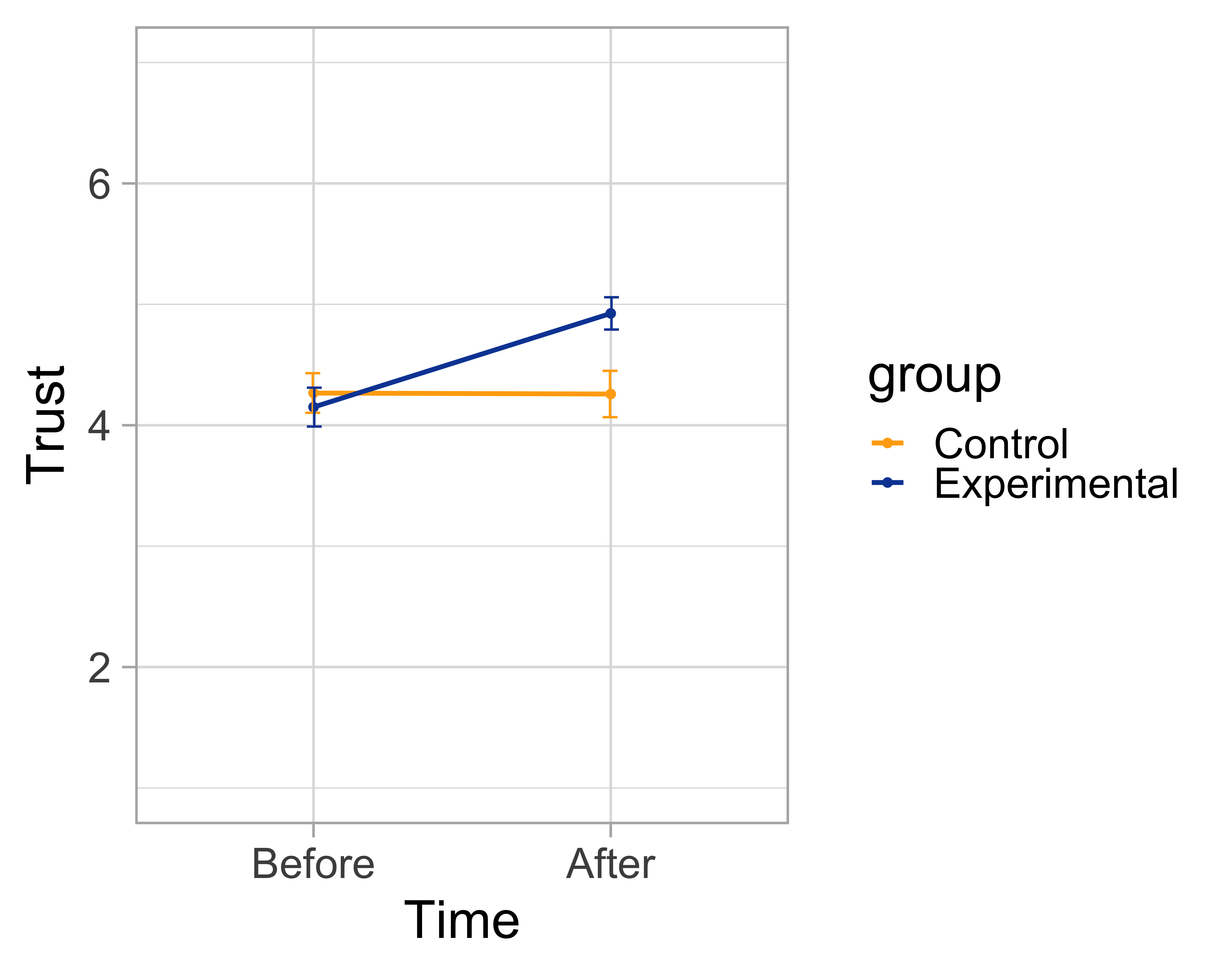}}\hspace{0.0\textwidth}
\subfloat[Grad-CAM]{\includegraphics[width=0.9\linewidth]{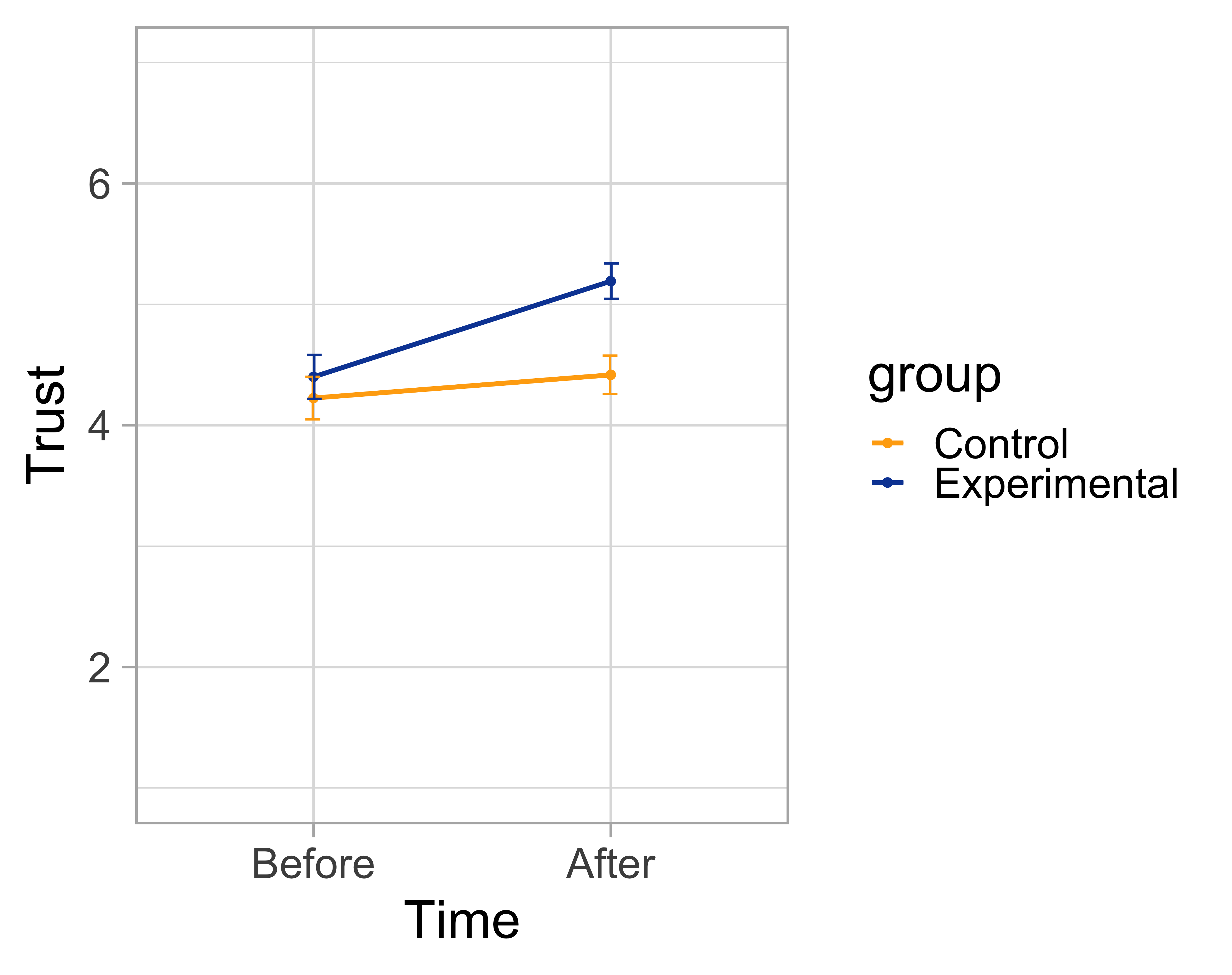}}
\caption{Participants' trust for (a) LIME and (b) Grad-CAM before and after conditions. }
\label{fig:t_interaction}
\end{figure}

For the trust, results show significant main effects of group ($F(1,116) = 4.31,$ $ p=.04$) and time ($F(1,116) = 70.0, p<.001$). The experimental group and the after condition display a higher trust score of participants. We also find a significant interaction effect between group and time ($F(1,116) = 43.7, p<.001$), as displayed in the figure \ref{fig:t_interaction}. Initially, there were no significant differences in trust scores between the experimental and control conditions. During participants' final decision, those who interact with the XAI expert (i.e., experimental condition) report a higher trust score. The enhancements of the experimental group, contrasted with the unchanged trust score of the control group indicate that informativeness and clarity through conversations can help static explanations gain more trust from users. While there exist numerous studies on how explanations of AI predictions can influence users' trust in AI predictions \citep{cheng2019explaining, Yu2019doi, kunkel2019let, zhang2020effect, ma2023who, kim2023help}, to our knowledge, this is the first experiment designed explicitly to gauge the impact of conversations on enhancing participants' trust in explanations. 

\subsection{Analysis of Collected Conversations}
\label{section:analysis_conversation}

We collect 60 free-form conversations between XAI experts and participants from~4 different discipline groups. On average, each conversation had 27.4 turns, with each turn comprising approximately 14.4 tokens. By analyzing the users' questions, we divide them into six categories:
\begin{itemize}[leftmargin=3.5ex, topsep=4pt]
    \item Basic concepts in machine learning: Questions about basic terms and concepts in machine learning that lay people may not know, e.g., what is a deep learning model, what is accuracy, the model structure, and the training data, etc.
    \item Application and performance of machine learning models: Questions about the ability, accuracy, and limitations of machine learning.
    \item Diagram reading: Questions about the explanation diagram generated by Grad-CAM or LIME, e.g., what different colors represent in the heatmap.
    \item Basic concepts in explainable AI: Questions about basic concepts of explanation methods, e.g., what are explanation methods?
    \item Mechanism of explanation methods: Questions about how explanation methods work and how the provided explanation is generated.
    \item Other explanations: Questions that require the generation of other types of explanations on the current predictions, explanations for different predictions, or comparisons between the provided explanation and other explanation methods.
\end{itemize}

Based on this categorization, we build a repository for questions that could occur in the conversations. In total, we collected 397 questions from the four different categories. Table \ref{tab:questions} contains examples and the number of questions in each category.
As observed in Table~\ref{tab:questions}, the questions of participants mainly revolve around basic concepts in machine learning, the fundamentals of explanation methods, and their underlying mechanisms. This trend might be attributed to the multi-disciplinarity of the participants. It suggests that many participants may not be familiar with machine learning models and explanation methods, which is aligned with the real application of explanation methods. Therefore, it's crucial to tailor responses to these questions to help users better understand explanations.
Furthermore, we note a marked interest in new explanations. This could indicate that as users become more familiar with provided explanation examples, they exhibit curiosity about alternative explanation methods and how models might behave under specific scenarios. Concurrently, the diagram reading category contains only~16 questions, implying that explanations generated by Grad-CAM and LIME were relatively straightforward and easy to understand.
The diverse range of questions sourced from our conversations underscores that static, one-off explanations are often insufficient for users to understand them. Engaging in dialogue can provide more dynamic and tailored explanations to users, hence deepening their understanding of static explanations.

Having well internalized their knowledge, experts are often unable to estimate what laypeople know \citep{wittwer2008underestimation}. This phenomenon is also referred to as the ``curse of knowledge'' \citep{camerer1989curse}. As a result, experts tend to overlook potential areas of confusion or make unwarranted assumptions about what is ``common knowledge''. While analyzing the collected conversations, we often find ourselves unable to anticipate the user questions, which corroborates the literature. 
We describe a few examples below.

Several participants misunderstood the idea of the heatmap produced by Grad-CAM as depicting literal heat dissipating from objects. They infer that the model uses the temperature of objects to perform classification. In reality, a heatmap is just a metaphor that visualizes numerical values distributed spatially, which refers to the feature importance in our case.  
This misconception leads to questions about how the heat of objects is measured and why non-living objects are warmer than their environment.  Some example utterances from participants include: \shortquote{So the Grad-cam method basically just refers to the usage of generating a heatmap to capture living matters correct? ... based on the parts of the image that generate more heat?} --P36, \shortquote{basically using heat to predict what is the input right?...how will we know what is the animal or input simply based on heat?} -- P47, \shortquote{if these are pictures, how do they figure out the heat since the animal isn't generating heat} -- P49, \shortquote{So a heat sensor is not required? A heatmap is automatically generated from each photo and analyzed using the model.} -- P52.

A second common misconception is the conflation between the post-hoc explanation technique and the classification models. 
Some example user questions include: \shortquote{is the explanation method what the model uses to classify \& predict what the image is supposed to be?} -- P6, \shortquote{Swin transformer uses LIME model? ... what are the differences between lime model and Swin transformer?} -- P8.
Furthermore, participants face challenges in understanding certain terms commonly used in AI and XAI, even though these terms are frequently used and understood within academic communities. Many participants asked questions about basic concepts in machine learning, such as: \shortquote{what is the explanation method?} -- P7, \shortquote{how do you classify the image?} -- P17, \shortquote{what is the algorithm? does it mean lime? what are deep neural networks?} -- P32, \shortquote{How would you explain the term "perturbations of images" to a five-year-old?} -- P46.

The observations from the interactions between XAI experts and layperson users demonstrate the importance of conversations for users to understand static explanations as they bridge the knowledge gap between the two groups. Conversations can reveal the specific areas of misunderstanding, such as incorrect implicit assumptions the users make and knowledge they lack. 
Hence, conversational explanations may help the AI system communicate with and bring genuine understanding to the users.

\begin{table*}[ht]
    \caption{Overview of Collected Questions. Including categories of questions, examples, and the count of questions in each category.}
    \centering
    \resizebox{\linewidth}{!}{
    \begin{tabular}{>{\centering\arraybackslash}m{3cm}|m{13cm}|m{0.8cm}}
    \toprule
        \multicolumn{1}{>{\centering\arraybackslash}m{3cm}|}{Question Category} &\multicolumn{1}{>{\centering\arraybackslash}m{13cm}|}{Question Examples} & \multicolumn{1}{>{\centering\arraybackslash}m{0.8cm}}{Num}\\
        \midrule
        Basic concepts in machine learning & 
        \begin{itemize}[leftmargin=2ex, topsep=3pt, parsep=0.2ex, after=\vspace{2pt-\baselineskip}, itemsep=-0.25ex,label=\raisebox{0.25ex}{\tiny$\bullet$}]
            \item What is a deep learning model?
            \item What is the image classification task?
            \item How does the model know what features to extract?
        \end{itemize}
         & 85 \\
        \midrule
        Application, performance, and limitations of machine learning models & 
        \begin{itemize}[leftmargin=2ex, topsep=3pt, parsep=0.2ex, after=\vspace{2pt-\baselineskip}, itemsep=-0.25ex,label=\raisebox{0.25ex}{\tiny$\bullet$}]
            \item How about the precision of the classification model?
            \item Where has this Swin Transformer classification method been used in practical applications?
            \item Will the different species of an animal affect the classification model categorizing the animal?
        \end{itemize}
        & 68 \\
        \midrule
        Diagram reading & 
        \begin{itemize}[leftmargin=2ex, topsep=3pt, parsep=0.2ex, after=\vspace{2pt-\baselineskip}, itemsep=-0.25ex,label=\raisebox{0.25ex}{\tiny$\bullet$}]
            \item Are regions colored in red areas that have been identified as containing key features for the animal?
            \item What are the yellow line spots for (in LIME explanations)?
            \item What do the red and blue colors mean (in Grad-CAM explanations)?
        \end{itemize} & 16\\
        \midrule
        Basic concepts of explanation methods &
        \begin{itemize}[leftmargin=2ex, topsep=3pt, parsep=0.2ex, after=\vspace{2pt-\baselineskip}, itemsep=-0.25ex,label=\raisebox{0.25ex}{\tiny$\bullet$}]
            \item What is the explanation model used for?
            \item Can LIME be used without the internet?
            \item What are some limitations of the Grad-CAM (LIME) method?
        \end{itemize} & 95 \\
        \midrule
        Mechanism of explanation methods &
        \begin{itemize}[leftmargin=2ex, topsep=3pt, parsep=0.2ex, after=\vspace{2pt-\baselineskip}, itemsep=-0.25ex,label=\raisebox{0.25ex}{\tiny$\bullet$}]
            \item Why does the (LIME) explanation not highlight all the parts of the leopard?
            \item How LIME model recognize the most important parts for the model prediction?
            \item Seems like the Classification Model and the Explanation Model are trained separately - how can we be sure that the underlying logic of making a prediction is the same for both models?
        \end{itemize} & 91\\
        \midrule
        Other explanations &
        \begin{itemize}[leftmargin=2ex, topsep=3pt, parsep=0.2ex, after=\vspace{2pt-\baselineskip}, itemsep=-0.25ex,label=\raisebox{0.25ex}{\tiny$\bullet$}]
            \item Can you list other visualization methods?
            \item Is there anything special about the Grad-CAM (or LIME) method that is different from others?
            \item What if there are both fishes and humans in an image? How should this image be classified, and can you provide such explanations?
        \end{itemize} & 42\\
    \bottomrule
    \end{tabular}}
    \label{tab:questions}
\end{table*}

\subsection{Implications for building dialogue systems to explain static explanations}
Our study indicates the impact of conversational explanations on user comprehension, acceptance, and trust of static explanations. Static explanations, while informative, may not cater to users with varied backgrounds and expertise. Engaging in conversational explanations provides a dynamic and interactive medium for users to seek clarifications, ask questions, and thereby facilitate a deeper and more personalized understanding. 

The emergence of advanced conversational agents \citep{zhang-etal-2022-history, ni2023recent, shuster2022blenderbot}, especially knowledge-based question-answering \citep{ijcai2021p0611, luo-etal-2023-end, zhang-etal-2023-fc} powered by large language models \citep{touvron2023llama, ouyang2022training, LLMSurvey} paves the way toward conversational agents that can explain model decisions and discuss static explanations. Our study suggests the following desiderata for such agents.

\begin{itemize}[leftmargin=3.5ex, topsep=4pt]
    \item \textit{Extensive knowledge of AI and XAI.} As observed in our study, a large portion of user questions are related to core concepts of machine learning models and explanation methods. To answer those questions, conversational agents need to be trained on a comprehensive corpus encompassing AI and XAI concepts. Besides, in our study, participants also are curious about the applications, performances, and limitations of machine learning models and explanation methods. Therefore, besides answering abstract questions, dialogue systems also should relate them to real-world applications and limitations. 
    
    \item \textit{Capability to generate new explanations as needed.} As an improved understanding of the provided explanations, participants in our study exhibit curiosity about alternative explanation methods and explaining different predictions. Dialogue systems should provide new explanations to users when requested. For instance, if a user is curious about how changing a feature would affect the model output, the system should generate a new explanation with the new feature, which showcases the effect.
    
    \item \textit{Capability to interpret scientific diagrams and visualizations.} A significant portion of AI and XAI explanations often comes in the form of diagrams \citep{selvaraju2017grad, ribeiro2016should}, such as heatmaps or feature importance visualizations. Our study reveals that users have questions related to understanding these diagrams. Answering these questions usually requires an understanding of specific regions of the diagrams, such as answering what parts of the object are highlighted by the yellow line in LIME explanations. Therefore, future dialogue systems should have visual processing capabilities, understanding and interpreting diagrams contextually. For instance, they should be able to recognize colors, patterns, and other graphical elements in heatmaps or charts and relate them to users' questions. The recent development in multimodal large language models \citep{zhu2023minigpt, driess2023palme, gong2023multimodalgpt} is a promising direction to achieve this goal.
    
\end{itemize}

\subsection{Limitations}
Despite the insights gained, there are several limitations that should be acknowledged.
First, the static explanations used in our study are limited. Our experiments focused on feature attribution explanation methods.
The applicability of our findings to other explanation methods, such as example-based explanation methods, remains an open question.
Second, as our main objective was to discern the effects of free-form conversational explanations, we did not delve into the comparative performance of different explanation methods. In our experiments, we intentionally selected explanation examples where the best classification model yielded the most reasonable explanations. The explanation examples discussed by participants and XAI experts were chosen such that they reasonably explain the predictions of the classification model. Future work would be to extend these conversations to include explanations that might be less reliable. 
Third, we explore how conversations foster user trust in explanations in our study. Nevertheless, previous studies \citep{wang2021are, zhang2020effect, taehyun2023improving} have shown that humans may trust AI models even if they make wrong decisions. We do not explore whether users' trust in our study is misplaced, which we leave for future work.
Fourth, we use AI to classify the images. Previous studies \citep{Bankins2022wf, formosa2022medical} found that participants favor humans over AI decision-makers when their decisions directly affect participant welfare. In our study, AI decisions do not directly affect participant welfare. We also did not investigate if the participants preferred conversations with humans or AI chatbots or if their trust in the explanations was affected by that variable.  
Finally, our research is confined to one geographical region and includes only students and staff from the university. Factors such as cultural backgrounds and age-related differences could potentially influence user interactions with XAI and how they seek to clarify confusion. Future studies could involve recruiting participants from diverse countries, regions, and age groups.

\section{Conclusion}
In our work, we conduct Wizard-of-Oz experiments to investigate how free-form conversations assist users in understanding static explanations, promoting trust, and making informed decisions about AI models. Participants engage in conversational explanations with XAI experts to understand how the provided static explanation explains the model decision. To evaluate the effects of conversations, we design objective and subjective measurements. We observe a notable improvement in users' comprehension, acceptance, trust, and collaboration after conversations. From collected conversations, we find that participants’ questions and confusions are diverse and unanticipated. Our findings advocate for the integration of dialogue systems in future XAI designs to ensure more personalized explanations. 


\bibliographystyle{ACM-Reference-Format}
\bibliography{sample-base}


\begin{thebibliography}{149}


\ifx \showCODEN    \undefined \def \showCODEN     #1{\unskip}     \fi
\ifx \showDOI      \undefined \def \showDOI       #1{#1}\fi
\ifx \showISBNx    \undefined \def \showISBNx     #1{\unskip}     \fi
\ifx \showISBNxiii \undefined \def \showISBNxiii  #1{\unskip}     \fi
\ifx \showISSN     \undefined \def \showISSN      #1{\unskip}     \fi
\ifx \showLCCN     \undefined \def \showLCCN      #1{\unskip}     \fi
\ifx \shownote     \undefined \def \shownote      #1{#1}          \fi
\ifx \showarticletitle \undefined \def \showarticletitle #1{#1}   \fi
\ifx \showURL      \undefined \def \showURL       {\relax}        \fi
\providecommand\bibfield[2]{#2}
\providecommand\bibinfo[2]{#2}
\providecommand\natexlab[1]{#1}
\providecommand\showeprint[2][]{arXiv:#2}

\bibitem[Abdul et~al\mbox{.}(2018)]%
        {Abdul2018trends}
\bibfield{author}{\bibinfo{person}{Ashraf Abdul}, \bibinfo{person}{Jo Vermeulen}, \bibinfo{person}{Danding Wang}, \bibinfo{person}{Brian~Y. Lim}, {and} \bibinfo{person}{Mohan Kankanhalli}.} \bibinfo{year}{2018}\natexlab{}.
\newblock \showarticletitle{Trends and Trajectories for Explainable, Accountable and Intelligible Systems: An {HCI} Research Agenda}. In \bibinfo{booktitle}{\emph{Proceedings of the 2018 CHI Conference on Human Factors in Computing Systems}}. \bibinfo{publisher}{Association for Computing Machinery}, \bibinfo{pages}{1–18}.
\newblock


\bibitem[Adadi and Berrada(2018)]%
        {adadi2018peeking}
\bibfield{author}{\bibinfo{person}{Amina Adadi} {and} \bibinfo{person}{Mohammed Berrada}.} \bibinfo{year}{2018}\natexlab{}.
\newblock \showarticletitle{Peeking inside the black-box: A survey on explainable artificial intelligence {(XAI)}}.
\newblock \bibinfo{journal}{\emph{IEEE access}}  \bibinfo{volume}{6} (\bibinfo{year}{2018}), \bibinfo{pages}{52138--52160}.
\newblock
\showISSN{2169-3536}


\bibitem[Adebayo et~al\mbox{.}(2018)]%
        {Adebayo2018SanityChecks}
\bibfield{author}{\bibinfo{person}{Julius Adebayo}, \bibinfo{person}{Justin Gilmer}, \bibinfo{person}{Michael Muelly}, \bibinfo{person}{Ian Goodfellow}, \bibinfo{person}{Moritz Hardt}, {and} \bibinfo{person}{Been Kim}.} \bibinfo{year}{2018}\natexlab{}.
\newblock \showarticletitle{Sanity Checks for Saliency Maps}. In \bibinfo{booktitle}{\emph{Advances in Neural Information Processing Systems}}, Vol.~\bibinfo{volume}{21}. \bibinfo{publisher}{Curran Associates, Inc.}, \bibinfo{pages}{9525--9536}.
\newblock
\urldef\tempurl%
\url{https://doi.org/10.5555/3327546.3327621}
\showDOI{\tempurl}


\bibitem[Adebayo et~al\mbox{.}(2020)]%
        {adebayo2020debugging}
\bibfield{author}{\bibinfo{person}{Julius Adebayo}, \bibinfo{person}{Michael Muelly}, \bibinfo{person}{Ilaria Liccardi}, {and} \bibinfo{person}{Been Kim}.} \bibinfo{year}{2020}\natexlab{}.
\newblock \showarticletitle{Debugging tests for model explanations}. In \bibinfo{booktitle}{\emph{Proceedings of the 34th International Conference on Neural Information Processing Systems}}. \bibinfo{publisher}{Curran Associates Inc.}, \bibinfo{address}{Red Hook, NY, USA}, Article \bibinfo{articleno}{60}, \bibinfo{numpages}{13}~pages.
\newblock
\showISBNx{9781713829546}
\urldef\tempurl%
\url{https://doi.org/10.5555/3495724.3495784}
\showDOI{\tempurl}


\bibitem[Alkan et~al\mbox{.}(2022)]%
        {alkan2022frote}
\bibfield{author}{\bibinfo{person}{Oznur Alkan}, \bibinfo{person}{Dennis Wei}, \bibinfo{person}{Massimiliano Mattetti}, \bibinfo{person}{Rahul Nair}, \bibinfo{person}{Elizabeth Daly}, {and} \bibinfo{person}{Diptikalyan Saha}.} \bibinfo{year}{2022}\natexlab{}.
\newblock \showarticletitle{{FROTE}: Feedback rule-driven oversampling for editing models}. In \bibinfo{booktitle}{\emph{Proceedings of Machine Learning and Systems}}, Vol.~\bibinfo{volume}{4}. \bibinfo{pages}{276--301}.
\newblock


\bibitem[Alvarez-Melis and Jaakkola(2017)]%
        {alvarez2017causal}
\bibfield{author}{\bibinfo{person}{David Alvarez-Melis} {and} \bibinfo{person}{Tommi Jaakkola}.} \bibinfo{year}{2017}\natexlab{}.
\newblock \showarticletitle{A causal framework for explaining the predictions of black-box sequence-to-sequence models}. In \bibinfo{booktitle}{\emph{Proceedings of the 2017 Conference on Empirical Methods in Natural Language Processing}}. \bibinfo{publisher}{Association for Computational Linguistics}, \bibinfo{pages}{412--421}.
\newblock
\urldef\tempurl%
\url{https://doi.org/10.18653/v1/D17-1042}
\showDOI{\tempurl}


\bibitem[Amershi et~al\mbox{.}(2014)]%
        {Amershi_2014}
\bibfield{author}{\bibinfo{person}{Saleema Amershi}, \bibinfo{person}{Maya Cakmak}, \bibinfo{person}{William~Bradley Knox}, {and} \bibinfo{person}{Todd Kulesza}.} \bibinfo{year}{2014}\natexlab{}.
\newblock \showarticletitle{Power to the People: The Role of Humans in Interactive Machine Learning}.
\newblock \bibinfo{journal}{\emph{AI Magazine}} \bibinfo{volume}{35}, \bibinfo{number}{4} (\bibinfo{year}{2014}), \bibinfo{pages}{105--120}.
\newblock
\urldef\tempurl%
\url{https://doi.org/10.1609/aimag.v35i4.2513}
\showDOI{\tempurl}


\bibitem[Anderson and Burnham(2004)]%
        {anderson2004model}
\bibfield{author}{\bibinfo{person}{D Anderson} {and} \bibinfo{person}{K Burnham}.} \bibinfo{year}{2004}\natexlab{}.
\newblock \showarticletitle{Model selection and multi-model inference}.
\newblock \bibinfo{journal}{\emph{Springer-Verlag}}  \bibinfo{volume}{63} (\bibinfo{year}{2004}), \bibinfo{pages}{512}.
\newblock


\bibitem[Ashktorab et~al\mbox{.}(2020)]%
        {ashktorab2020humanai}
\bibfield{author}{\bibinfo{person}{Zahra Ashktorab}, \bibinfo{person}{Q.~Vera Liao}, \bibinfo{person}{Casey Dugan}, \bibinfo{person}{James Johnson}, \bibinfo{person}{Qian Pan}, \bibinfo{person}{Wei Zhang}, \bibinfo{person}{Sadhana Kumaravel}, {and} \bibinfo{person}{Murray Campbell}.} \bibinfo{year}{2020}\natexlab{}.
\newblock \showarticletitle{{Human-AI} Collaboration in a Cooperative Game Setting: Measuring Social Perception and Outcomes}.
\newblock \bibinfo{journal}{\emph{Proceedings of the ACM on Human-Computer Interaction}}  \bibinfo{volume}{4} (\bibinfo{year}{2020}), \bibinfo{numpages}{20}~pages.
\newblock


\bibitem[Bach et~al\mbox{.}(2022)]%
        {Tita2022Systematic}
\bibfield{author}{\bibinfo{person}{Tita~Alissa Bach}, \bibinfo{person}{Amna Khan}, \bibinfo{person}{Harry Hallock}, \bibinfo{person}{Gabriela Beltrão}, {and} \bibinfo{person}{Sonia Sousa}.} \bibinfo{year}{2022}\natexlab{}.
\newblock \showarticletitle{A Systematic Literature Review of User Trust in {AI}-Enabled Systems: An {HCI} Perspective}.
\newblock \bibinfo{journal}{\emph{International Journal of Human–Computer Interaction}} \bibinfo{volume}{0}, \bibinfo{number}{0} (\bibinfo{year}{2022}), \bibinfo{pages}{1--16}.
\newblock
\urldef\tempurl%
\url{https://doi.org/10.1080/10447318.2022.2138826}
\showDOI{\tempurl}


\bibitem[Bahdanau et~al\mbox{.}(2015)]%
        {bahdanau2014neural}
\bibfield{author}{\bibinfo{person}{Dzmitry Bahdanau}, \bibinfo{person}{Kyunghyun Cho}, {and} \bibinfo{person}{Yoshua Bengio}.} \bibinfo{year}{2015}\natexlab{}.
\newblock \showarticletitle{Neural machine translation by jointly learning to align and translate}. In \bibinfo{booktitle}{\emph{Proceedings of 3rd International Conference on Learning Representations}}. \bibinfo{publisher}{openreview}.
\newblock


\bibitem[Bankins et~al\mbox{.}(2022)]%
        {Bankins2022wf}
\bibfield{author}{\bibinfo{person}{Sarah Bankins}, \bibinfo{person}{Paul Formosa}, \bibinfo{person}{Yannick Griep}, {and} \bibinfo{person}{Deborah Richards}.} \bibinfo{year}{2022}\natexlab{}.
\newblock \showarticletitle{{AI} Decision Making with Dignity? Contrasting Workers' Justice Perceptions of Human and {AI} Decision Making in a Human Resource Management Context}.
\newblock \bibinfo{journal}{\emph{Information Systems Frontiers}} \bibinfo{volume}{24}, \bibinfo{number}{3} (\bibinfo{year}{2022}), \bibinfo{pages}{857--875}.
\newblock


\bibitem[Bansal et~al\mbox{.}(2021)]%
        {bansal2021does}
\bibfield{author}{\bibinfo{person}{Gagan Bansal}, \bibinfo{person}{Tongshuang Wu}, \bibinfo{person}{Joyce Zhou}, \bibinfo{person}{Raymond Fok}, \bibinfo{person}{Besmira Nushi}, \bibinfo{person}{Ece Kamar}, \bibinfo{person}{Marco~Tulio Ribeiro}, {and} \bibinfo{person}{Daniel Weld}.} \bibinfo{year}{2021}\natexlab{}.
\newblock \showarticletitle{Does the whole exceed its parts? {The} effect of {AI} explanations on complementary team performance}. In \bibinfo{booktitle}{\emph{Proceedings of the 2021 CHI Conference on Human Factors in Computing Systems}}. \bibinfo{publisher}{Association for Computing Machinery}, \bibinfo{pages}{1--16}.
\newblock


\bibitem[Bhat et~al\mbox{.}(2024)]%
        {bhat_evaluating_2024}
\bibfield{author}{\bibinfo{person}{Shreyas Bhat}, \bibinfo{person}{Joseph~B. Lyons}, \bibinfo{person}{Cong Shi}, {and} \bibinfo{person}{X.~Jessie Yang}.} \bibinfo{year}{2024}\natexlab{}.
\newblock \showarticletitle{Evaluating the Impact of Personalized Value Alignment in Human-Robot Interaction: {Insights} into Trust and Team Performance Outcomes}. In \bibinfo{booktitle}{\emph{Proceedings of the 2024 ACM/IEEE International Conference on Human-Robot Interaction}}. \bibinfo{publisher}{Association for Computing Machinery}, \bibinfo{pages}{32--41}.
\newblock
\urldef\tempurl%
\url{https://doi.org/10.1145/3610977.3634921}
\showDOI{\tempurl}


\bibitem[Bien and Tibshirani(2011)]%
        {bien2011prototype}
\bibfield{author}{\bibinfo{person}{Jacob Bien} {and} \bibinfo{person}{Robert Tibshirani}.} \bibinfo{year}{2011}\natexlab{}.
\newblock \showarticletitle{Prototype selection for interpretable classification}.
\newblock \bibinfo{journal}{\emph{The Annals of Applied Statistics}} \bibinfo{volume}{5}, \bibinfo{number}{4} (\bibinfo{year}{2011}), \bibinfo{pages}{2403--2424}.
\newblock


\bibitem[Biswas and Parikh(2013)]%
        {biswas2013simultaneous}
\bibfield{author}{\bibinfo{person}{Arijit Biswas} {and} \bibinfo{person}{Devi Parikh}.} \bibinfo{year}{2013}\natexlab{}.
\newblock \showarticletitle{Simultaneous active learning of classifiers \& attributes via relative feedback}. In \bibinfo{booktitle}{\emph{Proceedings of the IEEE Conference on Computer Vision and Pattern Recognition}}. \bibinfo{publisher}{IEEE}, \bibinfo{pages}{644--651}.
\newblock


\bibitem[Bodria et~al\mbox{.}(2023)]%
        {bodria2021benchmarking}
\bibfield{author}{\bibinfo{person}{Francesco Bodria}, \bibinfo{person}{Fosca Giannotti}, \bibinfo{person}{Riccardo Guidotti}, \bibinfo{person}{Francesca Naretto}, \bibinfo{person}{Dino Pedreschi}, {and} \bibinfo{person}{Salvatore Rinzivillo}.} \bibinfo{year}{2023}\natexlab{}.
\newblock \showarticletitle{Benchmarking and survey of explanation methods for black box models}.
\newblock \bibinfo{journal}{\emph{Data Mining and Knowledge Discovery}} \bibinfo{volume}{37}, \bibinfo{number}{5} (\bibinfo{year}{2023}), \bibinfo{pages}{1719--1778}.
\newblock
\urldef\tempurl%
\url{https://doi.org/10.1007/s10618-023-00933-9}
\showDOI{\tempurl}


\bibitem[Cai et~al\mbox{.}(2019)]%
        {Cai2019health}
\bibfield{author}{\bibinfo{person}{Carrie~J. Cai}, \bibinfo{person}{Samantha Winter}, \bibinfo{person}{David Steiner}, \bibinfo{person}{Lauren Wilcox}, {and} \bibinfo{person}{Michael Terry}.} \bibinfo{year}{2019}\natexlab{}.
\newblock \showarticletitle{{"Hello AI"}: Uncovering the Onboarding Needs of Medical Practitioners for Human-AI Collaborative Decision-Making}. In \bibinfo{booktitle}{\emph{Proceedings of the ACM on Human-Computer Interaction}}, Vol.~\bibinfo{volume}{3}. \bibinfo{numpages}{24}~pages.
\newblock
\urldef\tempurl%
\url{https://doi.org/10.1145/3359206}
\showDOI{\tempurl}


\bibitem[Camerer et~al\mbox{.}(1989)]%
        {camerer1989curse}
\bibfield{author}{\bibinfo{person}{Colin Camerer}, \bibinfo{person}{George Loewenstein}, {and} \bibinfo{person}{Martin Weber}.} \bibinfo{year}{1989}\natexlab{}.
\newblock \showarticletitle{The curse of knowledge in economic settings: An experimental analysis}.
\newblock \bibinfo{journal}{\emph{Journal of Political Economy}} \bibinfo{volume}{97}, \bibinfo{number}{5} (\bibinfo{year}{1989}), \bibinfo{pages}{1232--1254}.
\newblock


\bibitem[Carissoli et~al\mbox{.}(2023)]%
        {Claudia2023}
\bibfield{author}{\bibinfo{person}{Claudia Carissoli}, \bibinfo{person}{Luca Negri}, \bibinfo{person}{Marta Bassi}, \bibinfo{person}{Fabio~Alexander Storm}, {and} \bibinfo{person}{Antonella~Delle Fave}.} \bibinfo{year}{2023}\natexlab{}.
\newblock \showarticletitle{Mental Workload and Human-Robot Interaction in Collaborative Tasks: A Scoping Review}.
\newblock \bibinfo{journal}{\emph{International Journal of Human–Computer Interaction}} \bibinfo{volume}{0}, \bibinfo{number}{0} (\bibinfo{year}{2023}), \bibinfo{pages}{1--20}.
\newblock
\urldef\tempurl%
\url{https://doi.org/10.1080/10447318.2023.2254639}
\showDOI{\tempurl}


\bibitem[Carolin~Ebermann and Weibelzahl(2023)]%
        {Carolin2023ExplainableAI}
\bibfield{author}{\bibinfo{person}{Matthias~Selisky Carolin~Ebermann} {and} \bibinfo{person}{Stephan Weibelzahl}.} \bibinfo{year}{2023}\natexlab{}.
\newblock \showarticletitle{Explainable {AI}: The Effect of Contradictory Decisions and Explanations on Users’ Acceptance of {AI} Systems}.
\newblock \bibinfo{journal}{\emph{International Journal of Human–Computer Interaction}} \bibinfo{volume}{39}, \bibinfo{number}{9} (\bibinfo{year}{2023}), \bibinfo{pages}{1807--1826}.
\newblock
\urldef\tempurl%
\url{https://doi.org/10.1080/10447318.2022.2126812}
\showDOI{\tempurl}


\bibitem[Caruana et~al\mbox{.}(2015)]%
        {Caruana2015health}
\bibfield{author}{\bibinfo{person}{Rich Caruana}, \bibinfo{person}{Yin Lou}, \bibinfo{person}{Johannes Gehrke}, \bibinfo{person}{Paul Koch}, \bibinfo{person}{Marc Sturm}, {and} \bibinfo{person}{Noemie Elhadad}.} \bibinfo{year}{2015}\natexlab{}.
\newblock \showarticletitle{Intelligible Models for HealthCare: Predicting Pneumonia Risk and Hospital 30-Day Readmission}. In \bibinfo{booktitle}{\emph{Proceedings of the 21th ACM SIGKDD International Conference on Knowledge Discovery and Data Mining}}. \bibinfo{publisher}{ACM}, \bibinfo{pages}{1721–1730}.
\newblock
\urldef\tempurl%
\url{https://doi.org/10.1145/2783258.2788613}
\showDOI{\tempurl}


\bibitem[Chen et~al\mbox{.}(2019)]%
        {chen2019looks}
\bibfield{author}{\bibinfo{person}{Chaofan Chen}, \bibinfo{person}{Oscar Li}, \bibinfo{person}{Chaofan Tao}, \bibinfo{person}{Alina~Jade Barnett}, \bibinfo{person}{Jonathan Su}, {and} \bibinfo{person}{Cynthia Rudin}.} \bibinfo{year}{2019}\natexlab{}.
\newblock \bibinfo{booktitle}{\emph{This Looks like That: Deep Learning for Interpretable Image Recognition}}.
\newblock \bibinfo{publisher}{Curran Associates Inc.}
\newblock


\bibitem[Chen et~al\mbox{.}(2021)]%
        {chen2021hydra}
\bibfield{author}{\bibinfo{person}{Yuanyuan Chen}, \bibinfo{person}{Boyang Li}, \bibinfo{person}{Han Yu}, \bibinfo{person}{Pengcheng Wu}, {and} \bibinfo{person}{Chunyan Miao}.} \bibinfo{year}{2021}\natexlab{}.
\newblock \showarticletitle{Hydra: Hypergradient data relevance analysis for interpreting deep neural networks}. In \bibinfo{booktitle}{\emph{Proceedings of the AAAI Conference on Artificial Intelligence}}, Vol.~\bibinfo{volume}{35}. \bibinfo{publisher}{MIT Press}, \bibinfo{pages}{7081--7089}.
\newblock


\bibitem[Cheng et~al\mbox{.}(2019)]%
        {cheng2019explaining}
\bibfield{author}{\bibinfo{person}{Hao-Fei Cheng}, \bibinfo{person}{Ruotong Wang}, \bibinfo{person}{Zheng Zhang}, \bibinfo{person}{Fiona O'connell}, \bibinfo{person}{Terrance Gray}, \bibinfo{person}{F~Maxwell Harper}, {and} \bibinfo{person}{Haiyi Zhu}.} \bibinfo{year}{2019}\natexlab{}.
\newblock \showarticletitle{Explaining decision-making algorithms through {UI}: Strategies to help non-expert stakeholders}. In \bibinfo{booktitle}{\emph{Proceedings of the 2019 CHI conference on human factors in computing systems}}. \bibinfo{publisher}{Association for Computing Machinery}, \bibinfo{pages}{1--12}.
\newblock


\bibitem[Clark and Brennan(1991)]%
        {clark1991grounding}
\bibfield{author}{\bibinfo{person}{Herbert~H. Clark} {and} \bibinfo{person}{Susan~E. Brennan}.} \bibinfo{year}{1991}\natexlab{}.
\newblock \showarticletitle{Grounding in Communication}.
\newblock In \bibinfo{booktitle}{\emph{Perspectives on Socially Shared Cognition}}, \bibfield{editor}{\bibinfo{person}{Lauren~B Resnick}, \bibinfo{person}{John~M Levine}, {and} \bibinfo{person}{Stephanie~D Teasley}} (Eds.). \bibinfo{publisher}{American Psychological Association}, \bibinfo{pages}{127--149}.
\newblock
\urldef\tempurl%
\url{https://doi.org/10.1037/10096-006}
\showDOI{\tempurl}


\bibitem[Clark and Marshall(1981)]%
        {clark1981definite}
\bibfield{author}{\bibinfo{person}{Herbert~H. Clark} {and} \bibinfo{person}{Catherine~R. Marshall}.} \bibinfo{year}{1981}\natexlab{}.
\newblock \showarticletitle{Definite Knowledge and Mutual Knowledge}.
\newblock In \bibinfo{booktitle}{\emph{Elements of Discourse Understanding}}, \bibfield{editor}{\bibinfo{person}{Aravind~K. Joshi}, \bibinfo{person}{Bonnie~L. Webber}, {and} \bibinfo{person}{Ivan~A. Sag}} (Eds.). \bibinfo{pages}{10--63}.
\newblock


\bibitem[Cortez and Embrechts(2013)]%
        {cortez2013using}
\bibfield{author}{\bibinfo{person}{Paulo Cortez} {and} \bibinfo{person}{Mark~J Embrechts}.} \bibinfo{year}{2013}\natexlab{}.
\newblock \showarticletitle{Using sensitivity analysis and visualization techniques to open black box data mining models}.
\newblock \bibinfo{journal}{\emph{Information Sciences}}  \bibinfo{volume}{225} (\bibinfo{year}{2013}), \bibinfo{pages}{1--17}.
\newblock


\bibitem[Croce et~al\mbox{.}(2019)]%
        {croce2019auditing}
\bibfield{author}{\bibinfo{person}{Danilo Croce}, \bibinfo{person}{Daniele Rossini}, {and} \bibinfo{person}{Roberto Basili}.} \bibinfo{year}{2019}\natexlab{}.
\newblock \showarticletitle{Auditing deep learning processes through kernel-based explanatory models}. In \bibinfo{booktitle}{\emph{Proceedings of the 2019 Conference on Empirical Methods in Natural Language Processing and the 9th International Joint Conference on Natural Language Processing (EMNLP-IJCNLP)}}. \bibinfo{publisher}{Association for Computational Linguistics}, \bibinfo{pages}{4037--4046}.
\newblock


\bibitem[Danilevsky et~al\mbox{.}(2020)]%
        {danilevsky-etal-2020-survey}
\bibfield{author}{\bibinfo{person}{Marina Danilevsky}, \bibinfo{person}{Kun Qian}, \bibinfo{person}{Ranit Aharonov}, \bibinfo{person}{Yannis Katsis}, \bibinfo{person}{Ban Kawas}, {and} \bibinfo{person}{Prithviraj Sen}.} \bibinfo{year}{2020}\natexlab{}.
\newblock \showarticletitle{A Survey of the State of Explainable {AI} for Natural Language Processing}. In \bibinfo{booktitle}{\emph{Proceedings of the 1st Conference of the Asia-Pacific Chapter of the Association for Computational Linguistics and the 10th International Joint Conference on Natural Language Processing}}. \bibinfo{publisher}{Association for Computational Linguistics}, \bibinfo{pages}{447--459}.
\newblock


\bibitem[Davis(1989)]%
        {davis1989perceived}
\bibfield{author}{\bibinfo{person}{Fred~D. Davis}.} \bibinfo{year}{1989}\natexlab{}.
\newblock \showarticletitle{Perceived Usefulness, Perceived Ease of Use, and User Acceptance of Information Technology}.
\newblock \bibinfo{journal}{\emph{MIS Quarterly}} \bibinfo{volume}{13}, \bibinfo{number}{3} (\bibinfo{year}{1989}), \bibinfo{pages}{319--340}.
\newblock


\bibitem[Davis et~al\mbox{.}(1989)]%
        {davis1989user}
\bibfield{author}{\bibinfo{person}{Fred~D Davis}, \bibinfo{person}{Richard~P Bagozzi}, {and} \bibinfo{person}{Paul~R Warshaw}.} \bibinfo{year}{1989}\natexlab{}.
\newblock \showarticletitle{User acceptance of computer technology: A comparison of two theoretical models}.
\newblock \bibinfo{journal}{\emph{Management science}} \bibinfo{volume}{35}, \bibinfo{number}{8} (\bibinfo{year}{1989}), \bibinfo{pages}{982--1003}.
\newblock


\bibitem[Deng et~al\mbox{.}(2009)]%
        {deng2009imagenet}
\bibfield{author}{\bibinfo{person}{Jia Deng}, \bibinfo{person}{Wei Dong}, \bibinfo{person}{Richard Socher}, \bibinfo{person}{Li-Jia Li}, \bibinfo{person}{Kai Li}, {and} \bibinfo{person}{Fei-Fei Li}.} \bibinfo{year}{2009}\natexlab{}.
\newblock \showarticletitle{ImageNet: A large-scale hierarchical image database}. In \bibinfo{booktitle}{\emph{Proceedings of the 2009 IEEE conference on computer vision and pattern recognition}}. \bibinfo{publisher}{IEEE}, \bibinfo{pages}{248--255}.
\newblock
\urldef\tempurl%
\url{https://doi.org/10.1109/CVPR.2009.5206848}
\showDOI{\tempurl}


\bibitem[Diop et~al\mbox{.}(2019)]%
        {diop2019extension}
\bibfield{author}{\bibinfo{person}{El~Bachir Diop}, \bibinfo{person}{Shengchuan Zhao}, {and} \bibinfo{person}{Tran~Van Duy}.} \bibinfo{year}{2019}\natexlab{}.
\newblock \showarticletitle{An extension of the technology acceptance model for understanding travelers’ adoption of variable message signs}.
\newblock \bibinfo{journal}{\emph{PLoS one}} \bibinfo{volume}{14}, \bibinfo{number}{4} (\bibinfo{year}{2019}).
\newblock


\bibitem[Doshi-Velez and Kim(2017)]%
        {doshi2017towards}
\bibfield{author}{\bibinfo{person}{Finale Doshi-Velez} {and} \bibinfo{person}{Been Kim}.} \bibinfo{year}{2017}\natexlab{}.
\newblock \showarticletitle{Towards a rigorous science of interpretable machine learning}.
\newblock \bibinfo{journal}{\emph{arXiv preprint arXiv:1702.08608}} (\bibinfo{year}{2017}).
\newblock


\bibitem[Doshi-Velez et~al\mbox{.}(2015)]%
        {doshi2015graph}
\bibfield{author}{\bibinfo{person}{Finale Doshi-Velez}, \bibinfo{person}{Byron~C Wallace}, {and} \bibinfo{person}{Ryan Adams}.} \bibinfo{year}{2015}\natexlab{}.
\newblock \showarticletitle{{Graph-Sparse LDA}: A topic model with structured sparsity}. In \bibinfo{booktitle}{\emph{Twenty-Ninth AAAI conference on artificial intelligence}}, Vol.~\bibinfo{volume}{29}. \bibinfo{publisher}{AAAI Press}, \bibinfo{pages}{2575--2581}.
\newblock
\urldef\tempurl%
\url{https://doi.org/10.1609/aaai.v29i1.9603}
\showDOI{\tempurl}


\bibitem[Driess et~al\mbox{.}(2023)]%
        {driess2023palme}
\bibfield{author}{\bibinfo{person}{Danny Driess}, \bibinfo{person}{Fei Xia}, \bibinfo{person}{Mehdi S.~M. Sajjadi}, \bibinfo{person}{Corey Lynch}, \bibinfo{person}{Aakanksha Chowdhery}, \bibinfo{person}{Brian Ichter}, \bibinfo{person}{Ayzaan Wahid}, \bibinfo{person}{Jonathan Tompson}, \bibinfo{person}{Quan Vuong}, \bibinfo{person}{Tianhe Yu}, \bibinfo{person}{Wenlong Huang}, \bibinfo{person}{Yevgen Chebotar}, \bibinfo{person}{Pierre Sermanet}, \bibinfo{person}{Daniel Duckworth}, \bibinfo{person}{Sergey Levine}, \bibinfo{person}{Vincent Vanhoucke}, \bibinfo{person}{Karol Hausman}, \bibinfo{person}{Marc Toussaint}, \bibinfo{person}{Klaus Greff}, \bibinfo{person}{Andy Zeng}, \bibinfo{person}{Igor Mordatch}, {and} \bibinfo{person}{Pete Florence}.} \bibinfo{year}{2023}\natexlab{}.
\newblock \showarticletitle{{PaLM-E}: an embodied multimodal language model}. In \bibinfo{booktitle}{\emph{Proceedings of the 40th International Conference on Machine Learning}}. \bibinfo{publisher}{JMLR.org}, Article \bibinfo{articleno}{340}, \bibinfo{numpages}{20}~pages.
\newblock


\bibitem[D’Avella et~al\mbox{.}(2022)]%
        {Avella2022}
\bibfield{author}{\bibinfo{person}{Salvatore D’Avella}, \bibinfo{person}{Gerardo Camacho-Gonzalez}, {and} \bibinfo{person}{Paolo Tripicchio}.} \bibinfo{year}{2022}\natexlab{}.
\newblock \showarticletitle{On Multi-Agent Cognitive Cooperation: Can Virtual Agents Behave like Humans?}
\newblock \bibinfo{journal}{\emph{Neurocomputing}} \bibinfo{volume}{480}, \bibinfo{number}{C} (\bibinfo{year}{2022}), \bibinfo{pages}{27–38}.
\newblock
\urldef\tempurl%
\url{https://doi.org/10.1016/j.neucom.2022.01.025}
\showDOI{\tempurl}


\bibitem[Ehsan et~al\mbox{.}(2024)]%
        {ehsan2021explainable}
\bibfield{author}{\bibinfo{person}{Upol Ehsan}, \bibinfo{person}{Samir Passi}, \bibinfo{person}{Q~Vera Liao}, \bibinfo{person}{Larry Chan}, \bibinfo{person}{I-Hsiang Lee}, \bibinfo{person}{Michael Muller}, {and} \bibinfo{person}{Mark~O Riedl}.} \bibinfo{year}{2024}\natexlab{}.
\newblock \showarticletitle{The who in explainable {AI}: How {AI} background shapes perceptions of {AI} explanations}. In \bibinfo{booktitle}{\emph{Proceedings of the CHI Conference on Human Factors in Computing Systems}}. \bibinfo{publisher}{Association for Computing Machinery}, \bibinfo{pages}{1--32}.
\newblock
\urldef\tempurl%
\url{https://doi.org/10.1145/3613904.3642474}
\showDOI{\tempurl}


\bibitem[Fails and Olsen~Jr(2003)]%
        {fails2003interactive}
\bibfield{author}{\bibinfo{person}{Jerry~Alan Fails} {and} \bibinfo{person}{Dan~R Olsen~Jr}.} \bibinfo{year}{2003}\natexlab{}.
\newblock \showarticletitle{Interactive machine learning}. In \bibinfo{booktitle}{\emph{Proceedings of the 8th international conference on Intelligent user interfaces}} (Miami, Florida, USA). \bibinfo{pages}{39--45}.
\newblock


\bibitem[Feldhus et~al\mbox{.}(2022)]%
        {feldhus2022mediators}
\bibfield{author}{\bibinfo{person}{Nils Feldhus}, \bibinfo{person}{Ajay~Madhavan Ravichandran}, {and} \bibinfo{person}{Sebastian M{\"o}ller}.} \bibinfo{year}{2022}\natexlab{}.
\newblock \showarticletitle{Mediators: Conversational agents explaining {NLP} model behavior}.
\newblock \bibinfo{journal}{\emph{arXiv preprint arXiv:2206.06029}} (\bibinfo{year}{2022}).
\newblock


\bibitem[Feng and Boyd-Graber(2019)]%
        {feng2019what}
\bibfield{author}{\bibinfo{person}{Shi Feng} {and} \bibinfo{person}{Jordan Boyd-Graber}.} \bibinfo{year}{2019}\natexlab{}.
\newblock \showarticletitle{What Can {AI} Do for Me? {E}valuating Machine Learning Interpretations in Cooperative Play}. In \bibinfo{booktitle}{\emph{Proceedings of the 24th International Conference on Intelligent User Interfaces}}. \bibinfo{publisher}{Association for Computing Machinery}, \bibinfo{pages}{229–239}.
\newblock
\showISBNx{9781450362726}
\urldef\tempurl%
\url{https://doi.org/10.1145/3301275.3302265}
\showDOI{\tempurl}


\bibitem[Flathmann et~al\mbox{.}(2023)]%
        {Christopher2023Acceptance}
\bibfield{author}{\bibinfo{person}{Christopher Flathmann}, \bibinfo{person}{Beau~G. Schelble}, \bibinfo{person}{Nathan~J. McNeese}, \bibinfo{person}{Bart Knijnenburg}, \bibinfo{person}{Anand~K. Gramopadhye}, {and} \bibinfo{person}{Kapil~Chalil Madathil}.} \bibinfo{year}{2023}\natexlab{}.
\newblock \showarticletitle{The Purposeful Presentation of {AI} Teammates: Impacts on Human Acceptance and Perception}.
\newblock \bibinfo{journal}{\emph{International Journal of Human–Computer Interaction}} \bibinfo{volume}{0}, \bibinfo{number}{0} (\bibinfo{year}{2023}), \bibinfo{pages}{1--18}.
\newblock
\urldef\tempurl%
\url{https://doi.org/10.1080/10447318.2023.2254984}
\showDOI{\tempurl}


\bibitem[Formosa et~al\mbox{.}(2022)]%
        {formosa2022medical}
\bibfield{author}{\bibinfo{person}{Paul Formosa}, \bibinfo{person}{Wendy Rogers}, \bibinfo{person}{Yannick Griep}, \bibinfo{person}{Sarah Bankins}, {and} \bibinfo{person}{Deborah Richards}.} \bibinfo{year}{2022}\natexlab{}.
\newblock \showarticletitle{Medical {AI} and human dignity: {Contrasting} perceptions of human and artificially intelligent {(AI)} decision making in diagnostic and medical resource allocation contexts}.
\newblock \bibinfo{journal}{\emph{Computers in Human Behavior}}  \bibinfo{volume}{133} (\bibinfo{year}{2022}), \bibinfo{pages}{107296}.
\newblock
\urldef\tempurl%
\url{https://doi.org/10.1016/j.chb.2022.107296}
\showDOI{\tempurl}


\bibitem[Gero et~al\mbox{.}(2020)]%
        {gero2020mental}
\bibfield{author}{\bibinfo{person}{Katy~Ilonka Gero}, \bibinfo{person}{Zahra Ashktorab}, \bibinfo{person}{Casey Dugan}, \bibinfo{person}{Qian Pan}, \bibinfo{person}{James Johnson}, \bibinfo{person}{Werner Geyer}, \bibinfo{person}{Maria Ruiz}, \bibinfo{person}{Sarah Miller}, \bibinfo{person}{David~R. Millen}, \bibinfo{person}{Murray Campbell}, \bibinfo{person}{Sadhana Kumaravel}, {and} \bibinfo{person}{Wei Zhang}.} \bibinfo{year}{2020}\natexlab{}.
\newblock \showarticletitle{Mental Models of {AI} Agents in a Cooperative Game Setting}. In \bibinfo{booktitle}{\emph{Proceedings of the 2020 CHI Conference on Human Factors in Computing Systems}}. \bibinfo{publisher}{Association for Computing Machinery}, \bibinfo{pages}{1–12}.
\newblock
\showISBNx{9781450367080}
\urldef\tempurl%
\url{https://doi.org/10.1145/3313831.3376316}
\showDOI{\tempurl}


\bibitem[Glass et~al\mbox{.}(2008)]%
        {Glass2009towards}
\bibfield{author}{\bibinfo{person}{Alyssa Glass}, \bibinfo{person}{Deborah~L. McGuinness}, {and} \bibinfo{person}{Michael Wolverton}.} \bibinfo{year}{2008}\natexlab{}.
\newblock \showarticletitle{Toward establishing trust in adaptive agents}. In \bibinfo{booktitle}{\emph{Proceedings of the 13th International Conference on Intelligent User Interfaces}} (Gran Canaria, Spain). \bibinfo{pages}{227–236}.
\newblock
\urldef\tempurl%
\url{https://doi.org/10.1145/1378773.1378804}
\showDOI{\tempurl}


\bibitem[Gong et~al\mbox{.}(2023)]%
        {gong2023multimodalgpt}
\bibfield{author}{\bibinfo{person}{Tao Gong}, \bibinfo{person}{Chengqi Lyu}, \bibinfo{person}{Shilong Zhang}, \bibinfo{person}{Yudong Wang}, \bibinfo{person}{Miao Zheng}, \bibinfo{person}{Qian Zhao}, \bibinfo{person}{Kuikun Liu}, \bibinfo{person}{Wenwei Zhang}, \bibinfo{person}{Ping Luo}, {and} \bibinfo{person}{Kai Chen}.} \bibinfo{year}{2023}\natexlab{}.
\newblock \bibinfo{title}{{MultiModal-GPT}: A Vision and Language Model for Dialogue with Humans}.
\newblock
\newblock
\showeprint[arxiv]{2305.04790}~[cs.CV]


\bibitem[Gonz{\'a}lez et~al\mbox{.}(2021)]%
        {gonzalez-etal-2021-explanations}
\bibfield{author}{\bibinfo{person}{Ana~Valeria Gonz{\'a}lez}, \bibinfo{person}{Gagan Bansal}, \bibinfo{person}{Angela Fan}, \bibinfo{person}{Yashar Mehdad}, \bibinfo{person}{Robin Jia}, {and} \bibinfo{person}{Srinivasan Iyer}.} \bibinfo{year}{2021}\natexlab{}.
\newblock \showarticletitle{Do Explanations Help Users Detect Errors in Open-Domain {QA}? {An} Evaluation of Spoken vs. Visual Explanations}. In \bibinfo{booktitle}{\emph{Findings of the Association for Computational Linguistics: ACL-IJCNLP 2021}}, \bibfield{editor}{\bibinfo{person}{Chengqing Zong}, \bibinfo{person}{Fei Xia}, \bibinfo{person}{Wenjie Li}, {and} \bibinfo{person}{Roberto Navigli}} (Eds.). \bibinfo{publisher}{Association for Computational Linguistics}, \bibinfo{pages}{1103--1116}.
\newblock
\urldef\tempurl%
\url{https://doi.org/10.18653/v1/2021.findings-acl.95}
\showDOI{\tempurl}


\bibitem[Guesmi et~al\mbox{.}(2023)]%
        {Mouadh2023interactive}
\bibfield{author}{\bibinfo{person}{Mouadh Guesmi}, \bibinfo{person}{Mohamed~Amine Chatti}, \bibinfo{person}{Shoeb Joarder}, \bibinfo{person}{Qurat~Ul Ain}, \bibinfo{person}{Rawaa Alatrash}, \bibinfo{person}{Clara Siepmann}, {and} \bibinfo{person}{Tannaz Vahidi}.} \bibinfo{year}{2023}\natexlab{}.
\newblock \showarticletitle{Interactive Explanation with Varying Level of Details in an Explainable Scientific Literature Recommender System}.
\newblock \bibinfo{journal}{\emph{International Journal of Human–Computer Interaction}} \bibinfo{volume}{0}, \bibinfo{number}{0} (\bibinfo{year}{2023}), \bibinfo{pages}{1--22}.
\newblock
\urldef\tempurl%
\url{https://doi.org/10.1080/10447318.2023.2262797}
\showDOI{\tempurl}


\bibitem[Guo et~al\mbox{.}(2023)]%
        {guo_enabling_2023}
\bibfield{author}{\bibinfo{person}{Yaohui Guo}, \bibinfo{person}{X.~Jessie Yang}, {and} \bibinfo{person}{Cong Shi}.} \bibinfo{year}{2023}\natexlab{}.
\newblock \showarticletitle{Enabling Team of Teams: {A} Trust Inference and Propagation {(TIP)} Model in Multi-Human Multi-Robot Teams}. In \bibinfo{booktitle}{\emph{Robotics: {Science} and {Systems} {XIX}}}. \bibinfo{publisher}{Association for Computing Machinery}, \bibinfo{pages}{639–643}.
\newblock
\urldef\tempurl%
\url{https://doi.org/10.15607/RSS.2023.XIX.003}
\showDOI{\tempurl}


\bibitem[Ha and Kim(2023)]%
        {taehyun2023improving}
\bibfield{author}{\bibinfo{person}{Taehyun Ha} {and} \bibinfo{person}{Sangyeon Kim}.} \bibinfo{year}{2023}\natexlab{}.
\newblock \showarticletitle{Improving Trust in {AI} with Mitigating Confirmation Bias: Effects of Explanation Type and Debiasing Strategy for Decision-Making with Explainable {AI}}.
\newblock \bibinfo{journal}{\emph{International Journal of Human–Computer Interaction}} \bibinfo{volume}{0}, \bibinfo{number}{0} (\bibinfo{year}{2023}), \bibinfo{pages}{1--12}.
\newblock
\urldef\tempurl%
\url{https://doi.org/10.1080/10447318.2023.2285640}
\showDOI{\tempurl}


\bibitem[H\"{a}uslschmid et~al\mbox{.}(2017)]%
        {supportingTrust2017}
\bibfield{author}{\bibinfo{person}{Renate H\"{a}uslschmid}, \bibinfo{person}{Max von B\"{u}low}, \bibinfo{person}{Bastian Pfleging}, {and} \bibinfo{person}{Andreas Butz}.} \bibinfo{year}{2017}\natexlab{}.
\newblock \showarticletitle{SupportingTrust in Autonomous Driving}. In \bibinfo{booktitle}{\emph{Proceedings of the 22nd International Conference on Intelligent User Interfaces}}. \bibinfo{publisher}{Association for Computing Machinery}, \bibinfo{pages}{319–329}.
\newblock
\showISBNx{9781450343480}
\urldef\tempurl%
\url{https://doi.org/10.1145/3025171.3025198}
\showDOI{\tempurl}


\bibitem[He et~al\mbox{.}(2023)]%
        {xin2023what}
\bibfield{author}{\bibinfo{person}{Xin He}, \bibinfo{person}{Yeyi Hong}, \bibinfo{person}{Xi Zheng}, {and} \bibinfo{person}{Yong Zhang}.} \bibinfo{year}{2023}\natexlab{}.
\newblock \showarticletitle{What Are the Users’ Needs? {Design} of a User-Centered Explainable Artificial Intelligence Diagnostic System}.
\newblock \bibinfo{journal}{\emph{International Journal of Human–Computer Interaction}} \bibinfo{volume}{39}, \bibinfo{number}{7} (\bibinfo{year}{2023}), \bibinfo{pages}{1519--1542}.
\newblock
\urldef\tempurl%
\url{https://doi.org/10.1080/10447318.2022.2095093}
\showDOI{\tempurl}


\bibitem[Herse et~al\mbox{.}(2023)]%
        {Sarita2023using}
\bibfield{author}{\bibinfo{person}{Sarita Herse}, \bibinfo{person}{Jonathan Vitale}, {and} \bibinfo{person}{Mary-Anne Williams}.} \bibinfo{year}{2023}\natexlab{}.
\newblock \showarticletitle{Using Agent Features to Influence User Trust, Decision Making and Task Outcome during Human-Agent Collaboration}.
\newblock \bibinfo{journal}{\emph{International Journal of Human–Computer Interaction}} \bibinfo{volume}{39}, \bibinfo{number}{9} (\bibinfo{year}{2023}), \bibinfo{pages}{1740--1761}.
\newblock
\urldef\tempurl%
\url{https://doi.org/10.1080/10447318.2022.2150691}
\showDOI{\tempurl}


\bibitem[Hoffman et~al\mbox{.}(2018)]%
        {hoffman2018metrics}
\bibfield{author}{\bibinfo{person}{Robert~R Hoffman}, \bibinfo{person}{Shane~T Mueller}, \bibinfo{person}{Gary Klein}, {and} \bibinfo{person}{Jordan Litman}.} \bibinfo{year}{2018}\natexlab{}.
\newblock \showarticletitle{Metrics for explainable {AI}: {Challenges} and prospects}.
\newblock \bibinfo{journal}{\emph{arXiv preprint arXiv:1812.04608}} (\bibinfo{year}{2018}).
\newblock


\bibitem[Hohman et~al\mbox{.}(2019)]%
        {hohman2019gamut}
\bibfield{author}{\bibinfo{person}{Fred Hohman}, \bibinfo{person}{Andrew Head}, \bibinfo{person}{Rich Caruana}, \bibinfo{person}{Robert DeLine}, {and} \bibinfo{person}{Steven~M. Drucker}.} \bibinfo{year}{2019}\natexlab{}.
\newblock \showarticletitle{Gamut: A Design Probe to Understand How Data Scientists Understand Machine Learning Models}. In \bibinfo{booktitle}{\emph{Proceedings of the 2019 CHI Conference on Human Factors in Computing Systems}} (Glasgow, Scotland Uk). \bibinfo{publisher}{Association for Computing Machinery}, \bibinfo{pages}{1–13}.
\newblock
\showISBNx{9781450359702}
\urldef\tempurl%
\url{https://doi.org/10.1145/3290605.3300809}
\showDOI{\tempurl}


\bibitem[Hu et~al\mbox{.}(2018)]%
        {hu2018locally}
\bibfield{author}{\bibinfo{person}{Linwei Hu}, \bibinfo{person}{Jie Chen}, \bibinfo{person}{Vijayan~N Nair}, {and} \bibinfo{person}{Agus Sudjianto}.} \bibinfo{year}{2018}\natexlab{}.
\newblock \showarticletitle{Locally interpretable models and effects based on supervised partitioning {(LIME-SUP)}}.
\newblock \bibinfo{journal}{\emph{arXiv preprint arXiv:1806.00663}} (\bibinfo{year}{2018}).
\newblock


\bibitem[Idahl et~al\mbox{.}(2021)]%
        {idahl2021towards}
\bibfield{author}{\bibinfo{person}{Maximilian Idahl}, \bibinfo{person}{Lijun Lyu}, \bibinfo{person}{Ujwal Gadiraju}, {and} \bibinfo{person}{Avishek Anand}.} \bibinfo{year}{2021}\natexlab{}.
\newblock \showarticletitle{Towards Benchmarking the Utility of Explanations for Model Debugging}. In \bibinfo{booktitle}{\emph{Proceedings of the First Workshop on Trustworthy Natural Language Processing}}. \bibinfo{publisher}{Association for Computational Linguistics}, \bibinfo{pages}{68--73}.
\newblock
\urldef\tempurl%
\url{https://doi.org/10.18653/v1/2021.trustnlp-1.8}
\showDOI{\tempurl}


\bibitem[Ignatiev et~al\mbox{.}(2019)]%
        {ignatiev2019abduction}
\bibfield{author}{\bibinfo{person}{Alexey Ignatiev}, \bibinfo{person}{Nina Narodytska}, {and} \bibinfo{person}{Joao Marques-Silva}.} \bibinfo{year}{2019}\natexlab{}.
\newblock \showarticletitle{Abduction-based explanations for machine learning models}. In \bibinfo{booktitle}{\emph{Proceedings of the AAAI Conference on Artificial Intelligence}}. \bibinfo{pages}{1511--1519}.
\newblock


\bibitem[Jacovi and Goldberg(2020)]%
        {jacovi2020towards}
\bibfield{author}{\bibinfo{person}{Alon Jacovi} {and} \bibinfo{person}{Yoav Goldberg}.} \bibinfo{year}{2020}\natexlab{}.
\newblock \showarticletitle{Towards Faithfully Interpretable {NLP} Systems: How Should We Define and Evaluate Faithfulness?}. In \bibinfo{booktitle}{\emph{Proceedings of the 58th Annual Meeting of the Association for Computational Linguistics}} (Online). \bibinfo{publisher}{Association for Computational Linguistics}, \bibinfo{pages}{4198--4205}.
\newblock


\bibitem[Jeyakumar et~al\mbox{.}(2020)]%
        {jeyakumar2020can}
\bibfield{author}{\bibinfo{person}{Jeya~Vikranth Jeyakumar}, \bibinfo{person}{Joseph Noor}, \bibinfo{person}{Yu-Hsi Cheng}, \bibinfo{person}{Luis Garcia}, {and} \bibinfo{person}{Mani Srivastava}.} \bibinfo{year}{2020}\natexlab{}.
\newblock \showarticletitle{How Can {I} Explain This to You? {An} Empirical Study of Deep Neural Network Explanation Methods}. In \bibinfo{booktitle}{\emph{Advances in Neural Information Processing Systems}}, Vol.~\bibinfo{volume}{33}. \bibinfo{pages}{4211--4222}.
\newblock
\urldef\tempurl%
\url{https://doi.org/10.5555/3495724.3496078}
\showDOI{\tempurl}


\bibitem[Karimi et~al\mbox{.}(2020)]%
        {karimi2020model}
\bibfield{author}{\bibinfo{person}{Amir-Hossein Karimi}, \bibinfo{person}{Gilles Barthe}, \bibinfo{person}{Borja Balle}, {and} \bibinfo{person}{Isabel Valera}.} \bibinfo{year}{2020}\natexlab{}.
\newblock \showarticletitle{Model-agnostic counterfactual explanations for consequential decisions}. In \bibinfo{booktitle}{\emph{International Conference on Artificial Intelligence and Statistics}}. \bibinfo{publisher}{PMLR}, \bibinfo{pages}{895--905}.
\newblock


\bibitem[Kelley(1984)]%
        {kelly1984wizard-of-oz}
\bibfield{author}{\bibinfo{person}{J.~F. Kelley}.} \bibinfo{year}{1984}\natexlab{}.
\newblock \showarticletitle{An Iterative Design Methodology for User-Friendly Natural Language Office Information Applications}.
\newblock \bibinfo{journal}{\emph{ACM Transactions on Information Systems}} \bibinfo{volume}{2}, \bibinfo{number}{1} (\bibinfo{year}{1984}), \bibinfo{pages}{26–41}.
\newblock
\urldef\tempurl%
\url{https://doi.org/10.1145/357417.357420}
\showDOI{\tempurl}


\bibitem[Kim et~al\mbox{.}(2016)]%
        {kim2016examples}
\bibfield{author}{\bibinfo{person}{Been Kim}, \bibinfo{person}{Rajiv Khanna}, {and} \bibinfo{person}{Oluwasanmi Koyejo}.} \bibinfo{year}{2016}\natexlab{}.
\newblock \showarticletitle{Examples Are Not Enough, Learn to Criticize! Criticism for Interpretability}. In \bibinfo{booktitle}{\emph{Proceedings of the 30th International Conference on Neural Information Processing Systems}}. \bibinfo{publisher}{Curran Associates Inc.}, \bibinfo{pages}{2288–2296}.
\newblock
\showISBNx{9781510838819}


\bibitem[Kim et~al\mbox{.}(2023)]%
        {kim2023help}
\bibfield{author}{\bibinfo{person}{Sunnie S.~Y. Kim}, \bibinfo{person}{Elizabeth~Anne Watkins}, \bibinfo{person}{Olga Russakovsky}, \bibinfo{person}{Ruth Fong}, {and} \bibinfo{person}{Andr\'{e}s Monroy-Hern\'{a}ndez}.} \bibinfo{year}{2023}\natexlab{}.
\newblock \showarticletitle{"Help Me Help the AI": Understanding How Explainability Can Support Human-AI Interaction}. In \bibinfo{booktitle}{\emph{Proceedings of the 2023 CHI Conference on Human Factors in Computing Systems}}. \bibinfo{publisher}{Association for Computing Machinery}, \bibinfo{numpages}{17}~pages.
\newblock
\showISBNx{9781450394215}
\urldef\tempurl%
\url{https://doi.org/10.1145/3544548.3581001}
\showDOI{\tempurl}


\bibitem[Kindermans et~al\mbox{.}(2019)]%
        {Kindermans2019}
\bibfield{author}{\bibinfo{person}{Pieter-Jan Kindermans}, \bibinfo{person}{Sara Hooker}, \bibinfo{person}{Julius Adebayo}, \bibinfo{person}{Maximilian Alber}, \bibinfo{person}{Kristof~T. Sch{\"u}tt}, \bibinfo{person}{Sven D{\"a}hne}, \bibinfo{person}{Dumitru Erhan}, {and} \bibinfo{person}{Been Kim}.} \bibinfo{year}{2019}\natexlab{}.
\newblock \bibinfo{booktitle}{\emph{The (Un)reliability of Saliency Methods}}.
\newblock \bibinfo{publisher}{Springer International Publishing}, \bibinfo{pages}{267--280}.
\newblock


\bibitem[Krizhevsky et~al\mbox{.}(2012)]%
        {alex2012imagenet}
\bibfield{author}{\bibinfo{person}{Alex Krizhevsky}, \bibinfo{person}{Ilya Sutskever}, {and} \bibinfo{person}{Geoffrey~E Hinton}.} \bibinfo{year}{2012}\natexlab{}.
\newblock \showarticletitle{Imagenet classification with deep convolutional neural networks}. In \bibinfo{booktitle}{\emph{Advances in neural information processing systems}}, Vol.~\bibinfo{volume}{25}. \bibinfo{pages}{84--90}.
\newblock
\urldef\tempurl%
\url{https://doi.org/10.1145/3065386}
\showDOI{\tempurl}


\bibitem[Kulesza et~al\mbox{.}(2015)]%
        {kulesza2015principles}
\bibfield{author}{\bibinfo{person}{Todd Kulesza}, \bibinfo{person}{Margaret Burnett}, \bibinfo{person}{Weng-Keen Wong}, {and} \bibinfo{person}{Simone Stumpf}.} \bibinfo{year}{2015}\natexlab{}.
\newblock \showarticletitle{Principles of Explanatory Debugging to Personalize Interactive Machine Learning}. In \bibinfo{booktitle}{\emph{Proceedings of the 20th International Conference on Intelligent User Interfaces}}. \bibinfo{publisher}{Association for Computing Machinery}, \bibinfo{pages}{126–137}.
\newblock
\showISBNx{9781450333061}
\urldef\tempurl%
\url{https://doi.org/10.1145/2678025.2701399}
\showDOI{\tempurl}


\bibitem[Kunkel et~al\mbox{.}(2019)]%
        {kunkel2019let}
\bibfield{author}{\bibinfo{person}{Johannes Kunkel}, \bibinfo{person}{Tim Donkers}, \bibinfo{person}{Lisa Michael}, \bibinfo{person}{Catalin-Mihai Barbu}, {and} \bibinfo{person}{J\"{u}rgen Ziegler}.} \bibinfo{year}{2019}\natexlab{}.
\newblock \showarticletitle{Let Me Explain: Impact of Personal and Impersonal Explanations on Trust in Recommender Systems}. In \bibinfo{booktitle}{\emph{Proceedings of the 2019 CHI Conference on Human Factors in Computing Systems}}. \bibinfo{publisher}{Association for Computing Machinery}, \bibinfo{pages}{1–12}.
\newblock
\showISBNx{9781450359702}
\urldef\tempurl%
\url{https://doi.org/10.1145/3290605.3300717}
\showDOI{\tempurl}


\bibitem[Lai and Tan(2019)]%
        {lai2019on}
\bibfield{author}{\bibinfo{person}{Vivian Lai} {and} \bibinfo{person}{Chenhao Tan}.} \bibinfo{year}{2019}\natexlab{}.
\newblock \showarticletitle{On Human Predictions with Explanations and Predictions of Machine Learning Models: A Case Study on Deception Detection}. In \bibinfo{booktitle}{\emph{Proceedings of the Conference on Fairness, Accountability, and Transparency}}. \bibinfo{publisher}{Association for Computing Machinery}, \bibinfo{pages}{29–38}.
\newblock
\showISBNx{9781450361255}
\urldef\tempurl%
\url{https://doi.org/10.1145/3287560.3287590}
\showDOI{\tempurl}


\bibitem[Lakkaraju et~al\mbox{.}(2016)]%
        {lakkaraju2016interpretable}
\bibfield{author}{\bibinfo{person}{Himabindu Lakkaraju}, \bibinfo{person}{Stephen~H Bach}, {and} \bibinfo{person}{Jure Leskovec}.} \bibinfo{year}{2016}\natexlab{}.
\newblock \showarticletitle{Interpretable decision sets: A joint framework for description and prediction}. In \bibinfo{booktitle}{\emph{Proceedings of the 22nd ACM SIGKDD international conference on knowledge discovery and data mining}}. \bibinfo{pages}{1675--1684}.
\newblock


\bibitem[Lakkaraju et~al\mbox{.}(2022)]%
        {lakkaraju2022rethinking}
\bibfield{author}{\bibinfo{person}{Himabindu Lakkaraju}, \bibinfo{person}{Dylan Slack}, \bibinfo{person}{Yuxin Chen}, \bibinfo{person}{Chenhao Tan}, {and} \bibinfo{person}{Sameer Singh}.} \bibinfo{year}{2022}\natexlab{}.
\newblock \showarticletitle{Rethinking Explainability as a Dialogue: A Practitioner's Perspective}.
\newblock \bibinfo{journal}{\emph{arXiv preprint arXiv:2202.01875}} (\bibinfo{year}{2022}).
\newblock


\bibitem[Lan et~al\mbox{.}(2021)]%
        {ijcai2021p0611}
\bibfield{author}{\bibinfo{person}{Yunshi Lan}, \bibinfo{person}{Gaole He}, \bibinfo{person}{Jinhao Jiang}, \bibinfo{person}{Jing Jiang}, \bibinfo{person}{Wayne~Xin Zhao}, {and} \bibinfo{person}{Ji-Rong Wen}.} \bibinfo{year}{2021}\natexlab{}.
\newblock \showarticletitle{A Survey on Complex Knowledge Base Question Answering: Methods, Challenges and Solutions}. In \bibinfo{booktitle}{\emph{Proceedings of the Thirtieth International Joint Conference on Artificial Intelligence, {IJCAI-21}}}. \bibinfo{publisher}{International Joint Conferences on Artificial Intelligence Organization}, \bibinfo{pages}{4483--4491}.
\newblock
\urldef\tempurl%
\url{https://doi.org/10.24963/ijcai.2021/611}
\showDOI{\tempurl}


\bibitem[Larasati et~al\mbox{.}(2020)]%
        {oro70421}
\bibfield{author}{\bibinfo{person}{Retno Larasati}, \bibinfo{person}{Anna~De Liddo}, {and} \bibinfo{person}{Enrico Motta}.} \bibinfo{year}{2020}\natexlab{}.
\newblock \showarticletitle{The effect of explanation styles on user's trust}. In \bibinfo{booktitle}{\emph{2020 Workshop on Explainable Smart Systems for Algorithmic Transparency in Emerging Technologies}}. \bibinfo{publisher}{Association for Computing Machinery}.
\newblock


\bibitem[Lertvittayakumjorn et~al\mbox{.}(2020)]%
        {lertvittayakumjorn-etal-2020-find}
\bibfield{author}{\bibinfo{person}{Piyawat Lertvittayakumjorn}, \bibinfo{person}{Lucia Specia}, {and} \bibinfo{person}{Francesca Toni}.} \bibinfo{year}{2020}\natexlab{}.
\newblock \showarticletitle{{FIND}: {H}uman-in-the-{L}oop {D}ebugging {D}eep {T}ext {C}lassifiers}. In \bibinfo{booktitle}{\emph{Proceedings of the 2020 Conference on Empirical Methods in Natural Language Processing (EMNLP)}} (online). \bibinfo{pages}{332--348}.
\newblock
\urldef\tempurl%
\url{https://doi.org/10.18653/v1/2020.emnlp-main.24}
\showDOI{\tempurl}


\bibitem[Liang et~al\mbox{.}(2020)]%
        {liang-etal-2020-alice}
\bibfield{author}{\bibinfo{person}{Weixin Liang}, \bibinfo{person}{James Zou}, {and} \bibinfo{person}{Zhou Yu}.} \bibinfo{year}{2020}\natexlab{}.
\newblock \showarticletitle{{ALICE}: Active Learning with Contrastive Natural Language Explanations}. In \bibinfo{booktitle}{\emph{Proceedings of the 2020 Conference on Empirical Methods in Natural Language Processing (EMNLP)}} (online). \bibinfo{pages}{4380--4391}.
\newblock
\urldef\tempurl%
\url{https://doi.org/10.18653/v1/2020.emnlp-main.355}
\showDOI{\tempurl}


\bibitem[Liao et~al\mbox{.}(2020)]%
        {liao2020questioning}
\bibfield{author}{\bibinfo{person}{Q.~Vera Liao}, \bibinfo{person}{Daniel Gruen}, {and} \bibinfo{person}{Sarah Miller}.} \bibinfo{year}{2020}\natexlab{}.
\newblock \showarticletitle{Questioning the {AI}: Informing Design Practices for Explainable {AI} User Experiences}. In \bibinfo{booktitle}{\emph{Proceedings of the 2020 CHI Conference on Human Factors in Computing Systems}} (Honolulu, HI, USA). \bibinfo{pages}{1–15}.
\newblock


\bibitem[Liao and Varshney(2021)]%
        {liao2021human}
\bibfield{author}{\bibinfo{person}{Q~Vera Liao} {and} \bibinfo{person}{Kush~R Varshney}.} \bibinfo{year}{2021}\natexlab{}.
\newblock \showarticletitle{Human-centered explainable {AI} ({XAI}): From algorithms to user experiences}.
\newblock \bibinfo{journal}{\emph{arXiv preprint arXiv:2110.10790}} (\bibinfo{year}{2021}).
\newblock


\bibitem[Lim et~al\mbox{.}(2009)]%
        {lim2009why}
\bibfield{author}{\bibinfo{person}{Brian~Y Lim}, \bibinfo{person}{Anind~K Dey}, {and} \bibinfo{person}{Daniel Avrahami}.} \bibinfo{year}{2009}\natexlab{}.
\newblock \showarticletitle{Why and why not explanations improve the intelligibility of context-aware intelligent systems}. In \bibinfo{booktitle}{\emph{Proceedings of the SIGCHI conference on human factors in computing systems}}. \bibinfo{publisher}{Association for Computing Machinery}, \bibinfo{pages}{2119--2128}.
\newblock


\bibitem[Liu et~al\mbox{.}(2021a)]%
        {liu2021understanding}
\bibfield{author}{\bibinfo{person}{Han Liu}, \bibinfo{person}{Vivian Lai}, {and} \bibinfo{person}{Chenhao Tan}.} \bibinfo{year}{2021}\natexlab{a}.
\newblock \showarticletitle{Understanding the effect of out-of-distribution examples and interactive explanations on human-ai decision making}.
\newblock \bibinfo{journal}{\emph{Proceedings of the ACM on Human-Computer Interaction}} \bibinfo{volume}{5}, \bibinfo{number}{CSCW2} (\bibinfo{year}{2021}), \bibinfo{pages}{1--45}.
\newblock


\bibitem[Liu et~al\mbox{.}(2022)]%
        {li2022Application}
\bibfield{author}{\bibinfo{person}{Li Liu}, \bibinfo{person}{Fu Guo}, \bibinfo{person}{Zishuai Zou}, {and} \bibinfo{person}{Vincent~G. Duffy}.} \bibinfo{year}{2022}\natexlab{}.
\newblock \showarticletitle{Application, Development and Future Opportunities of Collaborative Robots (Cobots) in Manufacturing: A Literature Review}.
\newblock \bibinfo{journal}{\emph{International Journal of Human–Computer Interaction}} \bibinfo{volume}{0}, \bibinfo{number}{0} (\bibinfo{year}{2022}), \bibinfo{pages}{1--18}.
\newblock
\urldef\tempurl%
\url{https://doi.org/10.1080/10447318.2022.2041907}
\showDOI{\tempurl}


\bibitem[Liu et~al\mbox{.}(2018)]%
        {liu2018interpretation}
\bibfield{author}{\bibinfo{person}{Ninghao Liu}, \bibinfo{person}{Xiao Huang}, \bibinfo{person}{Jundong Li}, {and} \bibinfo{person}{Xia Hu}.} \bibinfo{year}{2018}\natexlab{}.
\newblock \showarticletitle{On interpretation of network embedding via taxonomy induction}. In \bibinfo{booktitle}{\emph{Proceedings of the 24th ACM SIGKDD International Conference on Knowledge Discovery \& Data Mining}} (London, United Kingdom). \bibinfo{pages}{1812--1820}.
\newblock
\urldef\tempurl%
\url{https://doi.org/10.1145/3219819.3220001}
\showDOI{\tempurl}


\bibitem[Liu et~al\mbox{.}(2021b)]%
        {liu2021swin}
\bibfield{author}{\bibinfo{person}{Ze Liu}, \bibinfo{person}{Yutong Lin}, \bibinfo{person}{Yue Cao}, \bibinfo{person}{Han Hu}, \bibinfo{person}{Yixuan Wei}, \bibinfo{person}{Zheng Zhang}, \bibinfo{person}{Stephen Lin}, {and} \bibinfo{person}{Baining Guo}.} \bibinfo{year}{2021}\natexlab{b}.
\newblock \showarticletitle{Swin transformer: Hierarchical vision transformer using shifted windows}. In \bibinfo{booktitle}{\emph{Proceedings of the IEEE/CVF international conference on computer vision}}. \bibinfo{publisher}{Computer Vision Foundation}, \bibinfo{pages}{10012--10022}.
\newblock


\bibitem[Lombrozo(2006)]%
        {lombrozo2006structure}
\bibfield{author}{\bibinfo{person}{Tania Lombrozo}.} \bibinfo{year}{2006}\natexlab{}.
\newblock \showarticletitle{The structure and function of explanations}.
\newblock \bibinfo{journal}{\emph{Trends in cognitive sciences}} \bibinfo{volume}{10}, \bibinfo{number}{10} (\bibinfo{year}{2006}), \bibinfo{pages}{464--470}.
\newblock


\bibitem[Lundberg and Lee(2017)]%
        {lundberg2017unified}
\bibfield{author}{\bibinfo{person}{Scott~M Lundberg} {and} \bibinfo{person}{Su-In Lee}.} \bibinfo{year}{2017}\natexlab{}.
\newblock \showarticletitle{A unified approach to interpreting model predictions}.
\newblock \bibinfo{journal}{\emph{Advances in neural information processing systems}}  \bibinfo{volume}{30} (\bibinfo{year}{2017}).
\newblock
\urldef\tempurl%
\url{https://doi.org/10.5555/3295222.3295230}
\showDOI{\tempurl}


\bibitem[Luo et~al\mbox{.}(2023)]%
        {luo-etal-2023-end}
\bibfield{author}{\bibinfo{person}{Man Luo}, \bibinfo{person}{Zhiyuan Fang}, \bibinfo{person}{Tejas Gokhale}, \bibinfo{person}{Yezhou Yang}, {and} \bibinfo{person}{Chitta Baral}.} \bibinfo{year}{2023}\natexlab{}.
\newblock \showarticletitle{End-to-end Knowledge Retrieval with Multi-modal Queries}. In \bibinfo{booktitle}{\emph{Proceedings of the 61st Annual Meeting of the Association for Computational Linguistics (Volume 1: Long Papers)}} (Toronto, Canada). \bibinfo{publisher}{Association for Computational Linguistics}, \bibinfo{pages}{8573--8589}.
\newblock
\urldef\tempurl%
\url{https://doi.org/10.18653/v1/2023.acl-long.478}
\showDOI{\tempurl}


\bibitem[Luo et~al\mbox{.}(2022)]%
        {luo_evaluating_2022}
\bibfield{author}{\bibinfo{person}{Ruikun Luo}, \bibinfo{person}{Na Du}, {and} \bibinfo{person}{X.~Jessie Yang}.} \bibinfo{year}{2022}\natexlab{}.
\newblock \showarticletitle{Evaluating Effects of Enhanced Autonomy Transparency on Trust, Dependence, and Human-Autonomy Team Performance over Time}.
\newblock \bibinfo{journal}{\emph{International Journal of Human–Computer Interaction}} \bibinfo{volume}{38}, \bibinfo{number}{18-20} (\bibinfo{year}{2022}), \bibinfo{pages}{1962--1971}.
\newblock
\urldef\tempurl%
\url{https://doi.org/10.1080/10447318.2022.2097602}
\showDOI{\tempurl}


\bibitem[Ma et~al\mbox{.}(2023)]%
        {ma2023who}
\bibfield{author}{\bibinfo{person}{Shuai Ma}, \bibinfo{person}{Ying Lei}, \bibinfo{person}{Xinru Wang}, \bibinfo{person}{Chengbo Zheng}, \bibinfo{person}{Chuhan Shi}, \bibinfo{person}{Ming Yin}, {and} \bibinfo{person}{Xiaojuan Ma}.} \bibinfo{year}{2023}\natexlab{}.
\newblock \showarticletitle{Who Should {I} Trust: {AI} or Myself? {L}everaging Human and {AI} Correctness Likelihood to Promote Appropriate Trust in {AI}-Assisted Decision-Making}. In \bibinfo{booktitle}{\emph{Proceedings of the 2023 CHI Conference on Human Factors in Computing Systems}} (Hamburg, Germany). Article \bibinfo{articleno}{759}, \bibinfo{numpages}{19}~pages.
\newblock
\urldef\tempurl%
\url{https://doi.org/10.1145/3544548.3581058}
\showDOI{\tempurl}


\bibitem[McKnight et~al\mbox{.}(1998)]%
        {mcknight1998initial}
\bibfield{author}{\bibinfo{person}{D~Harrison McKnight}, \bibinfo{person}{Larry~L Cummings}, {and} \bibinfo{person}{Norman~L Chervany}.} \bibinfo{year}{1998}\natexlab{}.
\newblock \showarticletitle{Initial trust formation in new organizational relationships}.
\newblock \bibinfo{journal}{\emph{Academy of Management review}} \bibinfo{volume}{23}, \bibinfo{number}{3} (\bibinfo{year}{1998}), \bibinfo{pages}{473--490}.
\newblock
\urldef\tempurl%
\url{https://doi.org/10.2307/259290}
\showDOI{\tempurl}


\bibitem[Mikolov et~al\mbox{.}(2013)]%
        {mikolov2013distributed}
\bibfield{author}{\bibinfo{person}{Tomas Mikolov}, \bibinfo{person}{Ilya Sutskever}, \bibinfo{person}{Kai Chen}, \bibinfo{person}{Greg~S Corrado}, {and} \bibinfo{person}{Jeff Dean}.} \bibinfo{year}{2013}\natexlab{}.
\newblock \showarticletitle{Distributed representations of words and phrases and their compositionality}.
\newblock \bibinfo{journal}{\emph{Advances in neural information processing systems}}  \bibinfo{volume}{26} (\bibinfo{year}{2013}), \bibinfo{pages}{1--9}.
\newblock
\urldef\tempurl%
\url{https://doi.org/10.5555/2999792.2999959}
\showDOI{\tempurl}


\bibitem[Miller(2019)]%
        {miller2019explanation}
\bibfield{author}{\bibinfo{person}{Tim Miller}.} \bibinfo{year}{2019}\natexlab{}.
\newblock \showarticletitle{Explanation in artificial intelligence: Insights from the social sciences}.
\newblock \bibinfo{journal}{\emph{Artificial intelligence}}  \bibinfo{volume}{267} (\bibinfo{year}{2019}), \bibinfo{pages}{1--38}.
\newblock


\bibitem[Mothilal et~al\mbox{.}(2020)]%
        {mothilal2020explaining}
\bibfield{author}{\bibinfo{person}{Ramaravind~K Mothilal}, \bibinfo{person}{Amit Sharma}, {and} \bibinfo{person}{Chenhao Tan}.} \bibinfo{year}{2020}\natexlab{}.
\newblock \showarticletitle{Explaining machine learning classifiers through diverse counterfactual explanations}. In \bibinfo{booktitle}{\emph{Proceedings of the 2020 conference on fairness, accountability, and transparency}} (Barcelona, Spain). \bibinfo{pages}{607--617}.
\newblock


\bibitem[Muir and Moray(1996)]%
        {muir1996trust}
\bibfield{author}{\bibinfo{person}{Bonnie~M Muir} {and} \bibinfo{person}{Neville Moray}.} \bibinfo{year}{1996}\natexlab{}.
\newblock \showarticletitle{Trust in automation. Part {II}. {Experimental} studies of trust and human intervention in a process control simulation}.
\newblock \bibinfo{journal}{\emph{Ergonomics}} \bibinfo{volume}{39}, \bibinfo{number}{3} (\bibinfo{year}{1996}), \bibinfo{pages}{429--460}.
\newblock


\bibitem[Nguyen et~al\mbox{.}(2021)]%
        {nguyen2021effectiveness}
\bibfield{author}{\bibinfo{person}{Giang Nguyen}, \bibinfo{person}{Daeyoung Kim}, {and} \bibinfo{person}{Anh Nguyen}.} \bibinfo{year}{2021}\natexlab{}.
\newblock \showarticletitle{The effectiveness of feature attribution methods and its correlation with automatic evaluation scores}.
\newblock \bibinfo{journal}{\emph{Advances in Neural Information Processing Systems}}  \bibinfo{volume}{34} (\bibinfo{year}{2021}), \bibinfo{pages}{26422--26436}.
\newblock
\urldef\tempurl%
\url{https://doi.org/10.5555/3540261.3542284}
\showDOI{\tempurl}


\bibitem[Nguyen et~al\mbox{.}(2022)]%
        {nguyen2022visual}
\bibfield{author}{\bibinfo{person}{Giang Nguyen}, \bibinfo{person}{Mohammad~Reza Taesiri}, {and} \bibinfo{person}{Anh Nguyen}.} \bibinfo{year}{2022}\natexlab{}.
\newblock \showarticletitle{Visual correspondence-based explanations improve {AI} robustness and human-AI team accuracy}.
\newblock \bibinfo{journal}{\emph{Neural Information Processing Systems (NeurIPS)}}  \bibinfo{volume}{35} (\bibinfo{year}{2022}), \bibinfo{pages}{34287–34301}.
\newblock
\urldef\tempurl%
\url{https://doi.org/10.5555/3600270.3602755}
\showDOI{\tempurl}


\bibitem[Ni et~al\mbox{.}(2023)]%
        {ni2023recent}
\bibfield{author}{\bibinfo{person}{Jinjie Ni}, \bibinfo{person}{Tom Young}, \bibinfo{person}{Vlad Pandelea}, \bibinfo{person}{Fuzhao Xue}, {and} \bibinfo{person}{Erik Cambria}.} \bibinfo{year}{2023}\natexlab{}.
\newblock \showarticletitle{Recent advances in deep learning based dialogue systems: A systematic survey}.
\newblock \bibinfo{journal}{\emph{Artificial intelligence review}} \bibinfo{volume}{56}, \bibinfo{number}{4} (\bibinfo{year}{2023}), \bibinfo{pages}{3055--3155}.
\newblock


\bibitem[Numata et~al\mbox{.}(2020)]%
        {numata2020achieving}
\bibfield{author}{\bibinfo{person}{Takashi Numata}, \bibinfo{person}{Hiroki Sato}, \bibinfo{person}{Yasuhiro Asa}, \bibinfo{person}{Takahiko Koike}, \bibinfo{person}{Kohei Miyata}, \bibinfo{person}{Eri Nakagawa}, \bibinfo{person}{Motofumi Sumiya}, {and} \bibinfo{person}{Norihiro Sadato}.} \bibinfo{year}{2020}\natexlab{}.
\newblock \showarticletitle{Achieving affective human--virtual agent communication by enabling virtual agents to imitate positive expressions}.
\newblock \bibinfo{journal}{\emph{Scientific reports}} \bibinfo{volume}{10}, \bibinfo{number}{1} (\bibinfo{year}{2020}), \bibinfo{pages}{5977}.
\newblock


\bibitem[Ouyang et~al\mbox{.}(2022)]%
        {ouyang2022training}
\bibfield{author}{\bibinfo{person}{Long Ouyang}, \bibinfo{person}{Jeffrey Wu}, \bibinfo{person}{Xu Jiang}, \bibinfo{person}{Diogo Almeida}, \bibinfo{person}{Carroll Wainwright}, \bibinfo{person}{Pamela Mishkin}, \bibinfo{person}{Chong Zhang}, \bibinfo{person}{Sandhini Agarwal}, \bibinfo{person}{Katarina Slama}, \bibinfo{person}{Alex Ray}, {et~al\mbox{.}}} \bibinfo{year}{2022}\natexlab{}.
\newblock \showarticletitle{Training language models to follow instructions with human feedback}.
\newblock \bibinfo{journal}{\emph{Advances in Neural Information Processing Systems}}  \bibinfo{volume}{35} (\bibinfo{year}{2022}), \bibinfo{pages}{27730--27744}.
\newblock
\urldef\tempurl%
\url{https://doi.org/10.5555/3600270.3602281}
\showDOI{\tempurl}


\bibitem[Pieters(2011)]%
        {pieters2011explanation}
\bibfield{author}{\bibinfo{person}{Wolter Pieters}.} \bibinfo{year}{2011}\natexlab{}.
\newblock \showarticletitle{Explanation and trust: what to tell the user in security and AI?}
\newblock \bibinfo{journal}{\emph{Ethics and information technology}}  \bibinfo{volume}{13} (\bibinfo{year}{2011}), \bibinfo{pages}{53--64}.
\newblock


\bibitem[Poursabzi-Sangdeh et~al\mbox{.}(2021)]%
        {forough2021manipulating}
\bibfield{author}{\bibinfo{person}{Forough Poursabzi-Sangdeh}, \bibinfo{person}{Daniel~G Goldstein}, \bibinfo{person}{Jake~M Hofman}, \bibinfo{person}{Jennifer~Wortman Vaughan}, {and} \bibinfo{person}{Hanna Wallach}.} \bibinfo{year}{2021}\natexlab{}.
\newblock \showarticletitle{Manipulating and Measuring Model Interpretability}. In \bibinfo{booktitle}{\emph{Proceedings of the 2021 CHI Conference on Human Factors in Computing Systems}} (Yokohama, Japan). \bibinfo{numpages}{52}~pages.
\newblock
\urldef\tempurl%
\url{https://doi.org/10.1145/3411764.3445315}
\showDOI{\tempurl}


\bibitem[Powles and Hodson(2017)]%
        {Powles2017}
\bibfield{author}{\bibinfo{person}{Julia Powles} {and} \bibinfo{person}{Hal Hodson}.} \bibinfo{year}{2017}\natexlab{}.
\newblock \showarticletitle{Google DeepMind and healthcare in an age of algorithms}.
\newblock \bibinfo{journal}{\emph{Health and Technology}} \bibinfo{volume}{7}, \bibinfo{number}{4} (\bibinfo{year}{2017}), \bibinfo{pages}{351--367}.
\newblock
\urldef\tempurl%
\url{https://doi.org/10.1007/s12553-017-0179-1}
\showDOI{\tempurl}


\bibitem[Poyiadzi et~al\mbox{.}(2020)]%
        {poyiadzi2020face}
\bibfield{author}{\bibinfo{person}{Rafael Poyiadzi}, \bibinfo{person}{Kacper Sokol}, \bibinfo{person}{Raul Santos-Rodriguez}, \bibinfo{person}{Tijl De~Bie}, {and} \bibinfo{person}{Peter Flach}.} \bibinfo{year}{2020}\natexlab{}.
\newblock \showarticletitle{{FACE}: Feasible and actionable counterfactual explanations}. In \bibinfo{booktitle}{\emph{Proceedings of the AAAI/ACM Conference on AI, Ethics, and Society}}. \bibinfo{publisher}{Association for Computing Machinery}, \bibinfo{pages}{344--350}.
\newblock


\bibitem[Quinn et~al\mbox{.}(2021)]%
        {quinn2021trust}
\bibfield{author}{\bibinfo{person}{Thomas~P Quinn}, \bibinfo{person}{Manisha Senadeera}, \bibinfo{person}{Stephan Jacobs}, \bibinfo{person}{Simon Coghlan}, {and} \bibinfo{person}{Vuong Le}.} \bibinfo{year}{2021}\natexlab{}.
\newblock \showarticletitle{Trust and medical {AI}: The challenges we face and the expertise needed to overcome them}.
\newblock \bibinfo{journal}{\emph{Journal of the American Medical Informatics Association}} \bibinfo{volume}{28}, \bibinfo{number}{4} (\bibinfo{year}{2021}), \bibinfo{pages}{890--894}.
\newblock


\bibitem[Ribeiro et~al\mbox{.}(2016)]%
        {ribeiro2016should}
\bibfield{author}{\bibinfo{person}{Marco~Tulio Ribeiro}, \bibinfo{person}{Sameer Singh}, {and} \bibinfo{person}{Carlos Guestrin}.} \bibinfo{year}{2016}\natexlab{}.
\newblock \showarticletitle{"{Why} should {I} trust you?": Explaining the predictions of any classifier}. In \bibinfo{booktitle}{\emph{Proceedings of the 22nd ACM SIGKDD international conference on knowledge discovery and data mining}}. \bibinfo{publisher}{Association for Computing Machinery}, \bibinfo{pages}{1135--1144}.
\newblock


\bibitem[Rohlfing et~al\mbox{.}(2020)]%
        {rohlfing2020explanation}
\bibfield{author}{\bibinfo{person}{Katharina~J Rohlfing}, \bibinfo{person}{Philipp Cimiano}, \bibinfo{person}{Ingrid Scharlau}, \bibinfo{person}{Tobias Matzner}, \bibinfo{person}{Heike~M Buhl}, \bibinfo{person}{Hendrik Buschmeier}, \bibinfo{person}{Elena Esposito}, \bibinfo{person}{Angela Grimminger}, \bibinfo{person}{Barbara Hammer}, \bibinfo{person}{Reinhold H{\"a}b-Umbach}, {et~al\mbox{.}}} \bibinfo{year}{2020}\natexlab{}.
\newblock \showarticletitle{Explanation as a social practice: Toward a conceptual framework for the social design of {AI} systems}.
\newblock \bibinfo{journal}{\emph{IEEE Transactions on Cognitive and Developmental Systems}} \bibinfo{volume}{13}, \bibinfo{number}{3} (\bibinfo{year}{2020}), \bibinfo{pages}{717--728}.
\newblock


\bibitem[Ross et~al\mbox{.}(2017)]%
        {ijcai2017p371}
\bibfield{author}{\bibinfo{person}{Andrew~Slavin Ross}, \bibinfo{person}{Michael~C. Hughes}, {and} \bibinfo{person}{Finale Doshi-Velez}.} \bibinfo{year}{2017}\natexlab{}.
\newblock \showarticletitle{Right for the Right Reasons: Training Differentiable Models by Constraining their Explanations}. In \bibinfo{booktitle}{\emph{Proceedings of the Twenty-Sixth International Joint Conference on Artificial Intelligence, {IJCAI-17}}} (Melbourne, Australia). \bibinfo{pages}{2662--2670}.
\newblock
\urldef\tempurl%
\url{https://doi.org/10.24963/ijcai.2017/371}
\showDOI{\tempurl}


\bibitem[Rudzi{\'n}ski(2016)]%
        {rudzinski2016multi}
\bibfield{author}{\bibinfo{person}{Filip Rudzi{\'n}ski}.} \bibinfo{year}{2016}\natexlab{}.
\newblock \showarticletitle{A multi-objective genetic optimization of interpretability-oriented fuzzy rule-based classifiers}.
\newblock \bibinfo{journal}{\emph{Applied Soft Computing}}  \bibinfo{volume}{38} (\bibinfo{year}{2016}), \bibinfo{pages}{118--133}.
\newblock


\bibitem[Schaffer et~al\mbox{.}(2019)]%
        {james2019ican}
\bibfield{author}{\bibinfo{person}{James Schaffer}, \bibinfo{person}{John O'Donovan}, \bibinfo{person}{James Michaelis}, \bibinfo{person}{Adrienne Raglin}, {and} \bibinfo{person}{Tobias H\"{o}llerer}.} \bibinfo{year}{2019}\natexlab{}.
\newblock \showarticletitle{I Can Do Better than Your {AI}: Expertise and Explanations}. In \bibinfo{booktitle}{\emph{Proceedings of the 24th International Conference on Intelligent User Interfaces}}. \bibinfo{publisher}{Association for Computing Machinery}, \bibinfo{pages}{240–251}.
\newblock
\showISBNx{9781450362726}
\urldef\tempurl%
\url{https://doi.org/10.1145/3301275.3302308}
\showDOI{\tempurl}


\bibitem[Schmid and Wrede(2022)]%
        {schmid2022missing}
\bibfield{author}{\bibinfo{person}{Ute Schmid} {and} \bibinfo{person}{Britta Wrede}.} \bibinfo{year}{2022}\natexlab{}.
\newblock \showarticletitle{What is Missing in {XAI} So Far? {An} Interdisciplinary Perspective}.
\newblock \bibinfo{journal}{\emph{KI-K{\"u}nstliche Intelligenz}} \bibinfo{volume}{36}, \bibinfo{number}{3-4} (\bibinfo{year}{2022}), \bibinfo{pages}{303--315}.
\newblock
\urldef\tempurl%
\url{https://doi.org/10.1007/s13218-022-00786-2}
\showDOI{\tempurl}


\bibitem[Schramowski et~al\mbox{.}(2020)]%
        {schramowski2020making}
\bibfield{author}{\bibinfo{person}{Patrick Schramowski}, \bibinfo{person}{Wolfgang Stammer}, \bibinfo{person}{Stefano Teso}, \bibinfo{person}{Anna Brugger}, \bibinfo{person}{Franziska Herbert}, \bibinfo{person}{Xiaoting Shao}, \bibinfo{person}{Hans-Georg Luigs}, \bibinfo{person}{Anne-Katrin Mahlein}, {and} \bibinfo{person}{Kristian Kersting}.} \bibinfo{year}{2020}\natexlab{}.
\newblock \showarticletitle{Making deep neural networks right for the right scientific reasons by interacting with their explanations}.
\newblock \bibinfo{journal}{\emph{Nature Machine Intelligence}} \bibinfo{volume}{2}, \bibinfo{number}{8} (\bibinfo{year}{2020}), \bibinfo{pages}{476--486}.
\newblock


\bibitem[Seaborn et~al\mbox{.}(2021)]%
        {seaborn2021voice}
\bibfield{author}{\bibinfo{person}{Katie Seaborn}, \bibinfo{person}{Norihisa~P. Miyake}, \bibinfo{person}{Peter Pennefather}, {and} \bibinfo{person}{Mihoko Otake-Matsuura}.} \bibinfo{year}{2021}\natexlab{}.
\newblock \showarticletitle{Voice in Human–Agent Interaction: A Survey}.
\newblock \bibinfo{journal}{\emph{Comput. Surveys}} \bibinfo{volume}{54}, \bibinfo{number}{4} (\bibinfo{year}{2021}), \bibinfo{numpages}{43}~pages.
\newblock
\urldef\tempurl%
\url{https://doi.org/10.1145/3386867}
\showDOI{\tempurl}


\bibitem[Sebo et~al\mbox{.}(2020)]%
        {sebo2020influence}
\bibfield{author}{\bibinfo{person}{Sarah Sebo}, \bibinfo{person}{Ling~Liang Dong}, \bibinfo{person}{Nicholas Chang}, \bibinfo{person}{Michal Lewkowicz}, \bibinfo{person}{Michael Schutzman}, {and} \bibinfo{person}{Brian Scassellati}.} \bibinfo{year}{2020}\natexlab{}.
\newblock \showarticletitle{The influence of robot verbal support on human team members: Encouraging outgroup contributions and suppressing ingroup supportive behavior}.
\newblock \bibinfo{journal}{\emph{Frontiers in Psychology}} (\bibinfo{year}{2020}), \bibinfo{pages}{3584}.
\newblock


\bibitem[Selvaraju et~al\mbox{.}(2017)]%
        {selvaraju2017grad}
\bibfield{author}{\bibinfo{person}{Ramprasaath~R Selvaraju}, \bibinfo{person}{Michael Cogswell}, \bibinfo{person}{Abhishek Das}, \bibinfo{person}{Ramakrishna Vedantam}, \bibinfo{person}{Devi Parikh}, {and} \bibinfo{person}{Dhruv Batra}.} \bibinfo{year}{2017}\natexlab{}.
\newblock \showarticletitle{{Grad-CAM}: {Visual} explanations from deep networks via gradient-based localization}. In \bibinfo{booktitle}{\emph{Proceedings of the IEEE international conference on computer vision}}. \bibinfo{publisher}{IEEE}, \bibinfo{pages}{618--626}.
\newblock


\bibitem[Sharma et~al\mbox{.}(2020)]%
        {sharma2019certifai}
\bibfield{author}{\bibinfo{person}{Shubham Sharma}, \bibinfo{person}{Jette Henderson}, {and} \bibinfo{person}{Joydeep Ghosh}.} \bibinfo{year}{2020}\natexlab{}.
\newblock \showarticletitle{{CERTIFAI}: A Common Framework to Provide Explanations and Analyse the Fairness and Robustness of Black-box Models}. In \bibinfo{booktitle}{\emph{Proceedings of the AAAI/ACM Conference on AI, Ethics, and Society}} (New York, NY, USA). \bibinfo{publisher}{Association for Computing Machinery}, \bibinfo{pages}{166–172}.
\newblock
\showISBNx{9781450371100}
\urldef\tempurl%
\url{https://doi.org/10.1145/3375627.3375812}
\showDOI{\tempurl}


\bibitem[Shen et~al\mbox{.}(2023)]%
        {shen2023convxai}
\bibfield{author}{\bibinfo{person}{Hua Shen}, \bibinfo{person}{Chieh-Yang Huang}, \bibinfo{person}{Tongshuang Wu}, {and} \bibinfo{person}{Ting-Hao~Kenneth Huang}.} \bibinfo{year}{2023}\natexlab{}.
\newblock \showarticletitle{{ConvXAI}: Delivering Heterogeneous {AI} Explanations via Conversations to Support Human-AI Scientific Writing}. In \bibinfo{booktitle}{\emph{Companion Publication of the 2023 Conference on Computer Supported Cooperative Work and Social Computing}}. \bibinfo{publisher}{Association for Computing Machinery}, \bibinfo{pages}{384–387}.
\newblock
\showISBNx{9798400701290}
\urldef\tempurl%
\url{https://doi.org/10.1145/3584931.3607492}
\showDOI{\tempurl}


\bibitem[Shih et~al\mbox{.}(2018)]%
        {shih2018symbolic}
\bibfield{author}{\bibinfo{person}{Andy Shih}, \bibinfo{person}{Arthur Choi}, {and} \bibinfo{person}{Adnan Darwiche}.} \bibinfo{year}{2018}\natexlab{}.
\newblock \showarticletitle{A symbolic approach to explaining Bayesian network classifiers}. In \bibinfo{booktitle}{\emph{Proceedings of the 27th International Joint Conference on Artificial Intelligence}} (Stockholm, Sweden). \bibinfo{publisher}{AAAI Press}, \bibinfo{pages}{5103–5111}.
\newblock
\showISBNx{9780999241127}


\bibitem[Shuster et~al\mbox{.}(2022)]%
        {shuster2022blenderbot}
\bibfield{author}{\bibinfo{person}{Kurt Shuster}, \bibinfo{person}{Jing Xu}, \bibinfo{person}{Mojtaba Komeili}, \bibinfo{person}{Da Ju}, \bibinfo{person}{Eric~Michael Smith}, \bibinfo{person}{Stephen Roller}, \bibinfo{person}{Megan Ung}, \bibinfo{person}{Moya Chen}, \bibinfo{person}{Kushal Arora}, \bibinfo{person}{Joshua Lane}, {et~al\mbox{.}}} \bibinfo{year}{2022}\natexlab{}.
\newblock \showarticletitle{{BlenderBot} 3: a deployed conversational agent that continually learns to responsibly engage}.
\newblock \bibinfo{journal}{\emph{arXiv preprint arXiv:2208.03188}} (\bibinfo{year}{2022}).
\newblock


\bibitem[Silva et~al\mbox{.}(2023)]%
        {Andrew2023Explainable}
\bibfield{author}{\bibinfo{person}{Andrew Silva}, \bibinfo{person}{Mariah Schrum}, \bibinfo{person}{Erin Hedlund-Botti}, \bibinfo{person}{Nakul Gopalan}, {and} \bibinfo{person}{Matthew Gombolay}.} \bibinfo{year}{2023}\natexlab{}.
\newblock \showarticletitle{Explainable Artificial Intelligence: {Evaluating} the Objective and Subjective Impacts of {XAI} on Human-Agent Interaction}.
\newblock \bibinfo{journal}{\emph{International Journal of Human–Computer Interaction}} \bibinfo{volume}{39}, \bibinfo{number}{7} (\bibinfo{year}{2023}), \bibinfo{pages}{1390--1404}.
\newblock
\urldef\tempurl%
\url{https://doi.org/10.1080/10447318.2022.2101698}
\showDOI{\tempurl}


\bibitem[Simonyan et~al\mbox{.}(2013)]%
        {simonyan2013deep}
\bibfield{author}{\bibinfo{person}{Karen Simonyan}, \bibinfo{person}{Andrea Vedaldi}, {and} \bibinfo{person}{Andrew Zisserman}.} \bibinfo{year}{2013}\natexlab{}.
\newblock \showarticletitle{Deep inside convolutional networks: Visualising image classification models and saliency maps}.
\newblock \bibinfo{journal}{\emph{arXiv preprint arXiv:1312.6034}} (\bibinfo{year}{2013}).
\newblock


\bibitem[Simonyan and Zisserman(2015)]%
        {vgg16Z14a}
\bibfield{author}{\bibinfo{person}{Karen Simonyan} {and} \bibinfo{person}{Andrew Zisserman}.} \bibinfo{year}{2015}\natexlab{}.
\newblock \showarticletitle{Very Deep Convolutional Networks for Large-Scale Image Recognition}. In \bibinfo{booktitle}{\emph{Proceedings of 3rd International Conference on Learning Representations}}. \bibinfo{publisher}{openreview}.
\newblock


\bibitem[Slack et~al\mbox{.}(2023)]%
        {slack2023explaining}
\bibfield{author}{\bibinfo{person}{Dylan Slack}, \bibinfo{person}{Satyapriya Krishna}, \bibinfo{person}{Himabindu Lakkaraju}, {and} \bibinfo{person}{Sameer Singh}.} \bibinfo{year}{2023}\natexlab{}.
\newblock \showarticletitle{Explaining machine learning models with interactive natural language conversations using TalkToModel}.
\newblock \bibinfo{journal}{\emph{Nature Machine Intelligence}} \bibinfo{volume}{5}, \bibinfo{number}{8} (\bibinfo{year}{2023}), \bibinfo{pages}{873--883}.
\newblock
\urldef\tempurl%
\url{https://doi.org/10.1038/s42256-023-00692-8}
\showDOI{\tempurl}


\bibitem[Smith-Renner et~al\mbox{.}(2020)]%
        {smith2020no}
\bibfield{author}{\bibinfo{person}{Alison Smith-Renner}, \bibinfo{person}{Ron Fan}, \bibinfo{person}{Melissa Birchfield}, \bibinfo{person}{Tongshuang Wu}, \bibinfo{person}{Jordan Boyd-Graber}, \bibinfo{person}{Daniel~S Weld}, {and} \bibinfo{person}{Leah Findlater}.} \bibinfo{year}{2020}\natexlab{}.
\newblock \showarticletitle{No explainability without accountability: An empirical study of explanations and feedback in interactive ml}. In \bibinfo{booktitle}{\emph{Proceedings of the 2020 CHI Conference on Human Factors in Computing Systems}}. \bibinfo{publisher}{Association for Computing Machinery}, \bibinfo{pages}{1--13}.
\newblock


\bibitem[Springer and Whittaker(2019)]%
        {springer2019progressive}
\bibfield{author}{\bibinfo{person}{Aaron Springer} {and} \bibinfo{person}{Steve Whittaker}.} \bibinfo{year}{2019}\natexlab{}.
\newblock \showarticletitle{Progressive Disclosure: Empirically Motivated Approaches to Designing Effective Transparency}. In \bibinfo{booktitle}{\emph{Proceedings of the 24th International Conference on Intelligent User Interfaces}}. \bibinfo{pages}{107–120}.
\newblock
\urldef\tempurl%
\url{https://doi.org/10.1145/3301275.3302322}
\showDOI{\tempurl}


\bibitem[Sundararajan et~al\mbox{.}(2017)]%
        {sundararajan2017axiomatic}
\bibfield{author}{\bibinfo{person}{Mukund Sundararajan}, \bibinfo{person}{Ankur Taly}, {and} \bibinfo{person}{Qiqi Yan}.} \bibinfo{year}{2017}\natexlab{}.
\newblock \showarticletitle{Axiomatic attribution for deep networks}. In \bibinfo{booktitle}{\emph{International conference on machine learning}}. \bibinfo{publisher}{JMLR.org}, \bibinfo{pages}{3319--3328}.
\newblock


\bibitem[Taesiri et~al\mbox{.}(2022)]%
        {taesiri2022visual}
\bibfield{author}{\bibinfo{person}{Mohammad~Reza Taesiri}, \bibinfo{person}{Giang Nguyen}, {and} \bibinfo{person}{Anh Nguyen}.} \bibinfo{year}{2022}\natexlab{}.
\newblock \showarticletitle{Visual correspondence-based explanations improve {AI} robustness and human-AI team accuracy}.
\newblock \bibinfo{journal}{\emph{Advances in Neural Information Processing Systems}}  \bibinfo{volume}{35} (\bibinfo{year}{2022}), \bibinfo{pages}{34287--34301}.
\newblock


\bibitem[Tenney et~al\mbox{.}(2020)]%
        {tenney-etal-2020-language}
\bibfield{author}{\bibinfo{person}{Ian Tenney}, \bibinfo{person}{James Wexler}, \bibinfo{person}{Jasmijn Bastings}, \bibinfo{person}{Tolga Bolukbasi}, \bibinfo{person}{Andy Coenen}, \bibinfo{person}{Sebastian Gehrmann}, \bibinfo{person}{Ellen Jiang}, \bibinfo{person}{Mahima Pushkarna}, \bibinfo{person}{Carey Radebaugh}, \bibinfo{person}{Emily Reif}, {and} \bibinfo{person}{Ann Yuan}.} \bibinfo{year}{2020}\natexlab{}.
\newblock \showarticletitle{The Language Interpretability Tool: Extensible, Interactive Visualizations and Analysis for {NLP} Models}. In \bibinfo{booktitle}{\emph{Proceedings of the 2020 Conference on Empirical Methods in Natural Language Processing: System Demonstrations}}. \bibinfo{publisher}{Association for Computational Linguistics}, \bibinfo{address}{Online}, \bibinfo{pages}{107--118}.
\newblock
\urldef\tempurl%
\url{https://doi.org/10.18653/v1/2020.emnlp-demos.15}
\showDOI{\tempurl}


\bibitem[Teso et~al\mbox{.}(2021)]%
        {teso2021interactive}
\bibfield{author}{\bibinfo{person}{Stefano Teso}, \bibinfo{person}{Andrea Bontempelli}, \bibinfo{person}{Fausto Giunchiglia}, {and} \bibinfo{person}{Andrea Passerini}.} \bibinfo{year}{2021}\natexlab{}.
\newblock \showarticletitle{Interactive label cleaning with example-based explanations}.
\newblock \bibinfo{journal}{\emph{Advances in Neural Information Processing Systems}}  \bibinfo{volume}{34} (\bibinfo{year}{2021}), \bibinfo{pages}{12966--12977}.
\newblock


\bibitem[Teso and Kersting(2019)]%
        {teso2019explantory}
\bibfield{author}{\bibinfo{person}{Stefano Teso} {and} \bibinfo{person}{Kristian Kersting}.} \bibinfo{year}{2019}\natexlab{}.
\newblock \showarticletitle{Explanatory Interactive Machine Learning}. In \bibinfo{booktitle}{\emph{Proceedings of the 2019 AAAI/ACM Conference on AI, Ethics, and Society}} (Honolulu, HI, USA). \bibinfo{pages}{239–245}.
\newblock
\urldef\tempurl%
\url{https://doi.org/10.1145/3306618.3314293}
\showDOI{\tempurl}


\bibitem[Touvron et~al\mbox{.}(2023)]%
        {touvron2023llama}
\bibfield{author}{\bibinfo{person}{Hugo Touvron}, \bibinfo{person}{Thibaut Lavril}, \bibinfo{person}{Gautier Izacard}, \bibinfo{person}{Xavier Martinet}, \bibinfo{person}{Marie-Anne Lachaux}, \bibinfo{person}{Timoth{\'e}e Lacroix}, \bibinfo{person}{Baptiste Rozi{\`e}re}, \bibinfo{person}{Naman Goyal}, \bibinfo{person}{Eric Hambro}, \bibinfo{person}{Faisal Azhar}, {et~al\mbox{.}}} \bibinfo{year}{2023}\natexlab{}.
\newblock \showarticletitle{{LLaMA}: Open and efficient foundation language models}.
\newblock \bibinfo{journal}{\emph{arXiv preprint arXiv:2302.13971}} (\bibinfo{year}{2023}).
\newblock


\bibitem[Tran et~al\mbox{.}(2021)]%
        {tran2021counterfactual}
\bibfield{author}{\bibinfo{person}{Khanh~Hiep Tran}, \bibinfo{person}{Azin Ghazimatin}, {and} \bibinfo{person}{Rishiraj Saha~Roy}.} \bibinfo{year}{2021}\natexlab{}.
\newblock \showarticletitle{Counterfactual Explanations for Neural Recommenders}. In \bibinfo{booktitle}{\emph{Proceedings of the 44th International ACM SIGIR Conference on Research and Development in Information Retrieval}}. \bibinfo{publisher}{Association for Computing Machinery}, \bibinfo{pages}{1627--1631}.
\newblock


\bibitem[Verma et~al\mbox{.}(2020)]%
        {verma2020counterfactual}
\bibfield{author}{\bibinfo{person}{Sahil Verma}, \bibinfo{person}{John Dickerson}, {and} \bibinfo{person}{Keegan Hines}.} \bibinfo{year}{2020}\natexlab{}.
\newblock \showarticletitle{Counterfactual explanations for machine learning: A review}.
\newblock \bibinfo{journal}{\emph{arXiv preprint arXiv:2010.10596}} (\bibinfo{year}{2020}).
\newblock


\bibitem[Vorm and Combs(2022)]%
        {vorm2022Integrating}
\bibfield{author}{\bibinfo{person}{E.~S. Vorm} {and} \bibinfo{person}{David J.~Y. Combs}.} \bibinfo{year}{2022}\natexlab{}.
\newblock \showarticletitle{Integrating Transparency, Trust, and Acceptance: The Intelligent Systems Technology Acceptance Model (ISTAM)}.
\newblock \bibinfo{journal}{\emph{International Journal of Human–Computer Interaction}} \bibinfo{volume}{38}, \bibinfo{number}{18-20} (\bibinfo{year}{2022}), \bibinfo{pages}{1828--1845}.
\newblock
\urldef\tempurl%
\url{https://doi.org/10.1080/10447318.2022.2070107}
\showDOI{\tempurl}


\bibitem[Wachter et~al\mbox{.}(2017)]%
        {wachter2017counterfactual}
\bibfield{author}{\bibinfo{person}{Sandra Wachter}, \bibinfo{person}{Brent Mittelstadt}, {and} \bibinfo{person}{Chris Russell}.} \bibinfo{year}{2017}\natexlab{}.
\newblock \showarticletitle{Counterfactual explanations without opening the black box: Automated decisions and the {GDPR}}.
\newblock \bibinfo{journal}{\emph{Harvard Journal of Law \& Technology}}  \bibinfo{volume}{31} (\bibinfo{year}{2017}), \bibinfo{pages}{841}.
\newblock
Issue 2.
\urldef\tempurl%
\url{https://doi.org/objects/uuid:86dfcdac-10b5-4314-bbd1-08e6e78b9094}
\showDOI{\tempurl}


\bibitem[Wang and Yin(2021)]%
        {wang2021are}
\bibfield{author}{\bibinfo{person}{Xinru Wang} {and} \bibinfo{person}{Ming Yin}.} \bibinfo{year}{2021}\natexlab{}.
\newblock \showarticletitle{Are Explanations Helpful? {A} Comparative Study of the Effects of Explanations in {AI}-Assisted Decision-Making}. In \bibinfo{booktitle}{\emph{26th International Conference on Intelligent User Interfaces}}. \bibinfo{publisher}{Association for Computing Machinery}, \bibinfo{pages}{318–328}.
\newblock
\showISBNx{9781450380171}
\urldef\tempurl%
\url{https://doi.org/10.1145/3397481.3450650}
\showDOI{\tempurl}


\bibitem[Wilkesmann and Wilkesmann(2011)]%
        {wilkesmann2011knowledge}
\bibfield{author}{\bibinfo{person}{Maximiliane Wilkesmann} {and} \bibinfo{person}{Uwe Wilkesmann}.} \bibinfo{year}{2011}\natexlab{}.
\newblock \showarticletitle{Knowledge transfer as interaction between experts and novices supported by technology}.
\newblock \bibinfo{journal}{\emph{Vine}} \bibinfo{volume}{41}, \bibinfo{number}{2} (\bibinfo{year}{2011}), \bibinfo{pages}{96--112}.
\newblock


\bibitem[Wittwer et~al\mbox{.}(2008)]%
        {wittwer2008underestimation}
\bibfield{author}{\bibinfo{person}{J{\"o}rg Wittwer}, \bibinfo{person}{Matthias N{\"u}ckles}, {and} \bibinfo{person}{Alexander Renkl}.} \bibinfo{year}{2008}\natexlab{}.
\newblock \showarticletitle{Is underestimation less detrimental than overestimation? {The} impact of experts’ beliefs about a layperson’s knowledge on learning and question asking}.
\newblock \bibinfo{journal}{\emph{Instructional Science}}  \bibinfo{volume}{36} (\bibinfo{year}{2008}), \bibinfo{pages}{27--52}.
\newblock


\bibitem[Xu et~al\mbox{.}(2023)]%
        {wei2023Transitioning}
\bibfield{author}{\bibinfo{person}{Wei Xu}, \bibinfo{person}{Marvin~J. Dainoff}, \bibinfo{person}{Liezhong Ge}, {and} \bibinfo{person}{Zaifeng Gao}.} \bibinfo{year}{2023}\natexlab{}.
\newblock \showarticletitle{Transitioning to Human Interaction with {AI} Systems: New Challenges and Opportunities for HCI Professionals to Enable Human-Centered AI}.
\newblock \bibinfo{journal}{\emph{International Journal of Human–Computer Interaction}} \bibinfo{volume}{39}, \bibinfo{number}{3} (\bibinfo{year}{2023}), \bibinfo{pages}{494--518}.
\newblock
\urldef\tempurl%
\url{https://doi.org/10.1080/10447318.2022.2041900}
\showDOI{\tempurl}


\bibitem[Yang et~al\mbox{.}(2019)]%
        {yang2019evaluating}
\bibfield{author}{\bibinfo{person}{Fan Yang}, \bibinfo{person}{Mengnan Du}, {and} \bibinfo{person}{Xia Hu}.} \bibinfo{year}{2019}\natexlab{}.
\newblock \showarticletitle{Evaluating explanation without ground truth in interpretable machine learning}.
\newblock \bibinfo{journal}{\emph{arXiv preprint arXiv:1907.06831}} (\bibinfo{year}{2019}).
\newblock


\bibitem[Yang et~al\mbox{.}(2020)]%
        {yang2020how}
\bibfield{author}{\bibinfo{person}{Fumeng Yang}, \bibinfo{person}{Zhuanyi Huang}, \bibinfo{person}{Jean Scholtz}, {and} \bibinfo{person}{Dustin~L. Arendt}.} \bibinfo{year}{2020}\natexlab{}.
\newblock \showarticletitle{How Do Visual Explanations Foster End Users' Appropriate Trust in Machine Learning?}. In \bibinfo{booktitle}{\emph{Proceedings of the 25th International Conference on Intelligent User Interfaces}}. \bibinfo{publisher}{Association for Computing Machinery}, \bibinfo{pages}{189–201}.
\newblock
\showISBNx{9781450371186}
\urldef\tempurl%
\url{https://doi.org/10.1145/3377325.3377480}
\showDOI{\tempurl}


\bibitem[Yang et~al\mbox{.}(2017a)]%
        {yang2017scalable}
\bibfield{author}{\bibinfo{person}{Hongyu Yang}, \bibinfo{person}{Cynthia Rudin}, {and} \bibinfo{person}{Margo Seltzer}.} \bibinfo{year}{2017}\natexlab{a}.
\newblock \showarticletitle{Scalable Bayesian rule lists}. In \bibinfo{booktitle}{\emph{International conference on machine learning}} (Sydney, NSW, Australia). \bibinfo{publisher}{JMLR.org}, \bibinfo{pages}{3921--3930}.
\newblock
\urldef\tempurl%
\url{https://doi.org/10.5555/3305890.3306086}
\showDOI{\tempurl}


\bibitem[Yang et~al\mbox{.}(2017b)]%
        {Yang:2017:EEU:2909824.3020230}
\bibfield{author}{\bibinfo{person}{X.~Jessie Yang}, \bibinfo{person}{Vaibhav~V. Unhelkar}, \bibinfo{person}{Kevin Li}, {and} \bibinfo{person}{Julie~A. Shah}.} \bibinfo{year}{2017}\natexlab{b}.
\newblock \showarticletitle{Evaluating Effects of User Experience and System Transparency on Trust in Automation}. In \bibinfo{booktitle}{\emph{Proceedings of the 2017 ACM/IEEE International Conference on Human-Robot Interaction}}. \bibinfo{pages}{408--416}.
\newblock
\urldef\tempurl%
\url{https://doi.org/10.1145/2909824.3020230}
\showDOI{\tempurl}


\bibitem[Yoon et~al\mbox{.}(2020)]%
        {yoon2020data}
\bibfield{author}{\bibinfo{person}{Jinsung Yoon}, \bibinfo{person}{Sercan Arik}, {and} \bibinfo{person}{Tomas Pfister}.} \bibinfo{year}{2020}\natexlab{}.
\newblock \showarticletitle{Data valuation using reinforcement learning}. In \bibinfo{booktitle}{\emph{International Conference on Machine Learning}}. \bibinfo{publisher}{JMLR.org}, \bibinfo{pages}{10842--10851}.
\newblock
\urldef\tempurl%
\url{https://doi.org/10.5555/3524938.3525943}
\showDOI{\tempurl}


\bibitem[Yu et~al\mbox{.}(2019)]%
        {Yu2019doi}
\bibfield{author}{\bibinfo{person}{Kun Yu}, \bibinfo{person}{Shlomo Berkovsky}, \bibinfo{person}{Ronnie Taib}, \bibinfo{person}{Jianlong Zhou}, {and} \bibinfo{person}{Fang Chen}.} \bibinfo{year}{2019}\natexlab{}.
\newblock \showarticletitle{Do {I} Trust My Machine Teammate? {An} Investigation from Perception to Decision}. In \bibinfo{booktitle}{\emph{Proceedings of the 24th International Conference on Intelligent User Interfaces}}. \bibinfo{pages}{460–468}.
\newblock
\urldef\tempurl%
\url{https://doi.org/10.1145/3301275.3302277}
\showDOI{\tempurl}


\bibitem[Zhang et~al\mbox{.}(2023)]%
        {zhang-etal-2023-fc}
\bibfield{author}{\bibinfo{person}{Lingxi Zhang}, \bibinfo{person}{Jing Zhang}, \bibinfo{person}{Yanling Wang}, \bibinfo{person}{Shulin Cao}, \bibinfo{person}{Xinmei Huang}, \bibinfo{person}{Cuiping Li}, \bibinfo{person}{Hong Chen}, {and} \bibinfo{person}{Juanzi Li}.} \bibinfo{year}{2023}\natexlab{}.
\newblock \showarticletitle{{FC}-{KBQA}: A Fine-to-Coarse Composition Framework for Knowledge Base Question Answering}. In \bibinfo{booktitle}{\emph{Proceedings of the 61st Annual Meeting of the Association for Computational Linguistics (Volume 1: Long Papers)}} (Toronto, Canada). \bibinfo{publisher}{Association for Computational Linguistics}, \bibinfo{pages}{1002--1017}.
\newblock
\urldef\tempurl%
\url{https://doi.org/10.18653/v1/2023.acl-long.57}
\showDOI{\tempurl}


\bibitem[Zhang et~al\mbox{.}(2022)]%
        {zhang-etal-2022-history}
\bibfield{author}{\bibinfo{person}{Tong Zhang}, \bibinfo{person}{Yong Liu}, \bibinfo{person}{Boyang Li}, \bibinfo{person}{Zhiwei Zeng}, \bibinfo{person}{Pengwei Wang}, \bibinfo{person}{Yuan You}, \bibinfo{person}{Chunyan Miao}, {and} \bibinfo{person}{Lizhen Cui}.} \bibinfo{year}{2022}\natexlab{}.
\newblock \showarticletitle{History-Aware Hierarchical Transformer for Multi-session Open-domain Dialogue System}. In \bibinfo{booktitle}{\emph{Findings of the Association for Computational Linguistics: EMNLP 2022}} (Abu Dhabi, United Arab Emirates). \bibinfo{publisher}{Association for Computational Linguistics}, \bibinfo{pages}{3395--3407}.
\newblock
\urldef\tempurl%
\url{https://doi.org/10.18653/v1/2022.findings-emnlp.247}
\showDOI{\tempurl}


\bibitem[Zhang et~al\mbox{.}(2020)]%
        {zhang2020effect}
\bibfield{author}{\bibinfo{person}{Yunfeng Zhang}, \bibinfo{person}{Q.~Vera Liao}, {and} \bibinfo{person}{Rachel K.~E. Bellamy}.} \bibinfo{year}{2020}\natexlab{}.
\newblock \showarticletitle{Effect of Confidence and Explanation on Accuracy and Trust Calibration in {AI}-Assisted Decision Making}. In \bibinfo{booktitle}{\emph{Proceedings of the 2020 Conference on Fairness, Accountability, and Transparency}} (Barcelona, Spain). \bibinfo{publisher}{Association for Computing Machinery}, \bibinfo{pages}{295–305}.
\newblock
\showISBNx{9781450369367}
\urldef\tempurl%
\url{https://doi.org/10.1145/3351095.3372852}
\showDOI{\tempurl}


\bibitem[Zhao et~al\mbox{.}(2023)]%
        {LLMSurvey}
\bibfield{author}{\bibinfo{person}{Wayne~Xin Zhao}, \bibinfo{person}{Kun Zhou}, \bibinfo{person}{Junyi Li}, \bibinfo{person}{Tianyi Tang}, \bibinfo{person}{Xiaolei Wang}, \bibinfo{person}{Yupeng Hou}, \bibinfo{person}{Yingqian Min}, \bibinfo{person}{Beichen Zhang}, \bibinfo{person}{Junjie Zhang}, \bibinfo{person}{Zican Dong}, \bibinfo{person}{Yifan Du}, \bibinfo{person}{Chen Yang}, \bibinfo{person}{Yushuo Chen}, \bibinfo{person}{Zhipeng Chen}, \bibinfo{person}{Jinhao Jiang}, \bibinfo{person}{Ruiyang Ren}, \bibinfo{person}{Yifan Li}, \bibinfo{person}{Xinyu Tang}, \bibinfo{person}{Zikang Liu}, \bibinfo{person}{Peiyu Liu}, \bibinfo{person}{Jian-Yun Nie}, {and} \bibinfo{person}{Ji-Rong Wen}.} \bibinfo{year}{2023}\natexlab{}.
\newblock \showarticletitle{A Survey of Large Language Models}.
\newblock \bibinfo{journal}{\emph{arXiv preprint arXiv:2303.18223}} (\bibinfo{year}{2023}).
\newblock


\bibitem[Zheng et~al\mbox{.}(2023)]%
        {zheng_designing_2023}
\bibfield{author}{\bibinfo{person}{Yifan Zheng}, \bibinfo{person}{Brigid Rowell}, \bibinfo{person}{Qiyuan Chen}, \bibinfo{person}{Jin~Yong Kim}, \bibinfo{person}{Raed~Al Kontar}, \bibinfo{person}{X.~Jessie Yang}, {and} \bibinfo{person}{Corey~A. Lester}.} \bibinfo{year}{2023}\natexlab{}.
\newblock \showarticletitle{Designing Human-Centered {AI} to Prevent Medication Dispensing Errors: {Focus} Group Study With Pharmacists}.
\newblock \bibinfo{journal}{\emph{JMIR Formative Research}} \bibinfo{volume}{7}, \bibinfo{number}{1} (\bibinfo{year}{2023}), \bibinfo{pages}{e51921}.
\newblock
\urldef\tempurl%
\url{https://doi.org/10.2196/51921}
\showDOI{\tempurl}


\bibitem[Zhu et~al\mbox{.}(2024)]%
        {zhu2023minigpt}
\bibfield{author}{\bibinfo{person}{Deyao Zhu}, \bibinfo{person}{Jun Chen}, \bibinfo{person}{Xiaoqian Shen}, \bibinfo{person}{Xiang Li}, {and} \bibinfo{person}{Mohamed Elhoseiny}.} \bibinfo{year}{2024}\natexlab{}.
\newblock \showarticletitle{Mini{GPT}-4: Enhancing Vision-Language Understanding with Advanced Large Language Models}. In \bibinfo{booktitle}{\emph{The Twelfth International Conference on Learning Representations}}. \bibinfo{publisher}{openreview}.
\newblock
\urldef\tempurl%
\url{https://doi.org/forum?id=1tZbq88f27}
\showDOI{\tempurl}


\end{thebibliography}

\clearpage
\appendix

\section{Objective Evaluation – Decision-Making of classification models.}
\label{appendix_section}
The evaluation aims to objectively evaluate participants' understanding of static explanations. We ask participants to choose, from three classification models, the most accurate on unobserved test data. 
All three classification models make the same decisions on 5 images, accompanied by static explanations from the same explanation method. The only differences between the three networks lie in their explanations. Hence, to make the correct selection, the participants must understand the explanations.

Figure \ref{q1_grad_cam} presents the full set of images listed in the objective evaluation for Grad-CAM, while Figure \ref{q1_lime} showcases the same for LIME.

\begin{figure*}[!b]
    \centering
    \includegraphics[page=1,width=0.8\textwidth]{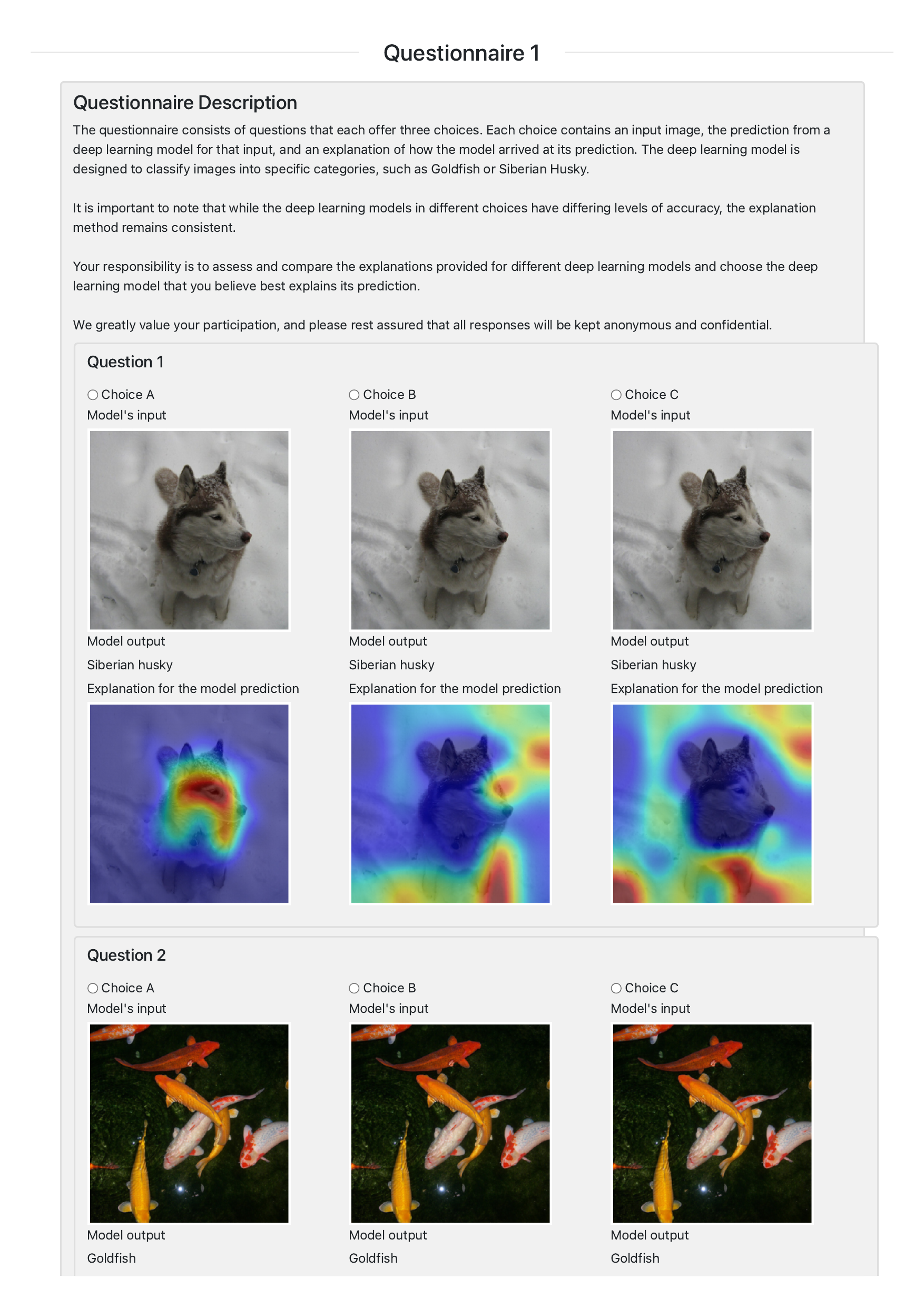}
    \caption{Objective evaluation questions used for Grad-CAM.}
\end{figure*}
\clearpage
\begin{figure*}[htb]\ContinuedFloat
    \centering
    \includegraphics[page=2,width=0.8\textwidth]{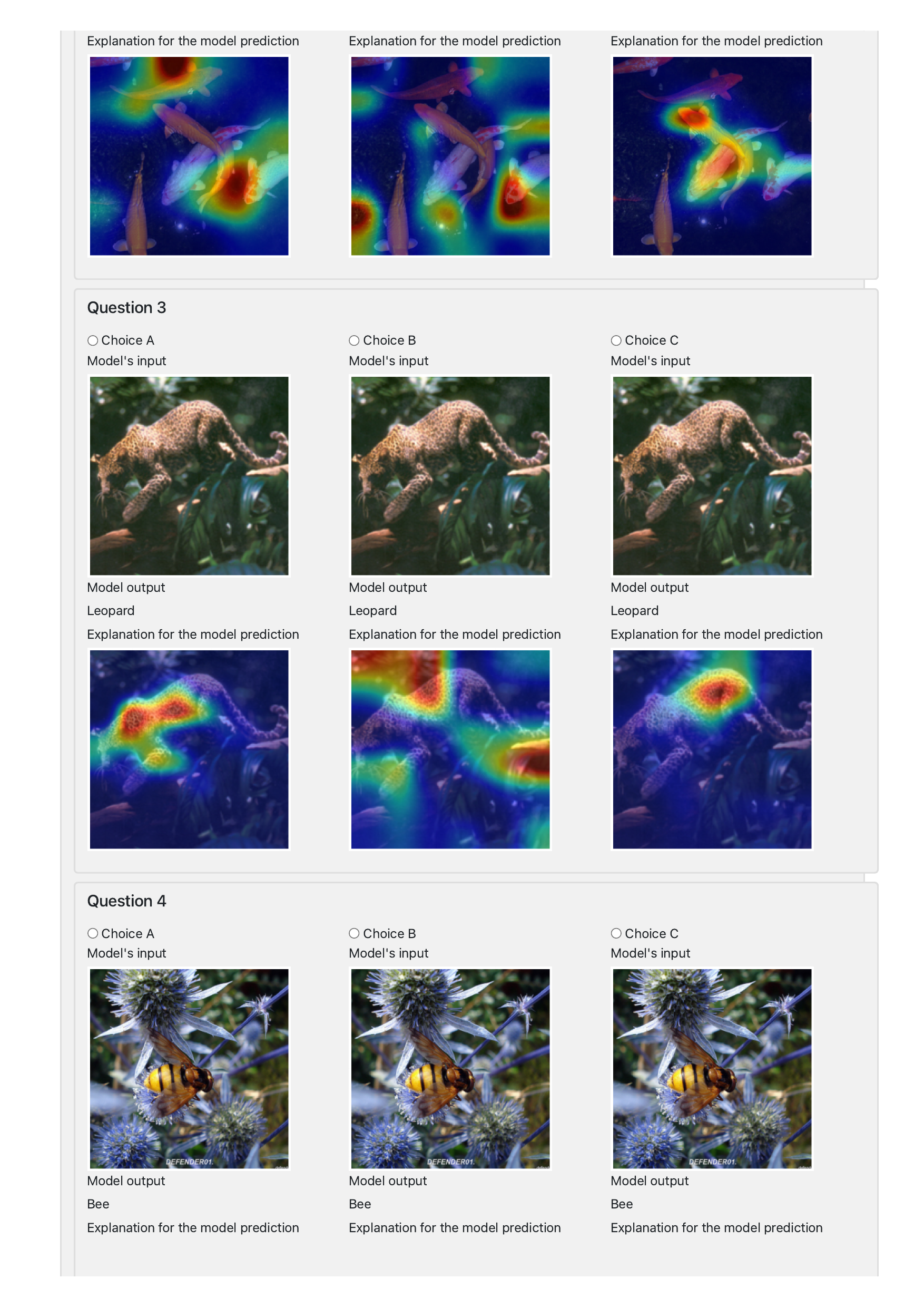}
    \caption{Objective evaluation questions used for Grad-CAM.}
\end{figure*}
\clearpage
\begin{figure*}[htb]\ContinuedFloat
    \centering
    \includegraphics[page=3,width=0.8\textwidth]{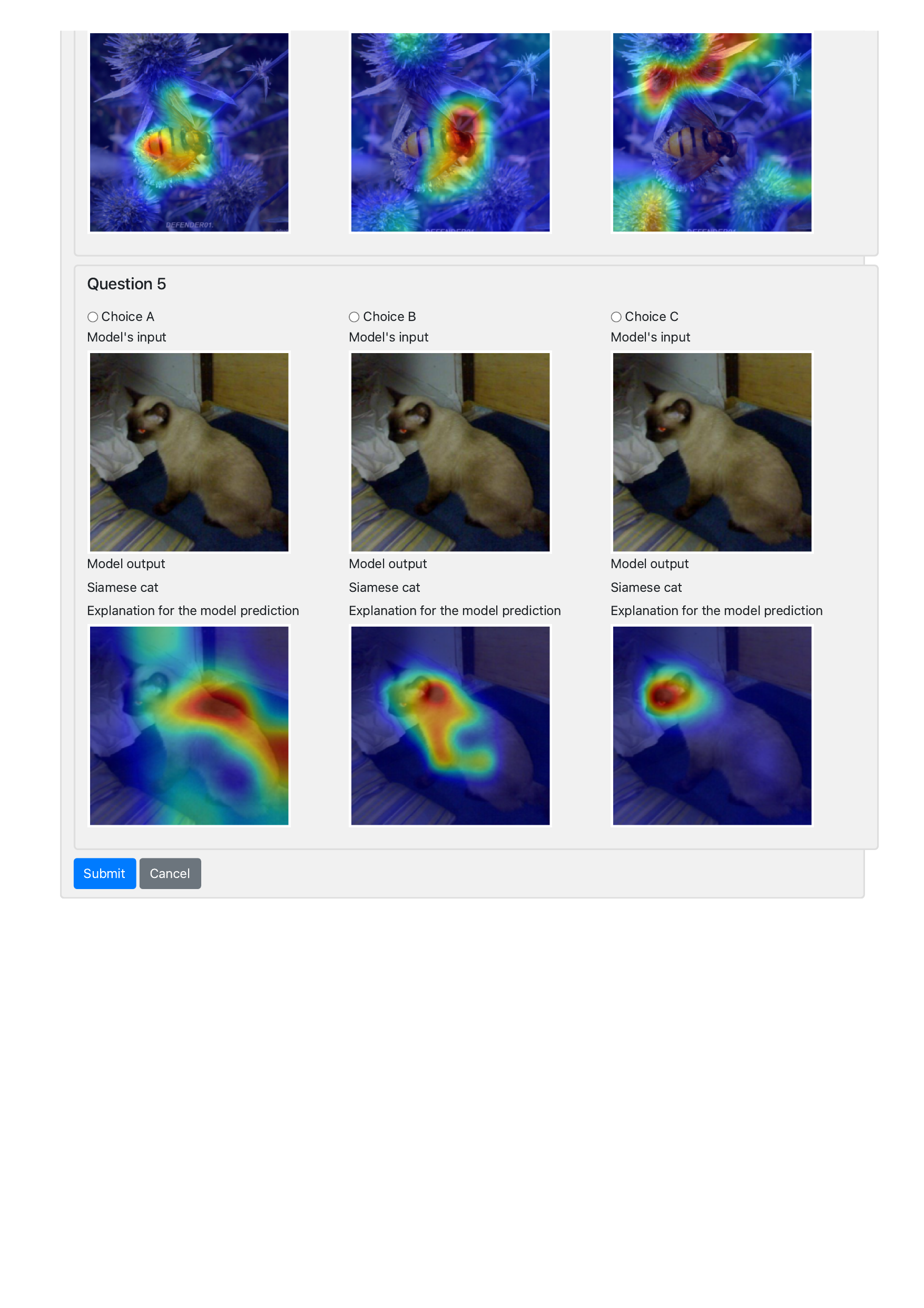}
    \caption{Objective evaluation questions used for Grad-CAM.}
    \label{q1_grad_cam}
\end{figure*}

\begin{figure*}[!b]
    \centering
    \includegraphics[page=1,width=0.8\textwidth]{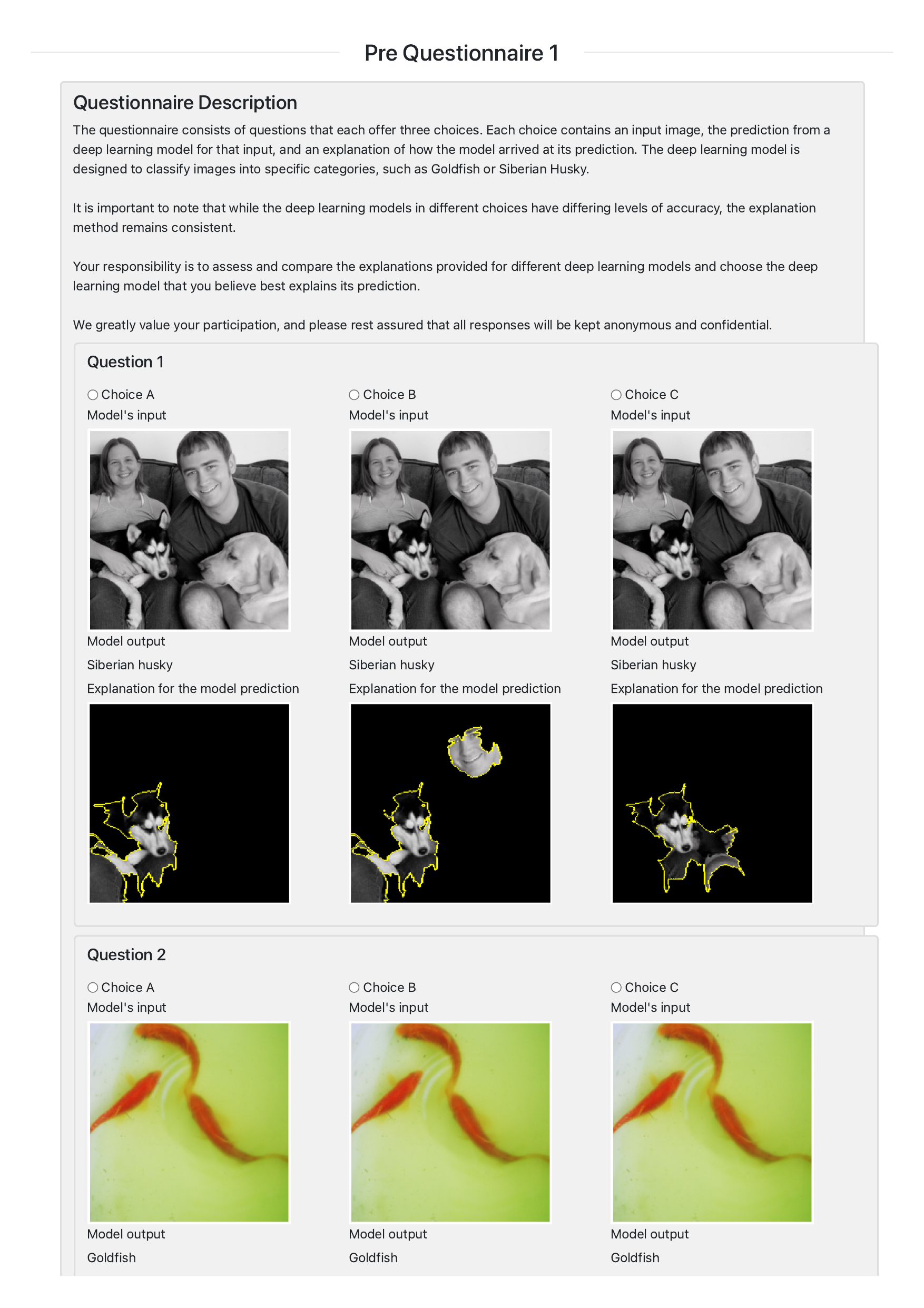}
    \caption{Objective evaluation questions used for LIME.}
\end{figure*}
\clearpage
\begin{figure*}[htb]\ContinuedFloat
    \centering
    \includegraphics[page=2,width=0.8\textwidth]{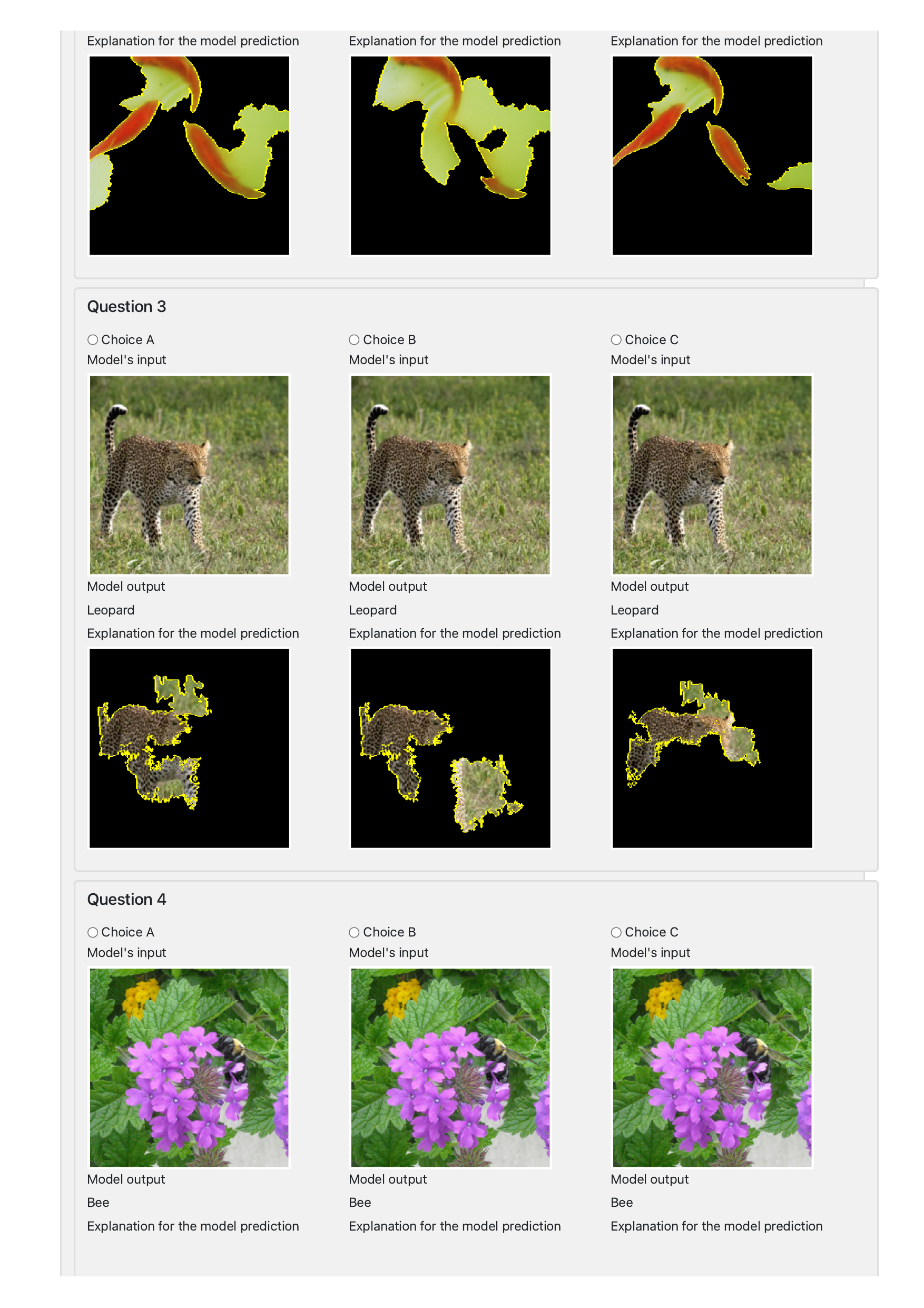}
    \caption{Objective evaluation questions used for LIME.}
\end{figure*}
\clearpage
\begin{figure*}[htb]\ContinuedFloat
    \centering
    \includegraphics[page=3,width=0.8\textwidth]{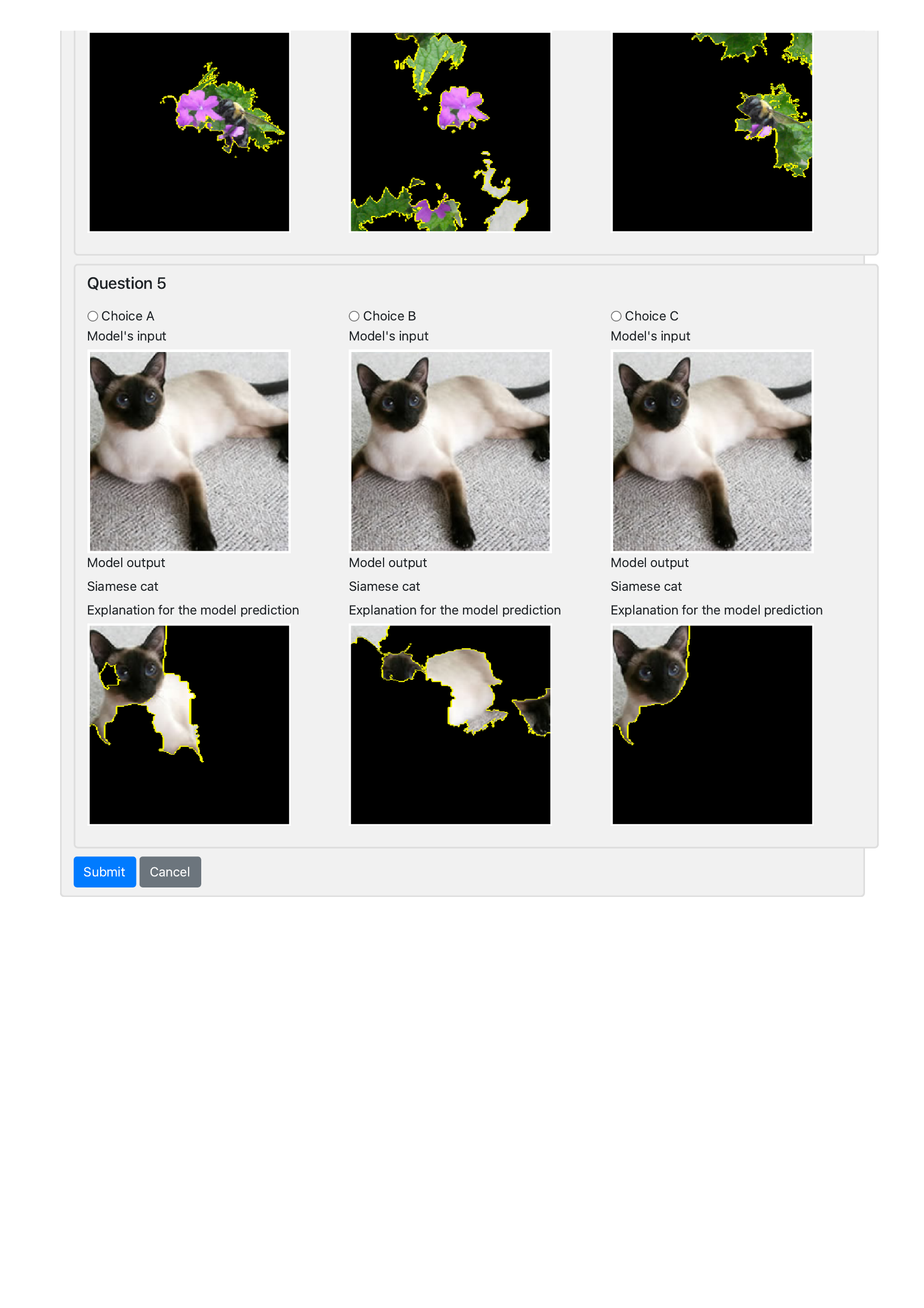}
    \caption{Objective evaluation questions used for LIME.}
    \label{q1_lime}
\end{figure*}

\end{document}